\documentclass[fleqn,usenatbib]{mnras}

\usepackage{newtxtext,newtxmath}

\usepackage[T1]{fontenc}

\DeclareRobustCommand{\VAN}[3]{#2}
\let\VANthebibliography\thebibliography
\def\thebibliography{\DeclareRobustCommand{\VAN}[3]{##3}\VANthebibliography}

\usepackage{graphicx}	
\usepackage{amsmath}	
\usepackage{hyperref}

\title[PG Quasars with the uGMRT]{Probing the origin of low-frequency radio emission in PG quasars with the uGMRT - I}

\author[S. Silpa et al.]{Silpa S.,$^{1}$\thanks{E$-$mail: silpa@ncra.tifr.res.in}
P. Kharb,$^{1}$
L. C. Ho,$^{2,3}$
C. H. Ishwara-Chandra,$^{1}$
M. E. Jarvis,$^{4,5,6}$
C. Harrison$^{7}$\\
$^{1}$National Centre for Radio Astrophysics $-$ Tata Institute of Fundamental Research, S. P. Pune University Campus, Ganeshkhind, Pune 411007, India \\
$^{2}$Kavli Institute for Astronomy and Astrophysics, Peking University, Beijing 100871, China\\
$^{3}$Department of Astronomy, School of Physics, Peking University, Beijing 100871, China\\
$^{4}$Max-Planck Institut f\"ur Astrophysik, Karl-Schwarzschild-Str. 1, 85748 Garching, Germany\\
$^{5}$European Southern Observatory, Karl-Schwarzschild-Str. 2, 85748 Garching, Germany\\
$^{6}$Ludwig Maximilian Universit\"at, Professor-Huber-Platz 2, 80539 Munich, Germany\\
$^{7}$School of Mathematics, Statistics and Physics, Newcastle University, Newcastle upon Tyne, NE1 7RU, UK}

\pubyear{2020}

\begin{document}
\label{firstpage}
\pagerange{\pageref{firstpage}$-$$-$\pageref{lastpage}}
\maketitle

\begin{abstract}
We present the results from 685 MHz observations with the upgraded Giant Metrewave Radio Telescope (uGMRT) of 22 quasars belonging to the Palomar-Green (PG) quasar sample. Only four sources reveal
extended radio structures on $\sim$10-30 kpc scales, while the rest are largely a combination of a radio core unresolved at the uGMRT resolution of $\sim$3-5$\arcsec$, surrounded by diffuse emission on few kpc to $\sim$10 kpc scales. A few sources reveal signatures of barely resolved jets and lobes in their spectral index images that are created using the uGMRT 685 MHz data and similar resolution GHz-frequency data from the Very Large Array. On the basis of their position on the radio-IR correlation as well as the spectral index images, we find that the radio emission in the two radio-loud (RL) quasars and nearly one-third of the radio-quiet (RQ) quasars is active galactic nucleus (AGN) dominated whereas the remaining sources appear to have significant contributions from stellar-related processes along with the AGN. While the two RL sources exhibit inverted spectral index in their cores, the RQ sources exhibit a range of spectral indices varying from flat to steep ($-0.1\gtrsim\alpha_{R}\gtrsim-1.1$) indicating the presence of unresolved jets/lobes or winds. Except for a significant correlation between the 685~MHz radio luminosity and the Eddington ratio, we do not find strong correlations between other 685 MHz radio properties and black hole (BH) properties in the RQ PG sources. This lack of correlations could be explained by the contribution of stellar-related emission, or radio emission from previous AGN activity episodes which may not be related to the current BH activity state. 
\end{abstract}
\begin{keywords}
techniques: interferometric $-$$-$ quasars: general $-$$-$ radio continuum: general 
\end{keywords}



\section{Introduction}
\label{Sec1}
Active galactic nuclei (AGN) are the energetic centres of galaxies that are believed to be powered by the gravitational energy released from accretion of matter on to supermassive black holes \citep[$10^5-10^{10}~M_\odot$; see review by][]{Rees84}. Bipolar outflows are launched perpendicular to the black hole-accretion disc systems. It is unclear why only 10-20\% of AGN launch powerful radio outflows 
\citep[e.g.,][]{Kellermann89, Urry03} that extend to hundreds of kiloparsecs or megaparsecs (the so called `radio-loud', RL AGN), while the vast majority of AGN launch small (parsec-scale) or weak outflows that are typically $\lesssim10$ kpc in extent and many a times bubble-/wind-like (the `radio-quiet', RQ AGN). The radio-loudness parameter (R) given by \citet{Kellermann89}, which is the ratio of radio flux density at 5~GHz to optical flux density in the optical B-band, is often used to separate these two classes: RL AGN have R$>$10 and RQ AGN have R$\le$10. The influence of host galaxy on the type of outflows observed is also unclear since RL AGN are mostly hosted by elliptical galaxies whereas RQ AGN reside in both spiral and elliptical galaxies. Differences in black hole (BH) masses, spins and accretion rates have often been suggested to explain this RL-RQ divide \citep{BlandfordRees74,WilsonColbert95,Tchekhovskoy10}. It is proposed in literature that high-mass BHs reside in RL sources whereas low-mass BHs reside in RQ sources \citep{Dunlop03,McLurejarvis04}. Also, radio-loudness decreases with increasing Eddington ratio \citep{Ho02, Sikora07}. However, while some authors suggest that the radio-loudness distribution is bimodal \citep{Kellermann89}, or at least not unimodal \citep{KratzerRichards15}, others suggest that the distribution is continuous \citep{White00,LaFranca10}. This RL-RQ dichotomy can be explained by understanding the different physical mechanisms contributing to radio emission in these two classes of AGN, namely RL and RQ.

In RL AGN, accretion on to BHs produces collimated relativistic jets that emit synchrotron radiation. The jets are generally accepted to be produced by the electromagnetic extraction of rotational energy from spinning BHs \citep{BlandfordZnajek77}. However, the origin of radio emission in RQ AGN is poorly understood \citep{Panessa19}. Possible origins include: (1) thermal free-free emission from H~II regions around young, massive stars like O- and B-type stars \citep{Terzian65}; (2) non-thermal synchrotron emission associated with starburst superwinds \citep{Condon13, Kellermann16}; (3) coronal emission due to magnetic heating of the corona, presumably located above the accretion disc \citep{LaorBehar08, RaginskiLaor16}; (4) optically thin free-free emission from a slow, dense accretion disc wind \citep{BlundellKuncic07} or from the torus \citep{Carilli19} \citep[however see][]{Ho99, LalHo10}; (5) compact or low-powered radio jets on $<\sim$kpc scales \citep{Falcke00, Jarvis19, Kharb19}; (6) a combination of both AGN and starburst-driven winds \citep{Cecil01,IrwinSaikia03,HotaSaikia06}; (7) AGN-driven winds, which can be launched and accelerated by the following mechanisms: (i) thermal-driving \citep{Begelman83,Woods96,Mizumoto19}; (ii) radiative-driving \citep{ProgaKallman04,ZakamskaGreene14,Nims15}; (iii) magnetic-driving \citep{BlandfordPayne82, KoniglKartje94, Everett05, Fukumura10}.

In this paper, we present a 685~MHz study of 22 Palomar-Green (PG) quasars using the upgraded Giant Metrewave Radio Telescope (uGMRT) to address the question of the origin of radio emission in RQ AGN. Our aim with the uGMRT was to observe and understand the origin of low-frequency diffuse and extended emission in the PG quasars at resolutions that are intermediate between the previous Very large Array (VLA) A and D-array 5 GHz observations \citep[0.5$\arcsec$ and 18$\arcsec$ respectively;][]{Kellermann89,Kellermann94}. The 685 MHz observations is the best compromise between resolution and sensitivity with the uGMRT. These observations provide an angular resolution of $\sim$4$\arcsec$ which corresponds to spatial extents of few kpc to tens of kpc in the PG sources. As these sources have extensive GHz frequency radio data available from the VLA, we aim to obtain information about spectral indices and determine the spectral ages of outflows using uGMRT and VLA data. Our goal is to also constrain the AGN duty cycle by detecting emission from multiple activity episodes of the AGN that is presumably steep spectrum and therefore bright at low frequencies. We have also obtained polarimetric data with the uGMRT at 685 MHz, which we will present in a forthcoming paper (Silpa et al. 2021, in preparation).

The paper is organized in the following manner: Section~\ref{Sec2} describes the PG sample and Section~\ref{Sec3} describes the uGMRT data reduction and analysis. We present our results in Section~\ref{Sec4} and conclusions in Section~\ref{Sec5}. Throughout this paper, we have assumed $\Lambda$ cold dark matter cosmology with $H_0$ = 73~km~s$^{-1}$~ Mpc$^{-1}$, $\Omega_{m}$ = 0.27 and $\Omega_{v}$ = 0.73. The spectral index $\alpha_{R}$ is defined such that flux density at frequency $\nu$, $S_\nu\propto\nu^{\alpha_{R}}$.

\section{The Sample}
\label{Sec2}
The PG catalogue of $\sim$1800 UV-excess (i.e., U$-$B$<-$0.44) objects is derived from a survey covering an area of $\sim$10,714~deg$^2$ in absolute galactic latitudes above 30$^{\circ}$, using 266 double U and B exposures of the Palomar 18 inch Schmidt telescope \citep{Green86}. The Palomar Bright Quasar Survey (BQS), which is a subset of the larger PG survey, included only those objects which fulfilled the
(1) morphological criteria: dominant star-like appearance;
(2) spectroscopic criteria: presence of broad emission lines.
The BQS sample consisted of 114 objects, which included 92 Quasars with $M_B$ < $-$23 and 22 Seyferts or low-luminosity quasars with $M_B$ > $-$23. The PG quasar sample, which is our sample of interest here, is composed of objects from BQS with $z<0.5$. These are 87 sources comprising of quasars and Seyfert type 1 galaxies \citep{BorosonGreen92}. Nearly 80\% of this sample is RQ, while $\sim$20\% is RL.

The PG quasar sample is one of the most well-studied samples of low-redshift AGN. The wealth of data that exists for this sample includes: {(a)} accurate BH masses measured from reverberation mapping \citep{Kaspi00} and single-epoch spectroscopy data \citep{VestergaardPeterson06}; {(b)} host galaxy morphologies and galaxy bulge/disk decompositions from HST imaging data \citep{Kim08, Kim17}; {(c)} complete broad-band spectral energy distributions (SEDs) and accurate bolometric luminosities from data covering radio to hard X-rays \citep{Shang11}; {(d)} observations of dust \citep{Petric15, Shangguan18} and gas \citep{Evans06, Shangguan20} properties, enabling examination of the interstellar medium of the host galaxies; and {(e)} IR data determining torus properties \citep{Zhuang18} and star formation rates (SFRs) \citep[Xie et al. 2020, submitted]{Shi14}. Despite the immense wealth of data, there is lack of high-resolution, high-sensitivity low-frequency radio data for this sample.

In this paper, we present results from 685 MHz observations of 22 (out of 87) PG quasars with the uGMRT. This sub-sample was a result of the limited approved time to carry out an initial pilot study with the uGMRT. We therefore chose targets that had recently been observed with the Atacama Large Millimeter/submillimeter Array (ALMA) by \citet{Shangguan20}, with the aim of looking at the low-frequency radio emission in tandem with CO(2-1) molecular emission. The results of this work will be presented in a future paper. 

\begin{table*}
\begin{center}
\caption{uGMRT observation details and VLA archival data details of PG quasars.}
\label{Table1}
{\begin{tabular}{cccccccccc}
\hline
Quasar & R.A. & Decl. & Redshift & Distance & uGMRT & uGMRT & VLA Project & VLA & VLA \\
& J2000.0 & J2000.0 & & & Phase cal & Phase cal & ID & telecope & frequency\\ 
& & & & & (IAU name) & flux density & & configuration & \\ 
& (hh mm ss.s)  & (+/-dd mm ss.ss) & & (Mpc)& & (Jy) & &  &(GHz) \\
\hline
PG 0003+199    &   00 06 19.5 & +20 12 10.49    &    0.02578     &   103         &   2330+110    & 1.28   & AL0418   & C & 8    \\
PG 0007+106    &   00 10 31.0 & +10 58 29.50    &    0.08934     &   387         &   2330+110    & 1.28    & AF0350   & C & 5   \\
PG 0049+171    &   00 51 54.7 & +17 25 58.50    &    0.06400     &   271         &   0119+321    & 3.47    & AL0418   & C & 8   \\
PG 0050+124    &   00 53 34.9 & +12 41 36.20    &    0.05890     &   248         &   0059+001    & 3.51    & AK0298   & C & 5    \\
PG 0923+129    &   09 26 03.2 & +12 44 03.63    &    0.02915     &   127         &   0842+185    & 1.07    & AK0298   & C & 5    \\
PG 0934+013    &   09 37 01.0 & +01 05 43.48    &    0.05034     &   220         &   0943-083    & 3.84   & AF0327   & C & 8   \\
PG 1011-040    &   10 14 20.6 & $-$04 18 40.30    &    0.05831     &   256         &   0943-083    & 3.84    & AS0455   & AB & 8   \\
PG 1119+120    &   11 21 47.1 & +11 44 18.26    &    0.05020     &   219         &   1120+143    & 3.66   & AK0298   & C & 5 \\
PG 1211+143    &   12 14 17.6 & +14 03 13.10    &    0.08090     &   358         &   1254+116    & 0.89    & AK0298   & C & 5    \\
PG 1229+204    &   12 32 03.6 & +20 09 29.21    &    0.06301     &   276         &   1254+116    & 0.89    & AK0298   & C & 5    \\
PG 1244+026    &   12 46 35.2 & +02 22 08.79    &    0.04818     &   210         &   1254+116    & 0.89    & AF0327   & C & 8   \\
PG 1310$-$108    &   13 13 05.7 & $-$11 07 42.40    &    0.03427     &   149         &   1248-199    & 7.84    & AK0298   & C & 5    \\
PG 1341+258    &   13 43 56.7 & +25 38 47.69    &    0.08656     &   383         &   1330+251    & 10.99   & -        &  - & -                  \\
PG 1351+236    &   13 54 06.4 & +23 25 49.09    &    0.05500     &   239         &   1330+251    & 10.99   & AF0327   & C & 8   \\
PG 1404+226    &   14 06 21.8 & +22 23 46.22    &    0.09800     &   437         &   1330+251    & 10.99   & AK0298   & C & 5    \\
PG 1426+015    &   14 29 06.5 & +01 17 06.48    &    0.08657     &   383         &   1445+099    & 2.34    & AK0298   & C & 5    \\
PG 1448+273    &   14 51 08.7 & +27 09 26.92    &    0.06500     &   283         &   3C 286       & 22.51   & AK0298   & C & 5   \\
PG 1501+106    &   15 04 01.2 & +10 26 16.15    &    0.03642     &   157         &   1445+099    & 2.34    & AK0298   & C & 5    \\
PG 2130+099    &   21 32 27.8 & +10 08 19.46    &    0.06298     &   266         &   2148+069    & 2.81    & AL0418   & C & 8    \\
PG 2209+184    &   22 11 53.8 & +18 41 49.86    &    0.07000     &   298         &   2148+069    & 2.81    & AL0418   & C & 8    \\
PG 2214+139    &   22 17 12.2 & +14 14 20.89    &    0.06576     &   279         &   2148+069    & 2.81   & AG0173   & B & 1.5    \\
PG 2304+042    &   23 07 02.9 & +04 32 57.22    &    0.04200     &   173         &   2330+110    & 1.28    & AL0418   & C & 8   \\
\hline
\end{tabular}}
\end{center}
\end{table*}

\section{Observations and data reduction}
\label{Sec3}
\subsection{uGMRT observations}
\label{Sec3.1}
The uGMRT observations of the PG quasar sample at 685 MHz (Band 4: 550-850 MHz) were carried out on 2018 November 23 and 2018 December 7 (project code 35\_042). The uGMRT has a seamless frequency coverage from 50 to 1500 MHz and instantaneous bandwidth of 400 MHz. The data sets obtained on both dates were individually reduced and analysed using the standard procedures in $\tt{CASA}$\footnote{Common Astronomy Software Applications; \citet{Shaw07}}. 27 antennas with 400 MHz bandwidth and 4096 channels were used; 3 antennas were not working during the observations. The details of these observations are given in Table~\ref{Table1}. 
Each source was observed for at least two scans of $\sim$20 min each for better {\it uv} coverage. The flux density calibrators (3C 48 and 3C 286) were observed for 5 min at the beginning and end of the observations and the phase calibrators (listed in Table~\ref{Table1}) were observed for 5 min once every $\sim$20 min of the on-source time.

\subsection{Data reduction and imaging}
\label{Sec3.2}
The LTA data files produced by GMRT were first converted into FITS format by $\tt{LISTSCAN}$ and $\tt{GVFITS}$ utilities. The usable bandwidth of uGMRT Band 4 data is 560-810 MHz only, since the sensitivity falls significantly beyond this range. Therefore, we omitted the bad channels and non-working antennas in the log file produced from $\tt{LISTSCAN}$ so as to produce a FITS file of the frequency range 560-810 MHz. The FITS file was then converted to Measurement Set (MS) using the task $\tt{IMPORTUVFITS}$ in $\tt{CASA}$. Subsequent flagging of data affected by radio frequency interference (RFI) at various stages were carried out using $\tt{TFCROP}$ and $\tt{RFLAG}$ modes of task $\tt{FLAGDATA}$ in $\tt{CASA}$. Manual flagging of all calibrators and individual target sources using $\tt{PLOTMS}$ in $\tt{CASA}$ significantly improved the data. 
\\
The calibration steps involved were: setting flux density values for amplitude calibrators, initial phase calibration to correct for phase variations with time in the bandpass, delay calibration to correct for antenna-based delays, bandpass calibration to correct for bandpass shapes and gain calibration to correct for complex antenna gains. The flux density values for amplitude calibrators were set using the task {\tt SETJY} in {\tt {CASA}} and were determined for phase calibrators using the task {\tt FLUXSCALE} in {\tt {CASA}}. The output of {\tt SETJY} was 32.18 Jy for 3C 48 and 22.51 Jy for 3C 286 (using Perley-Butler 2017 scale). The output of $\tt{FLUXSCALE}$ is listed in Table~\ref{Table1}. The calibrator solutions were then applied to the multisource target data set. The $\tt{CASA}$ task $\tt{SPLIT}$ was used to create visibility subsets for individual targets from the original data set while averaging the spectral channels such that the bandwidth smearing effects were negligible. Each target data set was finally imaged using the wide-band, wide-field imaging algorithm of $\tt{TCLEAN}$ task in $\tt{CASA}$. Faceted imaging was combined with multiterm imaging to achieve this ($\tt{specmode=mfs}$; $\tt{deconvolver=mtmfs}$; $\tt{gridder=widefiled}$). Several iterations of phase-only self-calibration and minimum two iterations of amplitude and phase self-calibration until convergence were performed to improve the sensitivity of the images\footnote{A Python script for reducing the total and polarized intensity uGMRT data is available at https://sites.google.com/view/silpasasikumar/}. The final synthesized beam of the images is typically $\sim$3-5$\arcsec$. The largest angular scale that can be measured through these observations is $\sim$25$\arcmin$. 

\subsection{Data reduction and imaging for PG 2209+184}
\label{Sec3.3}
The total intensity image for PG 2209+184 obtained using the standard procedures in $\tt{CASA}$ showed a lot of imaging artefacts. It neither improved with manual flagging nor with sub-band imaging. Therefore, we analysed this source using $\tt{SPAM}$\footnote{Source Peeling and Atmospheric Modeling; \citet{Intema14}}. $\tt{SPAM}$ cannot be directly run on GWB (GMRT Wideband Backend) data with a bandwidth of 400 MHz. Therefore, first, the GWB data were divided into sub-bands of data with bandwidth of 32 MHz, which is equal to the historical GSB (GMRT Software Backend) bandwidth. This was done while converting the LTA file to FITS file by editing the log file produced from $\tt{LISTSCAN}$. Then, $\tt{SPAM}$ was run on each sub-band data set independently, resulting in individual images from each data set. Several methods were tried to combine these images to yield one final image but none gave reasonable results, hence the sub-band UV data sets were combined. $\tt{SPAM}$ runs in 2 parts - (i) basic calibration; (ii) imaging. The first part (`pre\_calibrate\_targets') yields a file with extension CAL.RR.UVFITS, which is a calibrated stokes `RR' visibility data. We had multiple CAL.RR.UVFITS files from the different sub-band data sets, which were first converted to MS files by running the $\tt{CASA}$ task $\tt{IMPORTGMRT}$. Then, the multiple MS files were combined into a single MS file (taking two at a time) by running the $\tt{CASA}$ task $\tt{CONCAT}$. Finally, the combined MS file was imaged using $\tt{TCLEAN}$ task in $\tt{CASA}$. Also, before running $\tt{SPAM}$, three of the python scripts had to be edited to match with the GWB frequency ranges and calibrator scan lengths. An intriguing observation was the presence of extended radio emission on few tens of kpc-scales in the final image, which was not visible above noise in the individual images obtained from each sub-band data set. 

\subsection{Radio spectral index images}
\label{Sec3.4}
The spectral index images and spectral index noise images were made using the task $\tt{COMB}$ in $\tt{AIPS}$\footnote{Astronomical Image Processing System; \citet{Wells85}}. For this, we first obtained similar resolution ($\sim$3-5$\arcsec$) GHz-frequency raw visibility data from VLA Data Archive for 20 PG sources, excluding PG 0007+106 and PG 1341+258, which are discussed individually below (see Table~\ref{Table1} for VLA data details). The raw visibilities were first calibrated by running the task $\tt{VLARUN}$ in $\tt{AIPS}$, followed by imaging using the task $\tt{IMAGR}$ in $\tt{AIPS}$. Since, no flux calibrator was observed in the Projects - AK0298 and AG0173, we manually set it during the task $\tt{VLARUN}$ as the brightest phase calibrator from the respective projects. Also, their flux densities, taken from the VLA calibrator list\footnote{https://science.nrao.edu/facilities/vla/observing/callist} at respective frequencies, were set for the parameter $\tt{FLUX}$ in the task $\tt{VLARUN}$. The flux densities of the phase calibrators IAU 1252+119 (Project ID - AF0327), IAU 0922+005 (Project ID - AF0327) and IAU 2121+053 (Project ID - AL0418) obtained from the task $\tt{GETJY}$ in $\tt{VLARUN}$ did not agree with the values given in the VLA calibrator list at respective frequencies. However, these calibrators are suggested to be variable in the literature  \citep{StannardBentley77,Fanti81a,Hughes92,Tornikoski05}. 

For PG 0007+106, similar resolution calibrated visibility data were obtained from VLA Image Archive and imaged using the task $\tt{IMAGR}$ in $\tt{AIPS}$. Self-calibration failed for most of the faint sources (flux densities $\lesssim$ 2 mJy) and was therefore not carried out. The images of the sources at both frequencies (uGMRT and VLA) were then convolved with identical circular beams while setting the $\tt{ROBUST}$ parameter in $\tt{IMAGR}$ appropriately ($\tt{ROBUST}$=+5 to $-$5 for the higher and lower resolution images, respectively). Also, the uGMRT and VLA images were made spatially coincident using the task $\tt{OGEOM}$ in $\tt{AIPS}$. Finally, the spectral index images were created using the task $\tt{COMB}$ in $\tt{AIPS}$. This task gives a `two-point' spectral index for each pixel using the relation: $\alpha$ = log($\mathrm S_1$/$\mathrm S_2$)/log($\nu_1/\nu_2$) for flux densities $\mathrm S_1$ and $\mathrm S_2$ at frequencies $\nu_1$ and $\nu_2$. In the present work, $\nu_1$ = 685 MHz and $\nu_2$ is given in Table~\ref{Table1} for individual sources. Regions with flux densities below three times the {\it rms} noise were blanked while using $\tt{COMB}$. The visibility data for PG 1341+258 were available neither in the VLA Data Archive nor in the VLA Image Archive; therefore we could not produce its spectral index image. However, we obtained its spectral index value using the total flux density at 5 GHz from \citet{Kellermann94} and at 685 MHz from the current work. 

\section{Results}
\label{Sec4}
We present the results of our multifrequency radio study in the following subsections. Section~\ref{Sec4.1} presents the global properties of the PG sample. Section~\ref{Sec4.2} discusses the radio morphological features of individual sources. The later sections aim to determine the origin of radio emission based on spectral index analysis (Section~\ref{Sec4.3}), radio$-$IR correlation (Section~\ref{Sec4.4}) and correlations between radio and accretion properties (Section~\ref{Sec4.5}).

\begin{figure}
\centering{
\includegraphics[width=8cm]{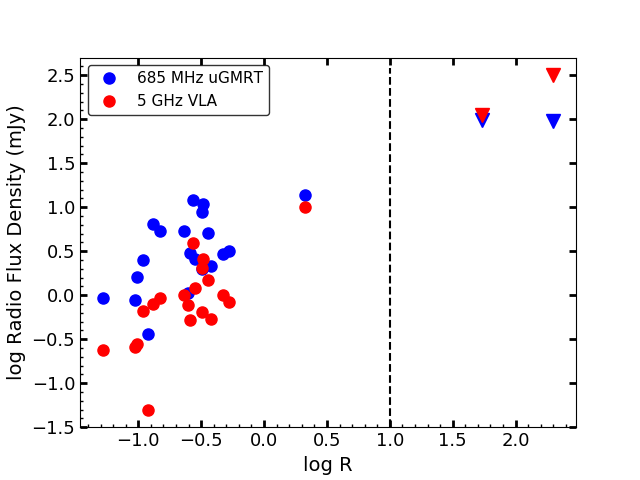}}
\caption{\small 685 MHz uGMRT and 5 GHz VLA flux density as a function of radio-loudness parameter for the PG sources. The values of 5 GHz flux density and radio-loudness are taken from \citet{Kellermann94}. The black dashed vertical line divides the RL and RQ PG sources.}
\label{Fig1}
\end{figure}

\subsection{Global properties of the PG sample
\label{Sec4.1}}The 685 MHz uGMRT contour images and radio-optical overlay images created using $\it{grizy}$\footnote{$\it{grizy}$ is a set of five broadband filters used in PanSTARRS1 observations. The mean wavelengths of g, r, i, z and y filters are 4866 {\AA}, 6215 {\AA}, 7545 {\AA}, 8679 {\AA} and 9633 {\AA} respectively (https://outerspace.stsci.edu/display/PANSTARRS/PS1+Filter+properties).}-color composite optical images taken from PanSTARRS \citep{Flewelling16} for individual sources (except PG 0007+106, which is provided in Figure~\ref{PG 0007}) are presented in APPENDIX B. Small positional offsets (typically $<$1$\arcsec$) between the 685~MHz uGMRT images and the optical images in seven sources were corrected using the HEASARC Fv software\footnote{https://heasarc.gsfc.nasa.gov/docs/software/ftools/fv/}. The 685 MHz peak flux density of individual sources were estimated using the task $\tt{TVMAXF}$ in $\tt{AIPS}$. It varies around a few mJy~beam$^{-1}$ to $\sim$100 mJy~beam$^{-1}$ and the {\it rms} noise is around $\sim$20$-$40 $\mu$Jy~beam$^{-1}$. The peak signal-to-noise ratio varies from $\sim$10 to $\sim$1700. The 685 MHz total flux density of all the sources, except PG 0007+106 and PG 1229+204, were estimated using the task $\tt{JMFIT}$ in $\tt{AIPS}$. PG 0007+106 and PG 1229+204 have multiple components, therefore their total flux densities were estimated using the task $\tt{TVSTAT}$ in $\tt{AIPS}$. The 685 MHz total and peak flux densities for individual sources are presented in Table~\ref{Table2}. The errors in the total flux densities were estimated from the calibration uncertainties (which is typically $\sim$10\% for uGMRT) and {\it rms} noise of the images.

We present a discussion on the optical morphology of the PG quasar host galaxies in APPENDIX A, based on their classification in NED\footnote{NASA/IPAC Extraglactic Database (https://ned.ipac.caltech.edu/)} (see Table~\ref{Table2}), HST image analysis \citep{Kim08, Kim17} and visual inspection of the PanSTARRS images. We find that the host galaxies of 50\% of the sources show signatures of galaxy interactions, such as tidal tails, rings, or other isophotal distortions (see APPENDIX A and the radio-optical overlay images in APPENDIX B). However, this disturbed fraction could be a lower limit due to lack of detailed analyses/high-resolution imaging across the sample (for e.g., the HST image analysis is available for only five sources in our sample). We also find that the radio emission in $\sim$30\% of these disturbed sources have AGN origin whereas the remaining $\sim$70\% have both AGN and stellar origin (see Sections~\ref{Sec4.3} and ~\ref{Sec4.4}). 

Figure~\ref{Fig1} clearly depicts the RL$-$RQ dichotomy in the PG sample using the new 685 MHz uGMRT data. The radio-loudness parameter values and 5 GHz VLA flux densities are taken from \citet{Kellermann94}. We see that the 685 MHz flux densities are higher than the 5 GHz flux densities. This suggests that the uGMRT 685 MHz observations have detected copious amounts of low frequency diffuse emission, missed out by the previous 5 GHz VLA observations.

\subsection{Radio Morphology of Individual Sources}
\label{Sec4.2}
In this Section, we briefly discuss the radio morphology of the PG sources. Nearly 80\% sources are only a combination of a radio core unresolved at the uGMRT resolution of $\sim$3-5$\arcsec$ surrounded by diffuse emission, together which we define as the `uGMRT core' or equivalently the `core' for our sources. The angular sizes of the cores are estimated using the Gaussian-fitting task $\tt{JMFIT}$ in $\tt{AIPS}$. We present the redshift-dependent linear sizes of the radio cores of individual sources in Table~\ref{Table2}. The error returned by the task $\tt{JMFIT}$ are taken as the error in the core size. The uGMRT cores of the PG sources are spatially resolved/barely resolved at $\sim$2-10 kpc scales. PG 0050+124 and PG 1501+106 show double-sided diffuse emission on few kpc-scales. PG 2214+139 is the only unresolved source in our sample. Only four sources - PG 0007+106, PG 1229+204, PG 1244+026 and PG 1448+273 (discussed individually below) show extended radio emission on $\sim$10-30 kpc scales. These sources extend beyond their host galaxies while the rest are largely confined to within their host galaxies.

\begin{figure*}
\centerline{
\includegraphics[width=7cm]{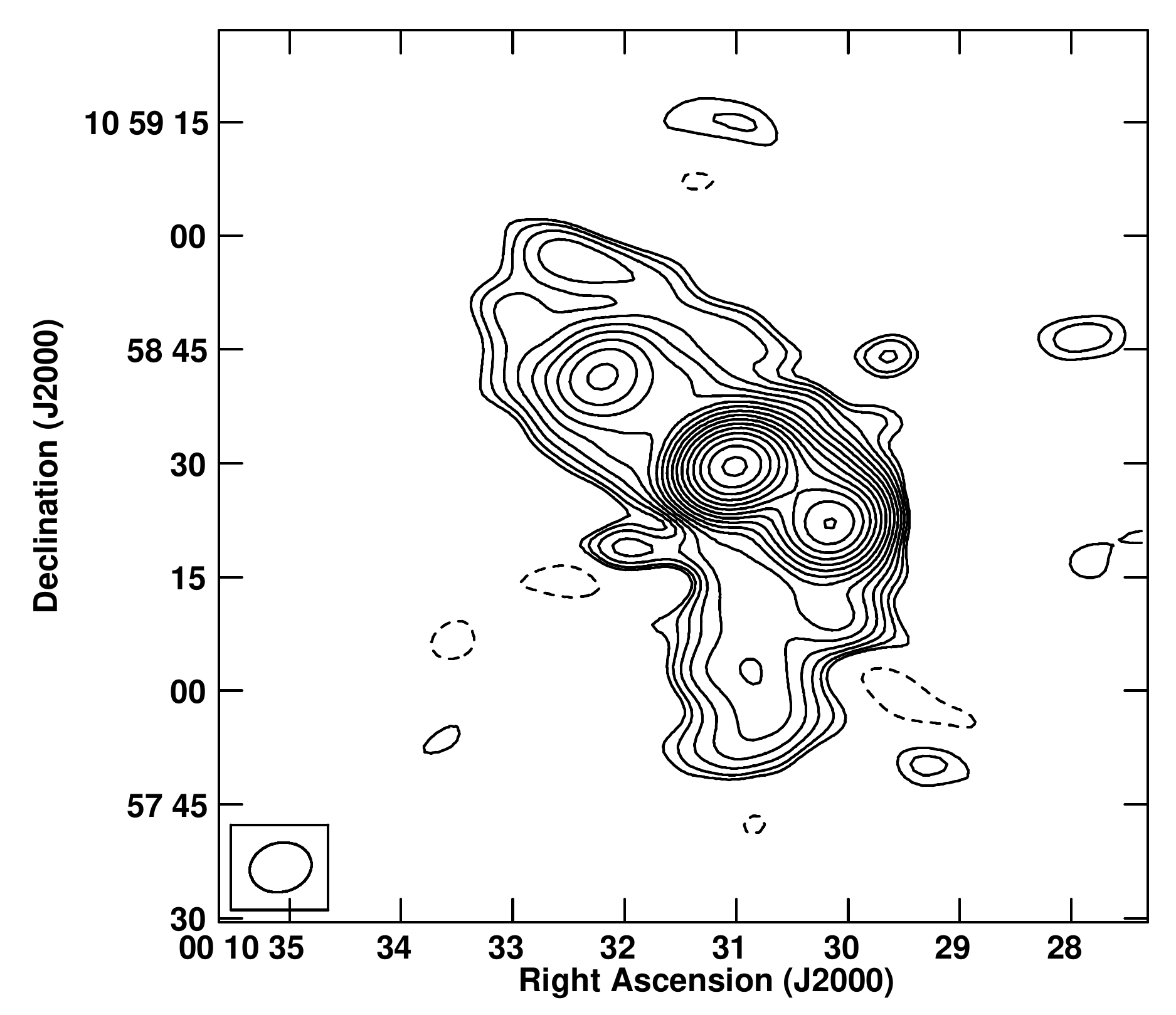}
\includegraphics[width=7cm]{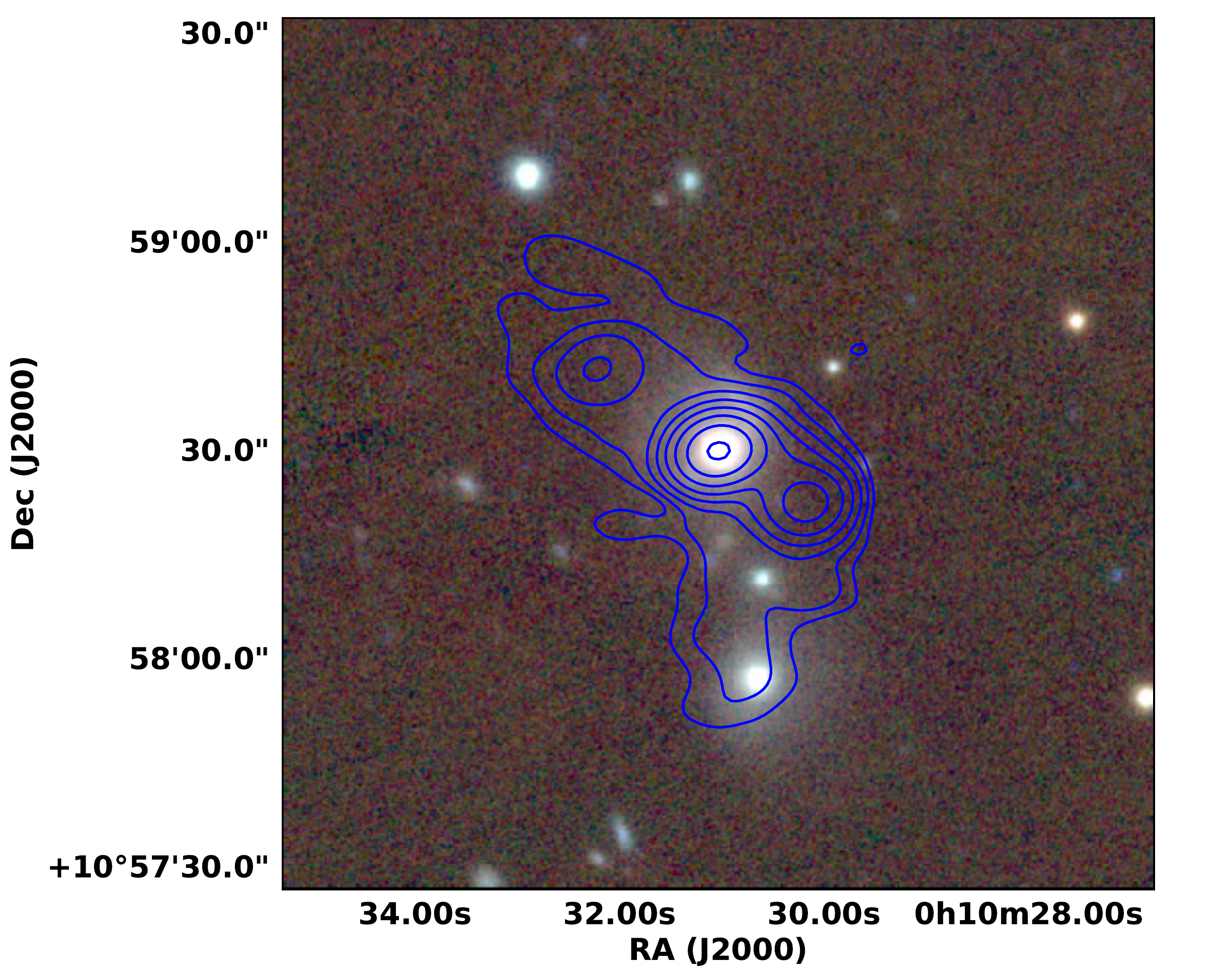}}
\caption{Left: 685 MHz uGMRT total intensity contour image of RL PG quasar PG 0007+106. The peak contour flux is 59 mJy beam$^{-1}$ and the contour levels are 0.21 $\times$ (-1, 1, 1.4, 2, 2.8, 4, 5.6, 8, 11.20, 16, 23, 32, 45, 64, 90, 128, 180, 256, 362, 512) mJy beam$^{-1}$. Right: 685 MHz total intensity contours in blue superimposed on PanSTARRS $\it{grizy}$-color composite optical image of PG 0007+106. The peak contour flux is 59 mJy beam$^{-1}$ and the contour levels are 0.21 $\times$ (2, 4, 8, 16, 32, 64, 128, 256, 512) mJy beam$^{-1}$.}
\label{PG 0007}
\end{figure*}

The RL quasar PG 0007+106 (a.k.a. III Zw 2) shows the most extended (projected) radio structure at 685 MHz. A triple radio source is clearly detected. The core-hotspot distance is $\sim$30 kpc on either side, with the south-western radio jet/lobe being on the approaching side. A misaligned system of low-frequency diffuse emission around the triple radio source is detected, which could be due to a previous episode of AGN activity or triggered by some merger event and misaligned due to a change in the jet direction. We present its lower resolution image in Figure~\ref{PG 0007}, which was produced by {\it uv}-tapering the visibility data at 15 k$\lambda$, in order to fully capture the diffuse extended emission. The 685 MHz peak flux density, {\it rms} noise and radio core size for PG 0007+106 reported in Table~\ref{Table2} were estimated from the original resolution image (i.e. before {\it uv}-tapering). Also, the 685 MHz total flux density reported in Table~\ref{Table2} and used in the calculation of luminosities, pertain to the hotspot-core-hotspot structure in the original resolution image and does not include the contribution of the misaligned lobe emission. Moreover, we find that the 685 MHz total flux density inclusive of the misaligned lobe emission estimated from the {\it uv}-tapered image of PG 0007+106 using the task $\tt{TVSTAT}$ in $\tt{AIPS}$ is $\sim$120 mJy.

Similar tapering of the visibility data in compact PG sources failed to detect any large-scale lobe-like diffuse emission. This suggests that our current uGMRT observations are not sensitive enough to detect $\sim$100 kpc-scale emission of the type detected in the X-shaped radio galaxy NGC 326 with  Low-Frequency Array (LOFAR) \citep{Hardcastle19}. Alternately, there is no large-scale remnant `relic' emission in the compact PG sources.

PG 1229+204 appears as a V-shaped triple radio source \citep[hotspot-core-hotspot; see e.g.,][]{Kharb12} comprising of a $\sim$kpc-scale extended core in the NE-SW direction and a $\sim$0.1 mJy component ($\sim$3$\sigma$ feature) at a distance of $\sim$15 kpc in the NW direction. This V-shaped morphology is possibly arising from the projection effects at small viewing angles. The radio morphology of PG 1244+026 resembles kpc-scale or 8-shaped radio structures in Seyfert galaxies \citep{Gallimore06}. A core with extended emission to the west in PG 1448+273 resembles one-sided core-jet structure. 

\begin{figure}
\centering{
\includegraphics[width=8cm]{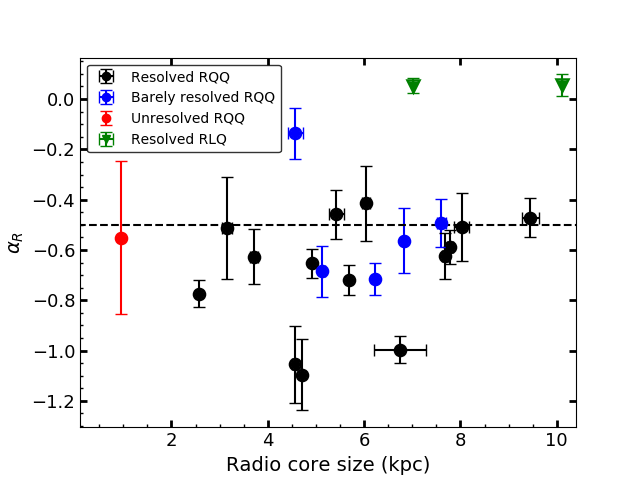}}
\caption{\small The mean core spectral index as a function of the uGMRT core size for the PG sources. The black dashed horizontal line marks $\alpha_{R}$=$-$0.5. We consider the sources with $\alpha_{R}$ $\geq-0.5$ to have flat spectrum cores and the sources with $\alpha_{R}$ $<-0.5$ to have steep spectrum cores.}
\label{Fig2}
\end{figure}

Figure~\ref{Fig2} reveals no correlation between the radio core size and the mean core spectral index, denoted by $\alpha_{R}$ (see Section~\ref{Sec4.3}) for the RQ sources; the Spearman rank correlation co-efficient $r_s$=0.256 and probability $\mathrm{p}_s$=0.277. This could suggest that the uGMRT cores of the sources with steep spectra contain unresolved/barely resolved jets and lobes. The flat spectra in the cores could either suggest the presence of synchrotron self-absorbed jet bases or free-free emission from accretion disk winds on scales of 100s of parsecs or from the tori.

\begin{table*}
\begin{center}
\caption{Summary of radio and optical properties of PG quasars}
\label{Table2}
{\begin{tabular}{ccccccccc}
\hline
Quasar & Host galaxy{$^{*}$} & Radio-loudness  & Classification & 685 MHz peak  & 685 MHz total & $\it{rms}$ noise & Radio core & Radio core \\ & morphology & R & by radio & flux density & flux density  & ($\mu$Jy~beam$^{-1}$)& resolved at &size\\ &(from NED){$^{**}$} & & & (mJy~beam$^{-1}$) & S$_\mathrm{685}$ (mJy) & & 685 MHz?  &(kpc)\\
\hline
PG 0003+199    &   S0/a         &    0.27  & RQ      &    11     & 12$\pm$1       & 31.68     &   Yes       &   2.578$\pm$0.006   \\
PG 0007+106    &   -            &    200   & RL      &    56     & 95$\pm$6       & 33.47     &   Yes       &   10.107$\pm$0.004  \\
PG 0049+171    &   -            &    0.32  & RQ      &    1.8    & 2.0$\pm$0.2    & 64.24     &   Yes       &   5.4$\pm$0.2       \\
PG 0050+124    &   Sa;Sbrst     &    0.33  & RQ      &    9.5    & 11.0$\pm$1.0   & 34.54     &   Yes       &   4.71$\pm$0.01     \\
PG 0923+129    &   S0?          &    2.1   & RQ      &    12     & 14$\pm$1       & 27.77     &   Yes       &   5.69$\pm$0.01     \\
PG 0934+013    &   SBab         &    0.38  & RQ      &    1.2    & 2.1$\pm$0.1    & 23.26     &   Yes       &   8.0$\pm$0.2       \\
PG 1011-040    &   SB(r)b: pec  &    0.097 & RQ      &    1.2    & 1.6$\pm$0.1    & 22.66     &   Yes       &   9.4$\pm$0.2       \\
PG 1119+120    &   SABa         &    0.15  & RQ      &    3.2    & 5.4$\pm$0.3    & 29.11     &   Yes       &   7.67$\pm$0.06     \\
PG 1211+143    &   -            &    0.13  & RQ      &    4.7    & 6.5$\pm$0.5    & 155.3     &   Yes       &   7.78$\pm$0.06     \\
PG 1229+204    &   SB0          &    0.11  & RQ      &    1.9    & 2.5$\pm$0.2    & 28.71     &   Yes       &   6.03$\pm$0.08     \\
PG 1244+026    &   E/S0         &    0.53  & RQ      &    2.9    & 3.1$\pm$0.3    & 26.48     &   Maybe     &   5.14$\pm$0.04     \\
PG 1310$-$108    &   -            &    0.095 & RQ      &    0.74   & 0.87$\pm$0.08  & 30.64     &   Yes       &   3.2$\pm$0.1       \\
PG 1341+258    &   -            &    0.12  & RQ      &    0.30   & 0.36$\pm$0.04  & 23.85     &   Yes       &   6.7$\pm$0.5       \\
PG 1351+236    &   -            &    0.26  & RQ      &    2.1    & 3.0$\pm$0.2    & 23.41     &   Yes       &   4.56$\pm$0.05     \\
PG 1404+226    &   -            &    0.47  & RQ      &    2.8    & 2.9$\pm$0.3    & 20.18     &   Maybe     &   6.83$\pm$0.04     \\
PG 1426+015    &   E/S0         &    0.28  & RQ      &    2.0    & 2.6$\pm$0.2    & 24.26     &   Maybe     &   7.61$\pm$0.09     \\
PG 1448+273    &   -            &    0.23  & RQ      &    5.3    & 5.3$\pm$0.5    & 26.97     &   Maybe     &   6.23$\pm$0.03     \\
PG 1501+106    &   E            &    0.36  & RQ      &    3.7    & 5.1$\pm$0.4    & 204.3     &   Yes       &   3.72$\pm$0.08     \\
PG 2130+099    &   (R)Sa        &    0.32  & RQ      &    7.5    & 8.7$\pm$0.7    & 24.44     &   Yes       &   4.93$\pm$0.01     \\
PG 2209+184    &   -            &    54    & RL      &    93     & 97$\pm$9       & 95.48     &   Yes       &   7.010$\pm$0.004   \\
PG 2214+139    &   -            &    0.052 & RQ      &    0.94   & 0.93$\pm$0.10  & 25.46     &   No        &   1.0         \\
PG 2304+042    &   -            &    0.25  & RQ      &    1.0   & 1.1$\pm$0.1  & 32.09     &   Maybe     &   4.6$\pm$0.2       \\
\hline
\end{tabular}}
$^{*}$ E=elliptical; S0=lenticular; S=spiral; Sbrst=starburst; pec=peculiar; ?=uncertainty; a,b represent the tightness of the spiral arms; A,B represent bars; (r) represents inner ring; (R) represents outer ring. $^{**}$ see APPENDIX A for the references
\end{center}
\end{table*}

\begin{table*}
\begin{center}
\caption{Summary of physical properties of PG quasars}
\label{Table3}
{\begin{tabular}{ccccccccccc}
\hline
Quasar & $\alpha_{R}^{a}$ & L$_\mathrm{685}^{b}$  & L$_\mathrm{1400}${$^{c}$} & L$_\mathrm{5000}${$^{d}$} & L$_\mathrm{IR,host}${$^{e}$}  & q$_\mathrm{IR}$ & log L$_\mathrm{X}^{f}$ & log M$_\mathrm{BH}${$^{g}$} & Edd. {$^{h}$} \\ &  & &  &  &  & &  &  & ratio\\ &  & (10$^{22}$ W Hz$^\mathrm{-1}$)  & (10$^{22}$ W Hz$^\mathrm{-1}$) & (10$^{28}$ erg s$^\mathrm{-1}$ Hz$^\mathrm{-1}$) & (10$^\mathrm{11}$ L$_{\sun}$) & & (erg s$^\mathrm{-1}$) & (M$_{\sun}$) & \\
\hline
PG 0003+199    &   $-$0.77$\pm$0.05    &    1.5$\pm$0.1   &  0.88$\pm$0.09   &   3.4$\pm$0.5    & 0.02   &   1.36$\pm$0.04   &   43.80  &    7.52   &   0.21   \\
PG 0007+106    &   +0.06$\pm$0.04         &    155$\pm$9    &  161$\pm$11     &   1732$\pm$174 & 0.48   &   0.48 $\pm$0.03    &   44.49  &    8.87   &   0.05   \\
PG 0049+171    &   $-$0.46$\pm$0.10    &    1.7$\pm$0.2   &  1.2$\pm$0.1    &   7$\pm$1    & 0.03   &   1.39$\pm$0.05   &   44.05  &    8.45   &   0.02   \\
PG 0050+124    &   $-$1.1$\pm$0.1      &    8.1$\pm$0.7   &  3.8$\pm$0.5        &   10$\pm$3        & 2.93   &   2.89$\pm$0.06     &   44.18  &    7.57   &   2.26   \\
PG 0923+129    &   $-$0.72$\pm$0.06    &    2.6$\pm$0.2   &  1.6$\pm$0.2    &   6.5$\pm$0.9        & 0.22   &   2.15$\pm$0.04   &   43.69  &    7.52   &   0.12   \\
PG 0934+013    &   $-$0.5$\pm$0.1    &    1.21$\pm$0.07 &  0.85$\pm$0.09  &   4$\pm$1    & 0.25   &   2.48$\pm$0.05   &   43.73  &    7.15   &   0.24   \\
PG 1011-040    &   $-$0.47$\pm$0.08    &    1.24$\pm$0.10 &  0.89$\pm$0.08  &   4.9$\pm$0.8    & 0.28   &   2.50$\pm$0.04   &   42.60  &    7.43   &   0.27   \\
PG 1119+120    &   $-$0.62$\pm$0.09      &    3.0$\pm$0.2   &  2.0$\pm$0.2    &   9$\pm$2        & 0.36   &   2.27$\pm$0.04   &   43.48  &    7.58   &   0.32   \\
PG 1211+143    &   $-$0.59$\pm$0.07      &    9.7$\pm$0.7   &  6.4$\pm$0.6    &   31$\pm$5       & 0.00   &   ...             &   44.54  &    8.10   &   0.46   \\
PG 1229+204    &   $-$0.4$\pm$0.1     &    2.2$\pm$0.2   &  1.6$\pm$0.2    &   10$\pm$3        & 0.26   &   2.20$\pm$0.06   &   44.00  &    8.26   &   0.08   \\
PG 1244+026    &   $-$0.7$\pm$0.1    &    1.6$\pm$0.1   &  1.0$\pm$0.1    &   4.3$\pm$0.9    & 0.16   &   2.20$\pm$0.05   &   43.58  &    6.62   &   1.15   \\
PG 1310$-$108    &   $-$0.5$\pm$0.2    &    0.23$\pm$0.02 &  0.16$\pm$0.03  &   0.8$\pm$0.3    & 0.04   &   2.41$\pm$0.07   &   43.30  &    7.99   &   0.04   \\
PG 1341+258    &   $-$1.00$\pm$0.05    &    0.64$\pm$0.07 &  0.32$\pm$0.04  &   0.9$\pm$0.1    & 0.22   &   2.85$\pm$0.05   &   43.83  &    8.15   &   0.06   \\
PG 1351+236    &   $-$1.0$\pm$0.2    &    2.0$\pm$0.1   &  1.0$\pm$0.1  &   2.6$\pm$0.8    & 0.50   &   2.71$\pm$0.06   &   43.11  &    8.67   &   0.006  \\
PG 1404+226    &   $-$0.6$\pm$0.1    &    6.4$\pm$0.6   &  4.3$\pm$0.6    &   21$\pm$6       & 0.21   &   1.69$\pm$0.06   &   43.77  &    7.01   &   0.95   \\
PG 1426+015    &   $-$0.49$\pm$0.10    &    4.3$\pm$0.3   &  3.1$\pm$0.3    &   16$\pm$3       & 0.51   &   2.23$\pm$0.04   &   44.58  &    9.15   &   0.04   \\
PG 1448+273    &   $-$0.71$\pm$0.06    &    5.0$\pm$0.5   &  3.1$\pm$0.3    &   12$\pm$2       & 0.23   &   1.88$\pm$0.05   &   43.86  &    7.09   &   0.95   \\
PG 1501+106    &   $-$0.6$\pm$0.1    &    1.5$\pm$0.1   &  1.0$\pm$0.1  &   4.4$\pm$1.0    & 0.20   &   2.32$\pm$0.05   &   43.65  &    8.64   &   0.02   \\
PG 2130+099    &   $-$0.65$\pm$0.06    &    7.2$\pm$0.6   &  4.6$\pm$0.4   &   20$\pm$3    & 0.50   &   2.04$\pm$0.04   &   44.35  &    8.04   &   0.31   \\
PG 2209+184    &   +0.05$\pm$0.03         &    96$\pm$9      &  100$\pm$10     &   1064$\pm$119   & 0.16   &   0.21$\pm$0.04   &   43.94  &    8.89   &   0.006  \\
PG 2214+139    &   $-$0.6$\pm$0.3      &    0.84$\pm$0.09 &  0.6$\pm$0.1  &   3$\pm$2    & 0.10   &   2.2$\pm$0.1   &   42.63  &    8.68   &   0.03   \\
PG 2304+042    &   $-$0.1$\pm$0.1      &    0.37$\pm$0.04 &  0.34$\pm$0.04  &   2.8$\pm$0.6        & 0.001  &   0.48$\pm$0.05   &   43.89  &    8.68   &   0.003  \\
\hline
\end{tabular}}
$^a$ Mean core spectral index. $^b$ 685 MHz uGMRT k-corrected rest-frame luminosity. $^c$ 1400 MHz k-corrected rest-frame luminosity derived using equations~\ref{2} and~\ref{3}. $^d$ 5000 MHz k-corrected rest-frame luminosity derived using equations~\ref{2} and~\ref{3}. $^e$ 8-1000 $\mu$m host galaxy IR luminosity from \citet{Lyu17}. $^f$ 0.2-20 keV X-ray luminosity from \citet{LaorBehar08}. $^g$ BH mass from \citet{Shangguan18}. $^h$ Eddington ratio (L$_\mathrm{bol}$/L$_\mathrm{Edd}$) estimated using M$_\mathrm{BH}$ and L$_\mathrm{bol}$ from \citet{Lyu17}. 
\end{center}
\end{table*}

\subsection{Radio Spectral Indices: Implications for Source Structure}
\label{Sec4.3}
Both stellar and AGN processes can contribute to radio emission in active galaxies. Each has a thermal and a non-thermal component. Although it is difficult to estimate the relative contributions of thermal and non-thermal emission, it is possible to infer the same from spectral index values. Thermal free-free radio emission from HII regions or accretion disc/torus has a spectral index of $-$0.1, whereas non-thermal radio emission from starburst superwinds or AGN jets/AGN-driven winds has much steeper spectral index.

We present the spectral index images created using 685 MHz uGMRT data and similar resolution ($\sim$3-5$\arcsec$) GHz-frequency VLA data in color superimposed with 685 MHz total intensity contours for individual sources in APPENDIX B. $\alpha_{R}$ is the mean spectral index value of the uGMRT core of a source estimated from its spectral index image using the task $\tt{IMSTAT}$ or $\tt{TVSTAT}$ in $\tt{AIPS}$ and the error in $\alpha_{R}$ is the mean value from the same region of the spectral index noise image. The values of $\alpha_{R}$ for individual sources are provided in Table~\ref{Table3}. The sources with $\alpha_{R}>0$ have inverted spectrum cores. We consider the sources with $\alpha_{R}$ $\geq-0.5$ to have flat spectrum cores and the sources with $\alpha_{R}$ $<-0.5$ to have steep spectrum cores.

The cores of the two RL sources, PG 0007+106 and PG 2209+184 exhibit inverted radio spectrum. The inverted (and flat) spectrum cores in RL sources are expected to be the unresolved optically thick synchrotron self-absorbed bases of relativistic jets. 

The jet and the south-western lobe of PG 0007+106 has a steep spectrum, with hints of diffuse $\alpha_{R}<-1$ emission typically associated with `relic' emission from a previous episode of AGN activity \citep[e.g.,][]{Roettiger94, Kharb16}. The south-western lobe and the core have also been detected in the similar resolution VLA image, which is used in making the spectral index image. However, the north-eastern lobe and the misaligned lobe have not been detected in this VLA image. We suggest that their non-detection is indicative of steep spectrum lobe emission rather the lack of VLA sensitivity based on our following calculations. Using the $\it{rms}$ noise in the VLA image and the intensity of the north-eastern and the misaligned lobes in the uGMRT image, we find that the north-eastern and the misaligned lobes should have been detected with $\sim$15$\sigma$ and $\sim$3.8$\sigma$ significance respectively, if their spectrum was say $-$0.7 (which is a typical value for optically thin synchrotron emission). Assuming the VLA image was adequately sensitive, we took a 3$\sigma$ flux limit for the north-eastern and the misaligned lobes and obtained a limit on their spectral index values, which turn out to be $\sim$ $-$1.5 and $-$0.8 respectively. The former is suggestive of `relic' emission in the north-eastern lobe. \citet{Brunthaler05} also found that both south-western and north-eastern lobes have steep spectral indices of $-$0.57 to $-$1.15. Also, it is worth noting that the lobe emission is not detected in the $\sim0.5\arcsec$ 5 GHz VLA image of \citet{Kellermann94}.

In PG 2209+184, there is no evidence of jets or lobes in the uGMRT image or signs of spectral steepening in the spectral index image. It is possibly confined in dense environments, causing a decrease in expansion losses and increase in radiative losses, therefore being RL and compact with self-absorbed inverted spectrum \citep[e.g., in Cygnus A,][]{Barthel96}. PG 2209+184 appears as an unresolved core with an elongated component in the 5 GHz VLA image (resolution $\sim0.5\arcsec$) of \citet{Kellermann94}. 

The remaining 20 sources are RQ and their cores reveal a range of spectral indices from flat to steep ($-0.1\gtrsim\alpha_{R}\gtrsim-1.1$), as was also found in \citet{Chiaraluce20} for a sample of hard-X-ray-selected AGN, although at higher frequencies. PG 0049+171, PG 1011$-$040, PG 1229+204, PG 1426+015 and PG 2304+042 have flat spectrum cores. PG 0934+013 and PG 1310$-$108 are at the cusp of flat-steep division. There is no evidence of jets or lobes in the uGMRT total intensity images of these sources except PG 1229+204 (see Section~\ref{Sec4.2}). The transition of uGMRT core spectral index from flat to steep in the SE-NW direction in the spectral index image of PG 1229+204, aligns with the double radio structure (possibly a core-jet/lobe structure) in the higher resolution ($\sim0.5\arcsec$) VLA image \citep{Kellermann94}. PG 2304+042 appears as an unresolved core with an elongated component in the higher resolution ($\sim0.5\arcsec$) VLA image \citep{Kellermann94}. PG 0049+171, PG 0934+013 and PG 1011$-$040 are unresolved in their higher resolution ($\sim0.5\arcsec$) VLA images \citep{Kellermann94}. This means that the diffuse emission detected by uGMRT in these sources is being resolved out by the VLA observations. These RQ flat cores can be interpreted as being the unresolved self-absorbed bases of low-powered radio jets (on milliarcsec or sub-arcsec scales) confined within the uGMRT cores or `frustrated' radio jets inside dense environments \citep{O'Dea91}. Within errors, the flat spectrum core in PG 2304+042 may also arise due to thermal free-free emission from the accretion disc winds or from the torus or HII regions. 

\begin{figure*}
\centering{
\includegraphics[height=6.5cm]{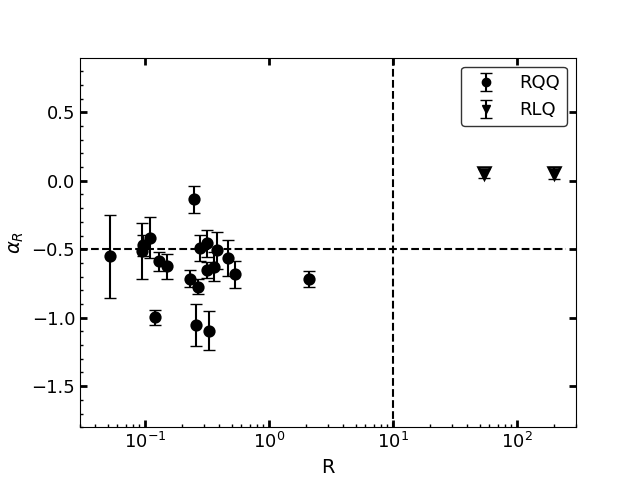}
\includegraphics[height=6.5cm]{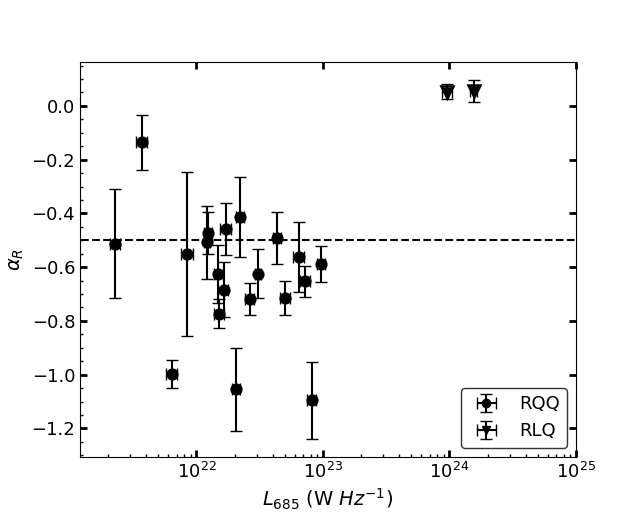}}
\caption{\small The mean core spectral index as a function of radio-loudness from \citet{Kellermann94} (left) and 685 MHz k-corrected rest-frame luminosity for the PG sources (right). The black dashed horizontal line marks $\alpha_{R}$=$-$0.5. We consider the sources with $\alpha_{R}$ $\geq-0.5$ to have flat spectrum cores and the sources with $\alpha_{R}$ $<-0.5$ to have steep spectrum cores.The black dashed vertical line in the left panel divides the RL and RQ PG sources.}
\label{Fig3}
\end{figure*}

The remaining 13 RQ sources have steep spectrum cores. These are PG 0003+199, PG 0050+124, PG 0923+129, PG 1119+120, PG 1211+143, PG 1244+026, PG 1341+258, PG 1351+236, PG 1404+226, PG 1448+273, PG 1501+106, PG 2130+099 and PG 2214+139. The steep spectrum cores are either due to unresolved/barely resolved jet/lobe emission or optically thin synchrotron emission from AGN-/starburst-driven winds \citep{Hwang18}. Some of these sources that reveal signatures of unresolved/barely resolved jets and lobes in their spectral index images are discussed individually below. We could rule out the possibility of these sources being steep Compact Symmetric Objects \citep[CSO;][]{Wilkinson94,Orienti16} as they are neither as compact ($<$1 kpc) nor as powerful (L$_\mathrm{1400}\ge10^{25}~\mathrm{W Hz}^{-1}$) as CSOs. 

The uGMRT core of PG 2130+099 is resolved into jet/lobe structure in the higher resolution ($\sim0.5\arcsec$) VLA image \citep{Kellermann94}, suggesting a jet origin for its steep spectrum, instead of AGN-/starburst-driven winds. PG 0003+199, PG 0050+124, PG 1404+226 and PG 1448+273 are unresolved in their higher resolution ($\sim0.5\arcsec$) VLA images \citep{Kellermann94}, which means that the VLA observations resolve out the diffuse emission detected by uGMRT in these sources. Moreover, the absence of jet/lobe structures at this VLA resolution in these sources could suggest that their steep spectrum is either due to jet/lobe emission on even smaller spatial scales (milliarcsec or sub-arcsec scales) or due to AGN-/starburst-driven winds. The total intensity images of PG 1244+026 and PG 1501+106 show kpc-scale extended emission (discussed in Section~\ref{Sec4.2}), which may suggest the presence of jet/lobe emission. However, this is not evident from their spectral index images, such as a gradient in their spectral index values away from the centre.

The spectral index image of PG 0050+124 suggests a barely resolved jet/lobe structure in the N-S direction. This coupled with $\alpha_{R}\sim-$1, may imply the presence of a pair of very steep lobes (may be `relic') inside the uGMRT core. The spectral index image of PG 0923+129 shows steep spectrum in the periphery of its core. This quasar is hosted by a ring galaxy. One of the suggested scenarios for the formation of ring galaxies is the merging of a very massive galaxy with a less massive one. The neighbouring galaxy in its radio-optical overlay image (see Figure~\ref{FigureB2} in APPENDIX B), seems to be its merger companion. This could suggest that the peripheral steep spectral emission arises from synchrotron radiation due to relativistic merger shocks \citep{BallardHeavens92}. A gradient in the spectral index from flat to steep towards the east in the spectral index image of PG 1119+120 could suggest a barely resolved jet/lobe structure in this direction. PG 1341+258 and PG 1351+236 have very steep spectrum of $\alpha_{R}\sim-$1, which could suggest the presence of unresolved lobe emission. An increasingly steeper spectral index towards the west coupled with a westward extension in the total intensity image of PG 1448+273 could suggest a barely resolved one-sided jet/lobe structure.  

Since one of the goals of the present study was to constrain the AGN duty cycle, we searched for signatures of diffuse `relic' emission in the PG sources. However, we did not detect as much `relic' emission as expected at low radio frequencies of the uGMRT. There are clear indications in one source - PG 0007+106, which is also supported by our calculations of the VLA sensitivity and missing flux. There are indications of the same in a few others based on their spectral index images. 

Figure~\ref{Fig3} (left-hand panel) reveals no correlation between $\alpha_{R}$ and the radio-loudness parameter (R) for the RQ sources (Spearman rank correlation coefficient $r_s$ = $-$0.283 and probability $\mathrm{p}_s$ = 0.227). Figure~\ref{Fig3} (right-hand panel) reveals no correlation between $\alpha_{R}$ and the 685 MHz k-corrected rest-frame luminosity (L$_\mathrm{685}$) for the RQ sources ($r_s$ = $-$0.296, $\mathrm{p}_s$ = 0.205). However, we see two distinct regions being occupied by RL and RQ sources in these plots. We also find that while the RQ sources cover a range of spectral indices from flat to steep ($-0.1\gtrsim\alpha_{R}\gtrsim-1.1$), the (two) RL sources do not possess steep spectrum cores. Albeit small number statistics, this could suggest that the radio emission mechanisms in RQ AGN is intrinsically different to those in RL AGN, such as star formation (SF)/jets/winds in RQ AGN versus mainly jets in RL AGN. 

We note that since the VLA and uGMRT observations were carried out at different epochs, variations in the flux density of the compact cores can affect the spectral index measurements. \citet{Barvainis05} studied radio variability of a subset of PG quasars at $\sim$8 GHz. These included six sources from our sample, namely, PG 0003+199, PG 0007+106, PG 0049+171, PG 2130+099, PG 2209+184 and PG 2304+042. We used 8 GHz VLA data and 685 MHz uGMRT data for making the spectral index images of PG 0003+199, PG 0049+171, PG 2130+099, PG 2209+184 and PG 2304+042. However, for PG 0007+106, we used 5 GHz VLA data and 685 MHz uGMRT data (see Table~\ref{Table1}). \citet{Barvainis05} found PG 0049+171, PG 2209+184 and PG 2304+042 to be variable. They estimated the maximum percentage change in the flux densities to be roughly 20\% for PG 0049+171, 20\% for PG 2209+184 and 100\% for PG 2304+042 over a time-scale of 0.5, 0.1 and 0.5 yr, respectively. Using this information, we find that the spectral index value varies between $-$0.37 and $-$0.62 for PG 0049+171 and between 0.1 and $-$0.8 for PG 2304+042. For PG 2209+184, the spectral index variation is of the order of 0.001, much smaller than the estimated error in $\alpha_R$.

\begin{figure}
\centering{
\includegraphics[width=8cm]{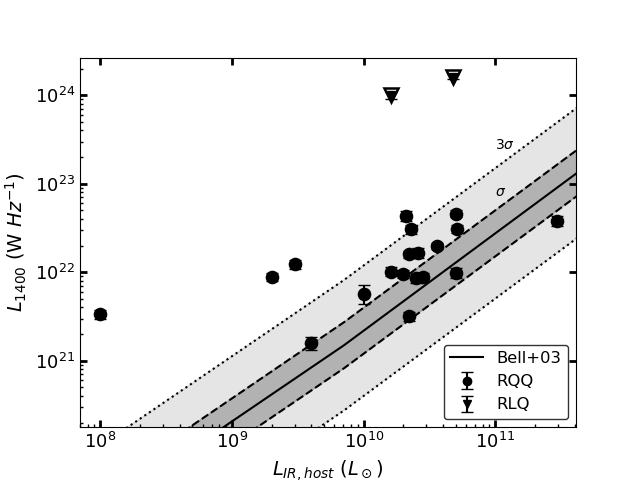}}
\caption{\small The radio$-$IR correlation for the PG sources. L$_\mathrm{1400}$ are the k-corrected rest-frame luminosities at 1400 MHz derived from 685 MHz uGMRT flux densities and $\alpha_{R}$ using equations~\ref{2} and~\ref{3} by substituting $\nu$=1400 MHz. L$_\mathrm{IR, host}$ are the 8-1000 $\mu$m host galaxy IR luminosities taken from \citet{Lyu17}.  The black solid line is the radio$-$IR correlation for star-forming galaxies from \citet{Bell03}. The black dashed lines mark the 1$\sigma$ regions and the black dotted lines mark the 3$\sigma$ regions.}
\label{Fig4}
\end{figure}

\begin{figure*}
\centering{
\includegraphics[width=8cm]{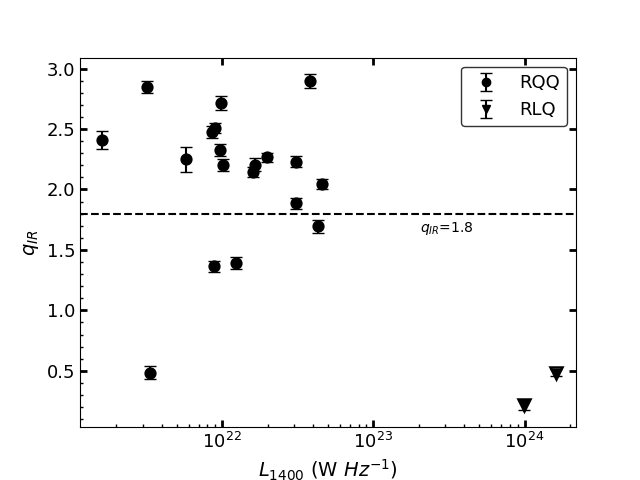}
\includegraphics[width=8cm]{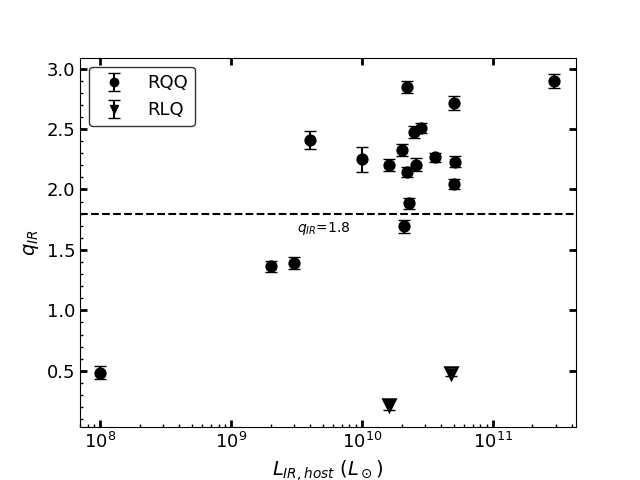}
\includegraphics[width=8cm]{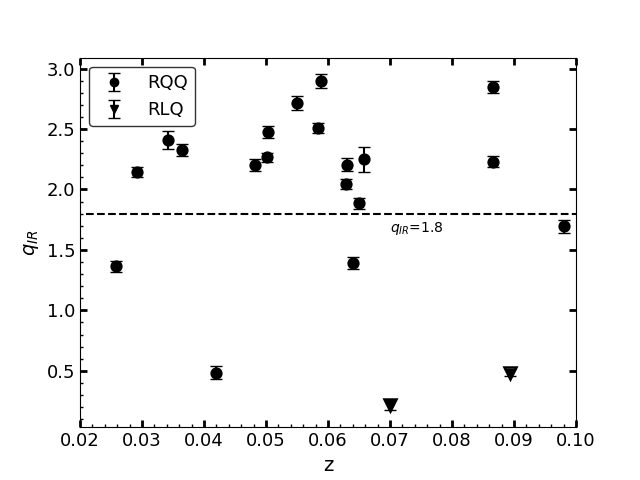}}
\caption{\small The q$_\mathrm{IR}$ as a function of 1400 MHz k-corrected rest-frame luminosity derived from 685 MHz uGMRT flux density and $\alpha_{R}$ using equations~\ref{2} and~\ref{3} by substituting $\nu$=1400 MHz (top left); 8$-$1000 $\mu$m host galaxy IR luminosity taken from \citet{Lyu17} (top right); and redshift for the PG sources (bottom). The black dashed line marks q$_\mathrm{IR}$ =1.8, which is the discriminator value between AGN versus SF contributions to the observed radio emission \citep{Condon02, Bell03}.}
\label{Fig5}
\end{figure*}

\subsection{The radio$-$IR correlation with uGMRT Data}
\label{Sec4.4}
The correlation between radio and infrared (IR) emission is one of the tightest and well-studied correlations in astrophysics \citep{Helou85, Condon92, Yun01, Bell03}. It helps to identify whether the observed radio emission is dominated by AGN activity or SF in the host galaxy. A useful parameter that quantifies the radio$-$IR correlation is the q$_\mathrm{IR}$ parameter \citep{Bell03}, which is defined as:
\begin{equation}
\mathrm{q_{IR}=log(L_{IR}/3.75\times10^{12} W) - log(L_{1400}/W~Hz^{-1})}
	\label{1}
\end{equation}The contribution of AGN to radio$-$IR correlation has been studied to some extent in the recent past. A low q$_\mathrm{IR}$ value and a significant positional offset in the radio$-$IR correlation plane suggests the dominance of AGN in a galaxy. \citet{Condon02} and \citet{Bell03} classify sources with q$_\mathrm{IR}$ $<$1.8 as being "radio-excess" due to dominant contribution of AGN in their radio emission. The radio$-$IR correlation has been studied as a function of galaxy types in the literature. The average q$_\mathrm{IR}$ values for star-forming galaxies, Seyfert galaxies and quasars are respectively 2.27, 2.14 and 2.04 \citep{Moric10}. The q$_\mathrm{IR}$ values for individual sources in our sample are listed in Table~\ref{Table3}. The average q$_\mathrm{IR}$ value for the PG sample is 1.95$\pm$0.01.

The flux density (S$_\mathrm{1400}$) and the k-corrected rest-frame luminosity at 1.4 GHz (L$_\mathrm{1400}$) for our sample are derived using the relations:
\begin{equation}
\label{2}
\mathrm{S_{\nu}=S_{685}(\nu/685)^{\alpha_{R}}}
\end{equation}
\begin{equation}
\label{3}
\mathrm{L_{\nu}=4\pi D_L^2 S_{\nu}(1+z)^{(-\alpha_{R}-1)}}
\end{equation}
where $\nu$ = 1400 MHz, S$_\mathrm{685}$ are the observed uGMRT 685 MHz total flux densities and $\alpha_{R}$ are the spectral index values taken from the images of individual sources. We use the 8-1000 $\mu$m IR luminosities (L$_\mathrm{IR, host}$) provided by \citet{Lyu17}, which include contributions from the host galaxy but not the AGN. \citet{Lyu17} used SED fitting from optical to far-IR to isolate IR luminosity associated with SF in the host galaxy from that associated with AGN. Separating the IR contributions from the host galaxy and AGN is important because if both IR and radio emission are dominated by AGN, then this could produce another correlation that can artificially cause AGN to follow the radio-IR correlation characteristic of star-forming galaxies \citep{Moric10, Zakamska16}.

Figure~\ref{Fig4} presents L$_\mathrm{1400}$ as a function of L$_\mathrm{IR, host}$ for the PG sources and compares them with the radio$-$IR correlation of star-forming galaxies from fig. 3 of \citet{Bell03}\footnote{The slope and intercept were obtained from the figure using: Mark Mitchell, Baurzhan Muftakhidinov and Tobias Winchen et al, "Engauge Digitizer Software." Webpage: http://markummitchell.github.io/engauge-digitizer.}:

\begin{equation}
 \label{4}
\mathrm{log~L_{1400} (W~Hz^{-1}) = 1.10~log~L_{IR, host} (L_\odot)+ 10.34} 
\end{equation}

The scatter in the radio$-$IR correlation of the star-forming galaxies in \citet{Bell03} is 0.26 dex, which was used to determine the 1$\sigma$ and 3$\sigma$ limits in Figure~\ref{Fig4}.

\begin{figure}
\centering{
\includegraphics[width=8cm]{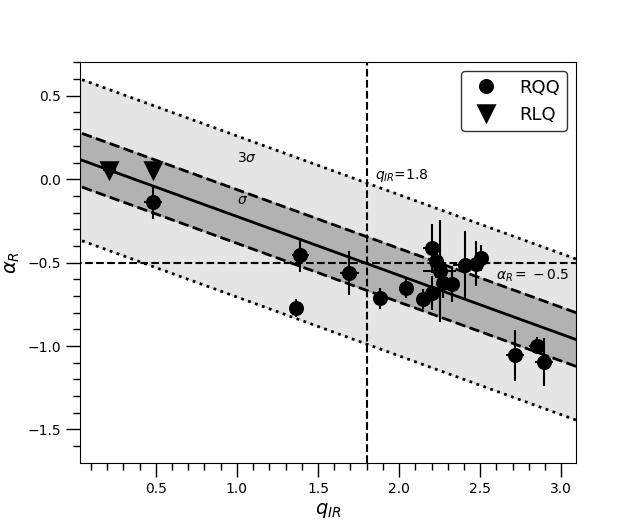}}
\caption{\small The mean core spectral index as a function of q$_\mathrm{IR}$ for the PG sources. The black dashed vertical line marks q$_\mathrm{IR}$=1.8, which is the discriminator value between AGN versus SF contributions to the observed radio emission \citep{Condon02, Bell03}. The black dashed horizontal line marks $\alpha_{R}$=$-$0.5. We consider the sources with $\alpha_{R}$ $\geq-0.5$ to have flat spectrum cores and the sources with $\alpha_{R}$ $<-0.5$ to have steep spectrum cores. The black solid line is the best-fitting line for the data. The black dashed slant lines mark the 1$\sigma$ regions and the black dotted slant lines mark the 3$\sigma$ regions.}
\label{Fig6}
\end{figure}

We see that both RL sources (PG 0007+106 \& PG 2209+184) and four RQ sources (PG 0003+199, PG 0049+171, PG 1404+226, PG 2304+042) clearly lie above the 3$\sigma$ limit of the radio$-$IR correlation. These sources also have q$_\mathrm{IR}$ $<$1.8. PG 1448+273 lies on the 3$\sigma$ limit and its q$_\mathrm{IR}$ is close to the AGN-SF discriminator value. PG 1211+143 is not shown in Figures~\ref{Fig4}-\ref{Fig6}. For this source the optical to far-IR SED fitting in \citet{Lyu17} indicated negligible contribution from host galaxy to the total IR luminosity compared to the AGN. This implies a negligible value of IR-based SFR following the \citet{Kennicutt98} SF law. The radio-based SFR can be taken as an upper limit to the SFR in this galaxy. Thus, consistent with \citet{Lyu17}, we conclude that AGN is the primary contributor to the radio emission in PG 1211+143.

Based on their position on the radio$-$IR correlation, we suggest that the radio emission in RL quasars and nearly one-third of the RQ quasars is AGN dominated. However, being on the radio$-$IR correlation does not rule out an AGN contribution in the remaining two-third of the RQ quasars \citep[e.g., in][]{Wong16}. In these sources, we suggest the possibility of a co-existence of AGN emission along with the stellar-related emission. There are signatures of unresolved/barely resolved jet/lobe structures on few arcsec-scales in the total intensity and the spectral index images of a few of these sources, such as, PG 0050+124, PG 1119+120, PG 1229+204, PG 1244+026, PG 1341+258, PG 1351+236, PG 1501+106 and PG 2130+099 (see Sections~\ref{Sec4.2} and~\ref{Sec4.3}). Besides, the possibility of AGN outflows on milli-arcsec or sub-arcsec scales cannot be ruled out in the remaining sources. Moreover, we find that in $\sim$70\% of our sample, the 1.4 GHz luminosities estimated from the uGMRT 685 MHz data are significantly higher than those predicted from their radio-based SFRs following the \citet{Bell03} SF law, suggesting that the excess emission is associated with AGN. The radio-based SFRs have been assumed to be the same as the IR-based SFRs derived from L$_\mathrm{IR, host}$ following the \citet{Kennicutt98} SF law. 

Figure~\ref{Fig5} provides q$_\mathrm{IR}$ as a function of L$_\mathrm{1400}$ (top left panel), L$_\mathrm{IR, host}$ (top right panel) and redshift (bottom panel) for the PG sources. Figure~\ref{Fig5} (top left panel) and (bottom panel) reveals no correlation for the RQ PG sources based on the Spearman rank correlation test with ($r_s$ ; $\mathrm{p}_s$ = $-$0.212 ; 0.383) and ($r_s$ ; $\mathrm{p}_s$ = 0.058 ; 0.814), respectively. However, in Figure~\ref{Fig5} top right panel, we note an increase in the q$_\mathrm{IR}$ values with increasing L$_\mathrm{IR, host}$ for the RQ PG sources, which is also implied as a positive correlation by the Spearman rank correlation test ($r_s$ = 0.544 and $\mathrm{p}_s$ = 0.016). Figure~\ref{Fig6} shows a 3$\sigma$ anticorrelation between q$_\mathrm{IR}$ and $\alpha_{R}$ for the PG sources. This figure represents the distribution of flat and steep spectrum core PG sources across the AGN$-$star-formation divide (q$_\mathrm{IR}$ =1.8). The best-fitting linear regression and the 1$\sigma$ and 3$\sigma$ limits are estimated using the $\tt{LTS\_LINEFIT}$ program described in \citet{Cappellari13}. This program computes Spearman rank and Pearson correlation co-efficients and does an iterative fit of the data assuming that the errors in both x- and y- co-ordinates are uncorrelated. It also returns the best fit slope and intercept values along with errors and plots the 1$\sigma$ and 3$\sigma$ limits based on the intrinsic scatter of the data. The Spearman rank and Pearson correlation tests imply an anti-correlation between q$_\mathrm{IR}$ and $\alpha_{R}$ for our full PG sample with $r_s$ = $-$0.48 and $\mathrm{p}_s$ = 0.027 and Pearson correlation coefficient $r_p$ = $-$0.81 and probability $\mathrm{p}_p$ = 8.9$\times10^{-6}$. However, for the RQ PG sample, only the Pearson test implies an anti-correlation ($r_p$ = $-$0.64 and $\mathrm{p}_p$ = 0.0033) whereas the Spearman test fails to show any correlation ($r_s$ = $-$0.3 and $\mathrm{p}_s$ = 0.21). Figure~\ref{Fig6} is discussed below in detail:

(i) As discussed above, the radio emission in RL quasars and nearly one-third of the RQ quasars is dominated by AGN activity. Among these, the RL cores and two RQ cores (PG 0049+171 and PG 2304+042) have flat spectrum, possibly arising from synchrotron self-absorption or free-free emission from accretion disc/torus. The radio emission in the two RQ cores with steep spectrum (PG 0003+199 and PG 1404+226) may arise from AGN jets/AGN-driven winds. PG 1448+273 which is at the cusp of AGN$-$SF divide, also has a steep spectrum core, possibly associated with AGN jets/AGN-driven winds. Morever, its total intensity image coupled with its spectral index image suggests AGN jet origin, instead of AGN winds for its steep spectrum (see Sections~\ref{Sec4.2} and~\ref{Sec4.3}). Note that PG 1211+143, which is not indicated in the plot, also has a steep spectrum core, possibly arising from AGN jets/AGN-driven winds. 

(ii) As mentioned earlier, the stellar-related emission is significant in two-third of the RQ quasars, based on their position on the radio$-$IR correlation and q$_\mathrm{IR}$ values. Among these, three sources (PG 1011$-$040, PG 1229+204 and PG 1426+015) have flat spectrum cores possibly arising from thermal free-free emission around HII regions or mixing of the thermal emission with stellar-origin synchrotron emission, nine sources (PG 0050+124, PG 0923+129, PG 1119+120, PG 1244+026, PG 1341+258, PG 1351+236, PG 1501+106, PG 2130+099 and PG 2214+139) have steep spectrum cores most likely due to synchrotron-emitting shocks associated with starburst-superwinds and two sources (PG 0934+013 and PG 1310$-$108) are at the cusp of flat-steep division. However, the co-existence of AGN emission along with such stellar-related processes cannot be ruled out in these sources.
\\
\\
Using a sample of $\sim$ 300 Type 1 and 2 quasars that included the complete PG sample, \citet{Zakamska16} concluded that the radio emission in the quasars is dominated by AGN activity. They used 1.4 GHz flux densities from Faint Images of Radio Sky at Twenty cm survey (FIRST) and NRAO VLA Sky Survey (NVSS) and assumed a spectral index value of $-$0.7 to estimate the luminosities. 20 out of 22 sources from our sample are covered by FIRST and NVSS surveys.\footnote{Out of these 20, 11 sources are detected in FIRST and 3 sources are detected in NVSS. The remaining 6 sources are neither detected in NVSS nor FIRST. Therefore, for these sources, we compared the 1.4 GHz peak flux densities estimated from the uGMRT 685 MHz data with their 5$\sigma$ values, where $\sigma$ is the FIRST {\it rms} noise level at the source position, except for PG 2214+139 where the position is near the edge of the FIRST survey area.} Note that the resolution of FIRST and NVSS is 5$\arcsec$ and 45$\arcsec$ respectively whereas the resolution of the uGMRT 685 MHz images is $\sim$3-5$\arcsec$. A comparison between the total 1.4 GHz flux densities estimated from the uGMRT 685 MHz data and those obtained from FIRST/NVSS reveals that $\sim$5-80\% of the total flux density has been resolved out across 12 sources by the uGMRT observations. For half of the remaining eight sources, the total flux densities estimated from the uGMRT data are about two times higher than their respective FIRST/NVSS values, whereas for the rest, they are nearly the same. This coupled with \citet{Zakamska16} work and the radio$-$IR correlation with uGMRT 685 MHz data could suggest that the emission resolved out by the uGMRT observations is associated with AGN and the excess emission detected by the uGMRT observations is associated with star formation. This is in agreement with our previous suggestion that the low frequency 685 MHz radio emission has significant contributions from both stellar-related processes and the AGN. 

\begin{figure}
\centering{
\includegraphics[width=8cm]{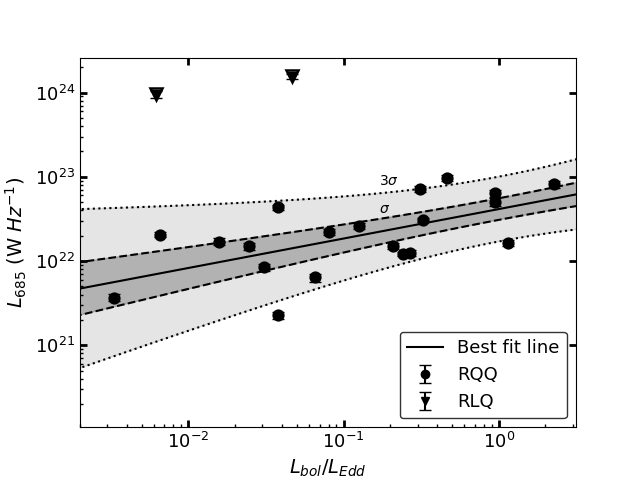}}
\caption{\small 685 MHz k-corrected rest-frame luminosity as a function of Eddington ratio for the PG sources. The black solid line is the best-fitting line for the data. The black dashed lines mark the 1$\sigma$ regions and the black dotted lines mark the 3$\sigma$ regions.}
\label{Fig7}
\end{figure}

\subsection{Relation between radio properties and accretion properties}
\label{Sec4.5}
\subsubsection{BH masses and Accretion rates}
\label{Sec4.5.1}
The correlations between radio properties and BH properties in the PG sources have been discussed in the recent past. However, these studies were carried out using high-frequency ($\sim$GHz) radio data. Using a combined sample of quasars from the FIRST Bright Quasar Survey and PG survey, \citet{Lacy01} found evidence for a strong dependence of 5 GHz rest-frame luminosity on BH mass (M$_\mathrm{BH}$) and Eddington ratio (L$_\mathrm{bol}$/L$_\mathrm{Edd}$). Using the 5 GHz and 8.4 GHz VLA data, \citet{Laor19} observed a highly significant anticorrelation between $\alpha_{R}$ and L$_\mathrm{bol}$/L$_\mathrm{Edd}$ and a marginal correlation between $\alpha_{R}$ and M$_\mathrm{BH}$ for a different sub-sample of the RQ PG sample. Using the uGMRT 685 MHz data, we searched for correlations between the radio properties, such as L$_\mathrm{685}$, q$_\mathrm{IR}$, radio core size and $\alpha_{R}$ and the BH properties, such as, M$_\mathrm{BH}$ and L$_\mathrm{bol}$/L$_\mathrm{Edd}$ for the RQ sources in our sample. We find a significant ($\sim3\sigma$) correlation between L$_\mathrm{685}$ and L$_\mathrm{bol}$/L$_\mathrm{Edd}$ with $r_s$ = 0.611 and $\mathrm{p}_s$ = 0.004. This is presented in Figure~\ref{Fig7}, where the best-fitting line was obtained using a $\tt{PYTHON}$ fitting procedure - $\tt{CURVE\_FIT}$ which also returns the optimal values of slope and intercept along with errors. The 1$\sigma$ and 3$\sigma$ scatter in the correlation were determined using these errors, assuming that they reflect the intrinsic scatter of the data. Moreover, the Spearman rank correlation test reveals no correlation between other radio and BH properties since the test yields $\mathrm{p}_s >$ 0.05. M$_\mathrm{BH}$ and AGN bolometric luminosity values are taken from \citet{Shangguan18} and \citet{Lyu17} respectively. M$_\mathrm{BH}$ are the single-epoch virial BH masses calculated using equation 4 of \citet{HoKim15} and L$_\mathrm{bol}$ = 5.29 L$_\mathrm{IR,AGN}$ where L$_\mathrm{IR,AGN}$ is the AGN IR luminosity from \citet{Lyu17}. L$_\mathrm{Edd}$ is calculated as L$_\mathrm{Edd}$ (L$_{\sun}$) = 3.2 $\times$ 10$^4$~(M$_\mathrm{BH}$/M$_{\sun}$).

The lack of strong correlations between the 685 MHz radio and BH properties (except for the one between L$_\mathrm{685}$ and L$_\mathrm{bol}$/L$_\mathrm{Edd}$) in the RQ sources could be explained by the contribution of stellar-related emission, or radio emission from previous AGN activity episodes that may not be directly related to the current BH activity state. The lack of correlations may also suggest additional contributions from BH spin, magnetic flux, and environment.

\subsubsection{Coronal emission with uGMRT Data}
\label{Sec4.5.2}
The emission from a magnetized corona likely located above the accretion disc was studied as a possible radio emission mechanism in RQ quasars by \citet{LaorBehar08} using the complete PG quasar sample. They observed a strong correlation between 5 GHz radio luminosity \citep{Kellermann89,Kellermann94} and 0.2-20 keV X-ray luminosity \citep{Brandt00,LaorBrandt02}: L$_\mathrm{R}\mathrm{(erg~s^{-1})}\sim 10^{-5}$ L$_\mathrm{X}\mathrm{(erg~s^{-1})}$, which is similar to the Guedel$-$Benz relation obeyed in coronally active stars for non-flaring radio and X-ray luminosity \citep{GuedelBenz93}: 
\begin{equation}
 \label{5}
\mathrm{log~L_{X} (erg~s^{-1}) \sim log~L_{R}(erg~s^{-1}~Hz^{-1}) + 15.5}
\end{equation}
The Guedel$-$Benz relation, along with correlated stellar X-ray and radio variability, indicates that the stellar corona is magnetically heated. Since RQ quasars and coronally active stars followed a similar L$_\mathrm{R}-\mathrm{L_X}$ relation, \citet{LaorBehar08} suggested that similar physical processes occurred in both objects. They proposed that the AGN corona is also magnetically heated and that both radio and X-ray emission could originate from a compact corona.

\begin{figure}
\centering{
\includegraphics[width=8cm]{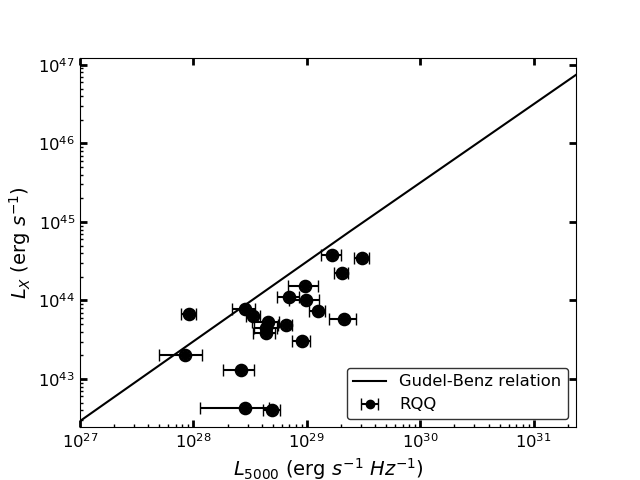}}
\caption{\small The 0.2-20 keV X-ray luminosity taken from \citet{LaorBehar08} as a function of 5000 MHz k-corrected rest-frame luminosity for the RQ PG sources. The latter values are derived from 685 MHz uGMRT flux densities and $\alpha_{R}$ using equations~\ref{2} and~\ref{3} by substituting $\nu$=5000 MHz. The black solid line marks the Guedel$-$Benz relation obeyed in coronally active stars \citep{GuedelBenz93}.}
\label{Fig8}
\end{figure}

The flux density (S$_\mathrm{5000}$) and the k-corrected rest-frame luminosity at 5~GHz (L$_\mathrm{5000}$) for the RQ sources in our PG sample are derived from the 685 MHz uGMRT flux densities and $\alpha_{R}$ using equations~\ref{2} and~\ref{3} by substituting $\nu$=5000 MHz. Figure~\ref{Fig8} shows L$_\mathrm{X}$ as a function of L$_\mathrm{5000}$ for our RQ PG sources and compares them with the Guedel$-$Benz relation (equation~\ref{5}). L$_\mathrm{X}$ values for individual sources are taken from \citet{LaorBehar08}. It can be seen that the sources are not evenly scattered across the correlation but are biased towards a negative residual, unlike that observed in \citet{LaorBehar08}. This could suggest that the coronal emission is not dominant at 685 MHz. This could also suggest that the 685~MHz-5~GHz spectral index is flat for coronal emission. This would be consistent with the work of \citet{RaginskiLaor16}.

\section{Conclusions}
\label{Sec5}
We have presented the first uGMRT 685 MHz images of a subsample of 22 quasars belonging to the PG quasar sample. Two are radio-loud and 20 are radio-quiet. 
\begin{enumerate}
\item
Nearly 80\% sources are only a combination of a radio core unresolved at the uGMRT resolution of $\sim$3-5$\arcsec$ surrounded by diffuse emission on few kpc to $\sim$10 kpc scales and only four sources show extended radio emission on $\sim$10-30 kpc scales.
\item
The RL quasar cores have an inverted radio spectrum whereas the RQ quasar cores exhibit a range of spectral indices from flat to steep ($-0.1\gtrsim\alpha_{R}\gtrsim-1.1$). A few of these sources show signatures of unresolved/barely resolved jet/lobe structures in their spectral index images which are created using the 685 MHz uGMRT data and similar resolution GHz-frequency VLA data.
\item
On the basis of their position on the radio-IR correlation as well as the spectral index images, we find that the radio emission in the two RL quasars and nearly one-third of the RQ quasars is AGN dominated while the remaining sources appear to have significant contributions from both stellar-related processes and the AGN. Also, on the basis of their positional offset from the Guedel$-$Benz relation, it appears that the radio emission in the RQ PG sources at 685 MHz does not have significant contribution from the magnetized coronal emission.
\item
We searched for signatures of `relic' emission in the PG sources. We find clear indications in one source, PG 0007+106, and suggestions of the same in a few others based on their spectral index images.
\item
We do not find strong correlations between the 685 MHz radio properties, such as L$_\mathrm{685}$, q$_\mathrm{IR}$, radio core size, and $\alpha_{R}$ and the BH properties, such as M$_\mathrm{BH}$ and L$_\mathrm{bol}$/L$_\mathrm{Edd}$ in the RQ sources except for the one between L$_\mathrm{685}$ and L$_\mathrm{bol}$/L$_\mathrm{Edd}$. This lack of correlations could be explained by the contribution of stellar-related emission, or radio emission from previous AGN activity episodes that may not be directly related to the current BH activity state. The lack of correlations may also suggest additional contributions from other physical parameters, such as BH spin, magnetic flux, and environment.
\end{enumerate}

\section*{Acknowledgements}
We thank the referee for their valuable comments. LCH was supported by the National Science Foundation of China (11721303, 11991052) and the National Key R\&D Program of China (2016YFA0400702). We thank the staff of the GMRT that made these observations possible. GMRT is run by the National Centre for Radio Astrophysics of the Tata Institute of Fundamental Research. We acknowledge the support of the Department of Atomic Energy, Government of India, under the project  12-R\&D-TFR-5.02-0700. This research has made use of the NASA/IPAC Extragalactic Database (NED), which is funded by the National Aeronautics and Space Administration and operated by the California Institute of Technology.

\section*{Data availability}
The data underlying this article will be shared on reasonable request to the corresponding author. The raw data can be obtained from the GMRT Data Archive: https://naps.ncra.tifr.res.in/goa/data/search




\bibliographystyle{mnras}
\bibliography{mnras} 




\appendix
\section{Optical Morphology}
Here, we discuss the optical morphology of the host galaxies. 

$\mathbf{PG~0003+199}$: The host could either be a lenticular galaxy or an unbarred spiral galaxy with tightly wound spiral arms (type S0/a; from NED). However, \citet{Crenshaw03}, \citet{Deo06} and \citet{Ohta07} classify this as a `point source' (type P), meaning the nucleus dominates the light distribution.
The PanSTARRS image reveals an optical jet-like feature emerging from the host galaxy that indicates that the galaxy is in some state of interaction. \citet{Ohta07} also suggests that this is an interacting or a peculiar galaxy. The HST image analysis reveals the presence of a bulge and an extended disc \citep{Kim17}.

$\mathbf{PG~0007+106}$: This source is classified as an Optically Violent Variable (OVV) quasar \citep{Veilleux09b}. The disturbed-looking host galaxy is part of an interconnected triplet of compact galaxies \citep{Zwicky71}. 

$\mathbf{PG~0050+124}$: The host is a starburst unbarred spiral galaxy with tightly wound spiral arms \citep[type Sa;Sbrst; from NED, see also][]{Hutchings90,Schinnerer98}. The two asymmetric spiral arms have knots of SF \citep{Surace01}. These tidal tails could be a result of merging \citep{Zheng99}.

$\mathbf{PG~0923+129}$: The host is classified as a lenticular galaxy, marked with an uncertainity in NED \citep[type S0?;][]{deVaucouleurs91}. However, \citet{Ohta07} classifies the host as a lenticular galaxy with an outer ring (type RS0). The presence of the outer ring is also evident from the PanSTARRS optical image (see Figure~\ref{FigureB2} in APPENDIX B). One of the scenarios suggested is that the ring galaxies are formed when a smaller galaxy passes through the center of a larger galaxy. The radio source in the south is associated with a neighbouring host galaxy, which could be the merger companion in the ring formation.

$\mathbf{PG~0934+013}$: The host is a barred spiral galaxy with weakly wound spiral arms \citep[type SBab; from NED and][]{Ohta07}.

$\mathbf{PG~1011-010}$: The host is a barred spiral galaxy with loosely wound spiral arms and possessing inner ring-like structures and morphological peculiarities \citep[type SB(r)b: pec; from NED and][]{Ohta07}.

$\mathbf{PG~1119+120}$: The host galaxy is an intermediate spiral galaxy with tightly wound spiral arms \citep[type SABa; from NED and][]{Ohta07}.

$\mathbf{PG~1211+143}$: The HST image analysis suggests the presence of a faint central feature resembling a bar or highly inclined disc-like structure. However, it is not clear if this is an artefact due to point-spread function mismatch \citep{Kim08}.

$\mathbf{PG~1229+204}$: The host is a barred lenticular galaxy (type SB0; from NED). However, \citep{Surace01} classifies the host as a barred spiral galaxy (type SB). An asymmetric nuclear bar runs in the NE-SW direction through the host \citep{Surace01}. The northern arm of the bar is wider than the southern arm. This luminous extension on one side of the bar is thought to be due to a small merging companion \citep{Hutchings92}. The HST image analysis reveals the presence of a faint tidal tail, which may be signature of interaction or minor merger \citep{Kim17}.

$\mathbf{PG~1244+026}$: The host could either be an elliptical galaxy or a lenticular galaxy \citep[type E/S0; from NED and][]{Ohta07}.

$\mathbf{PG~1310-108}$: \citet{Jahnke04} finds a clear disc with two asymmetric spiral or tidal arms. No luminous companion is visible as a source of tidal interaction; the closest ones being two galaxies with unknown redshifts, one which is 0.5 mag fainter than the host in V and 45 kpc to the north, and the other one is 2 mag fainter and 110 kpc east. We expect a case of minor merger here \citep{Sikora07}.

$\mathbf{PG~1426+015}$: The host could either be an elliptical galaxy or a lenticular galaxy (type E/S0; from NED). The HST image analysis reveals a tidal tail and a small companion \citep{Kim08} and also suggests that it is a disturbed elliptical galaxy \citep{Kim17}.

$\mathbf{PG~1448+273}$: The PanSTARRS image of the host galaxy reveals the presence of tidal tails. \citet{Ohta07} also suggests that this is an interacting or a peculiar galaxy.

$\mathbf{PG~1501+106}$: The host is an elliptical galaxy \citep[type E; from NED and][]{Tarenghi94}.

$\mathbf{PG~2130+099}$: The host is an unbarred spiral galaxy with tightly wound spiral arms and outer ring-like structures (type (R)Sa; from NED). The HST image analysis reveals a ring-like spiral disc \citep{Kim08} and also suggests the presence of a pseudo-bulge and a disturbed disc \citep{Kim17}. 

$\mathbf{PG~2214+139}$:
This source is classified as an Ultra Luminous Infrared Galaxy (ULIRG) by \citet{Veilleux09a}. 

\section{Supplementary data}
We present the 685 MHz uGMRT images, radio-optical overlays and radio spectral index images in this section.

\begin{figure*}
\centerline{
\includegraphics[width=7cm]{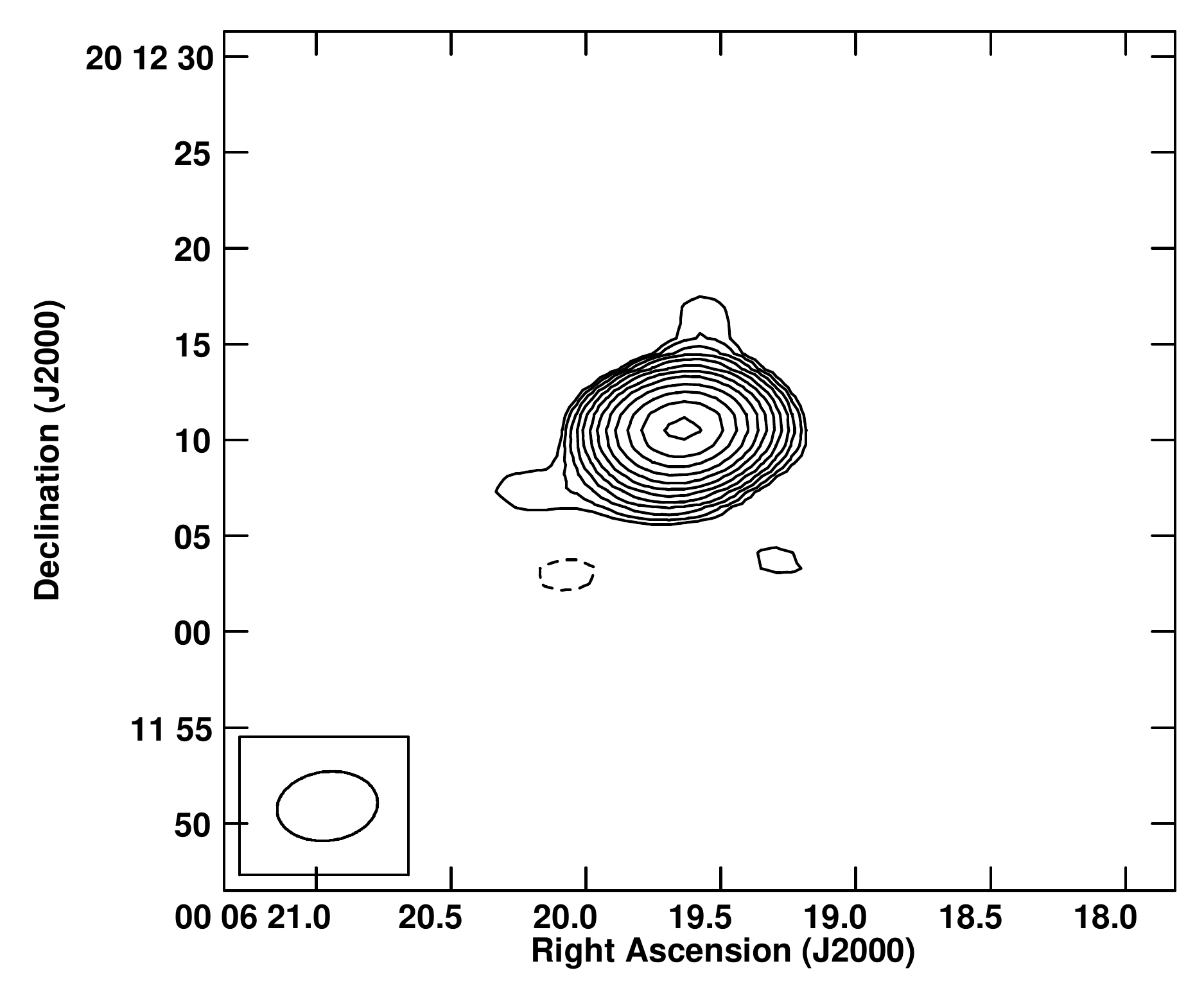}
\includegraphics[width=7cm]{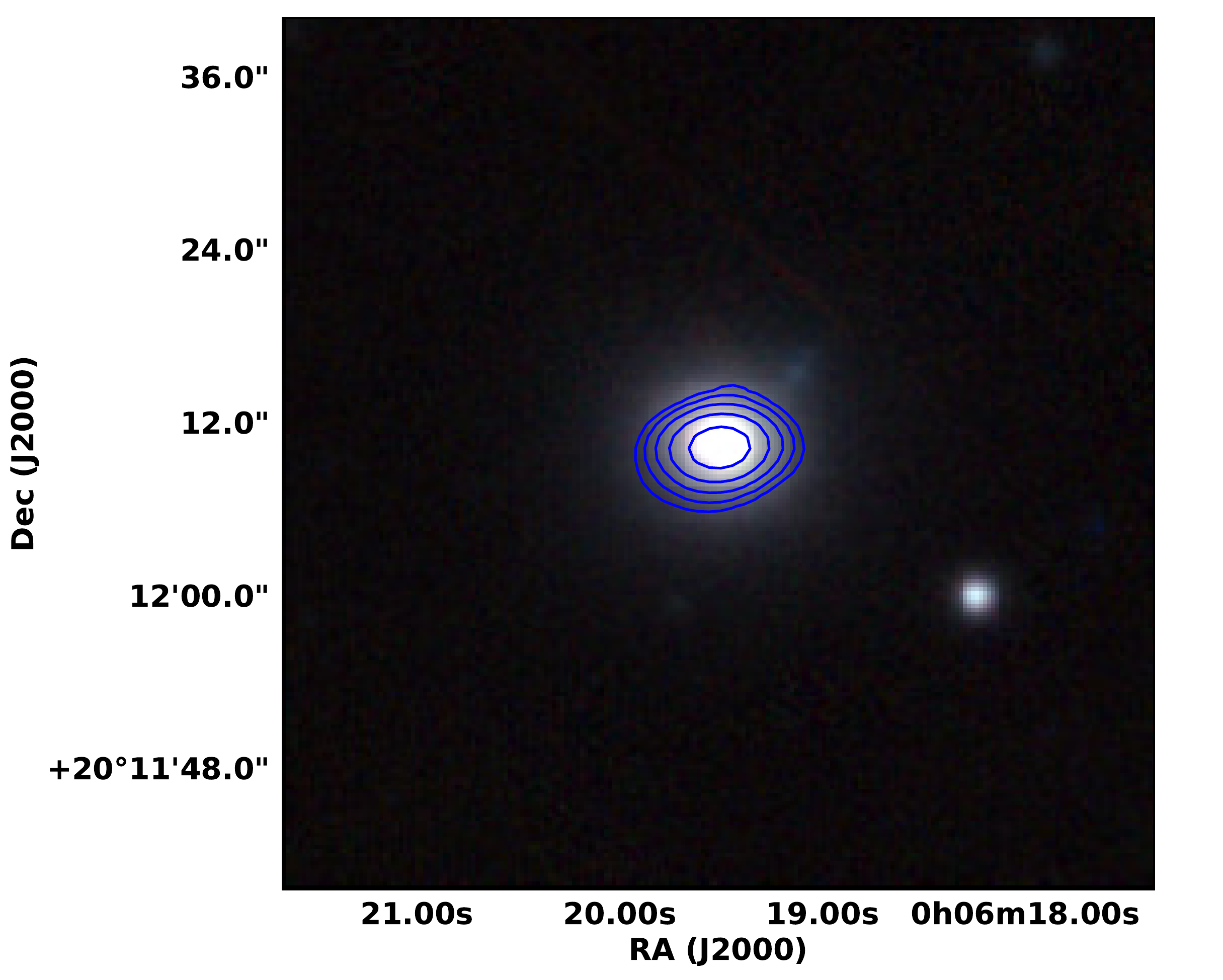}}
\end{figure*}

\begin{figure*}
\centerline{
\includegraphics[width=7cm]{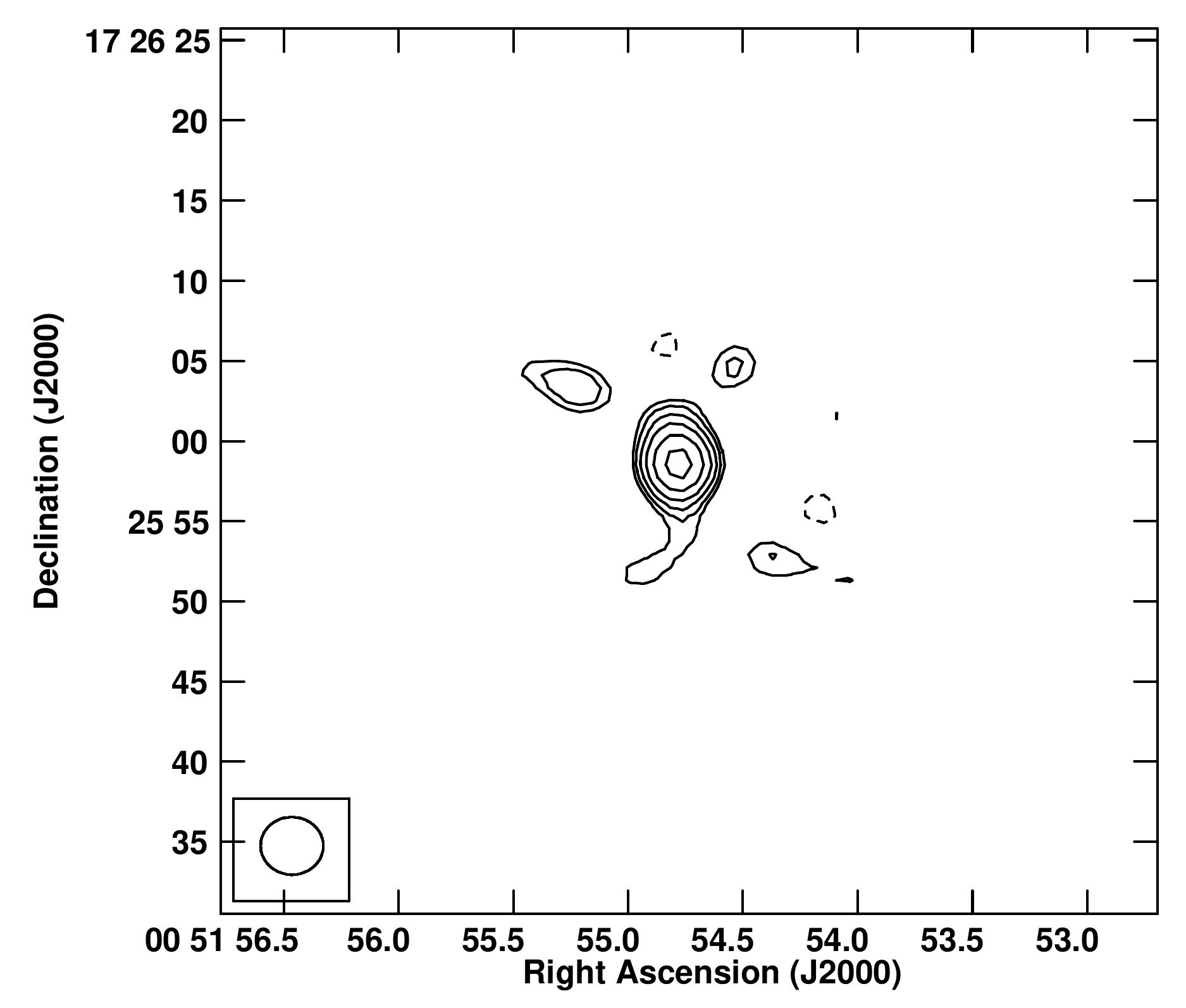}
\includegraphics[width=7cm]{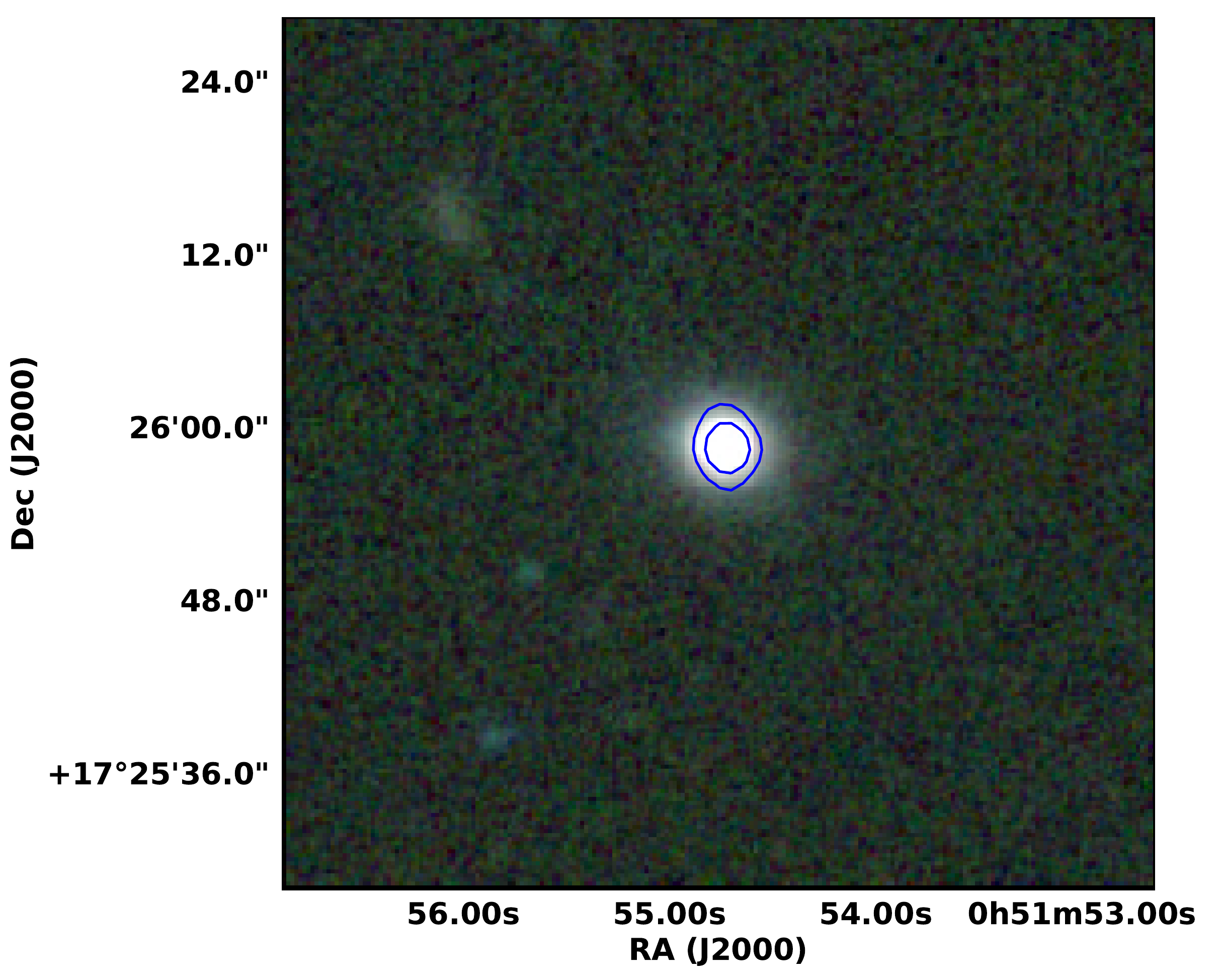}}
\end{figure*}

\begin{figure*}
\centerline{
\includegraphics[width=7cm]{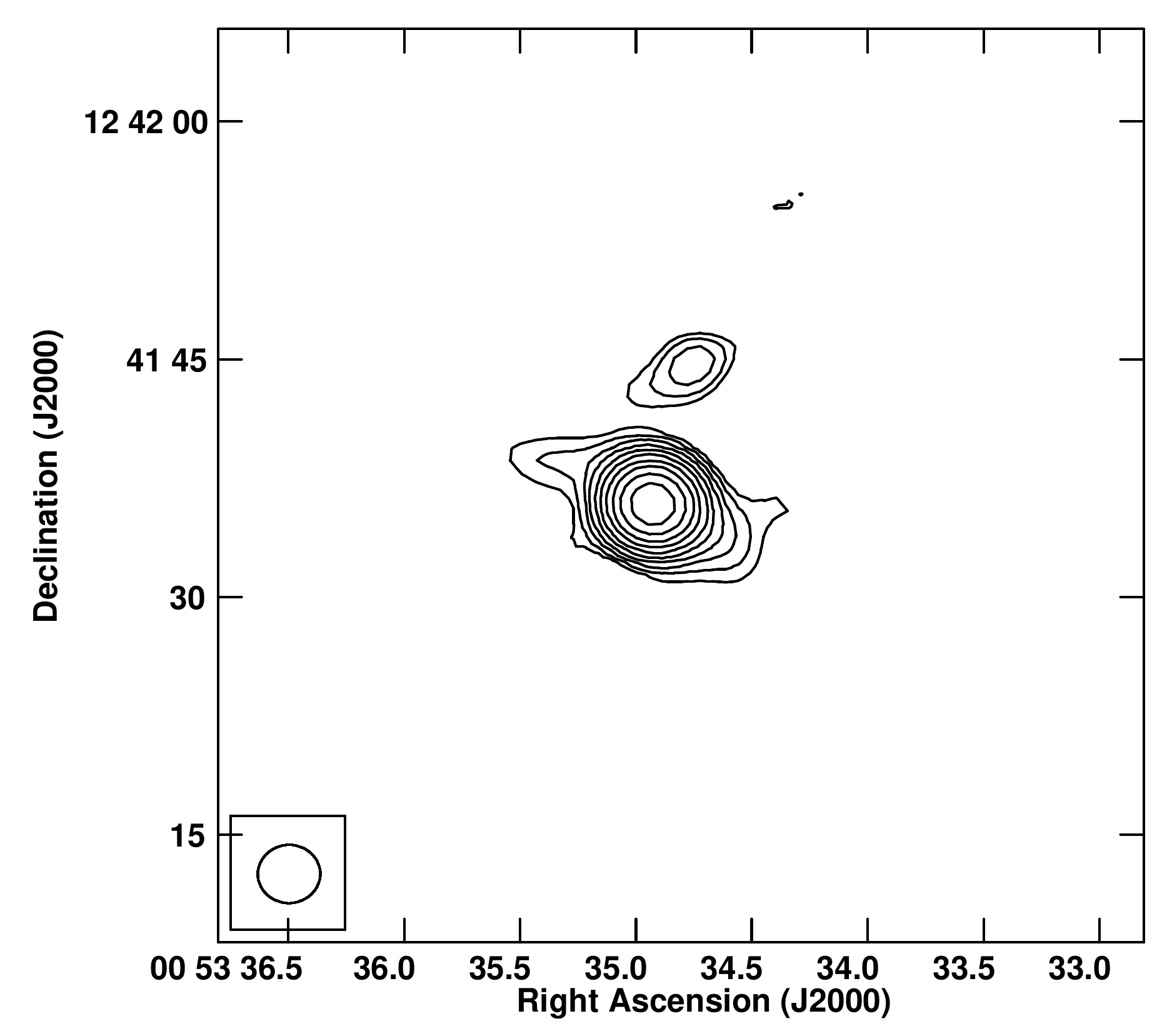}
\includegraphics[width=7cm]{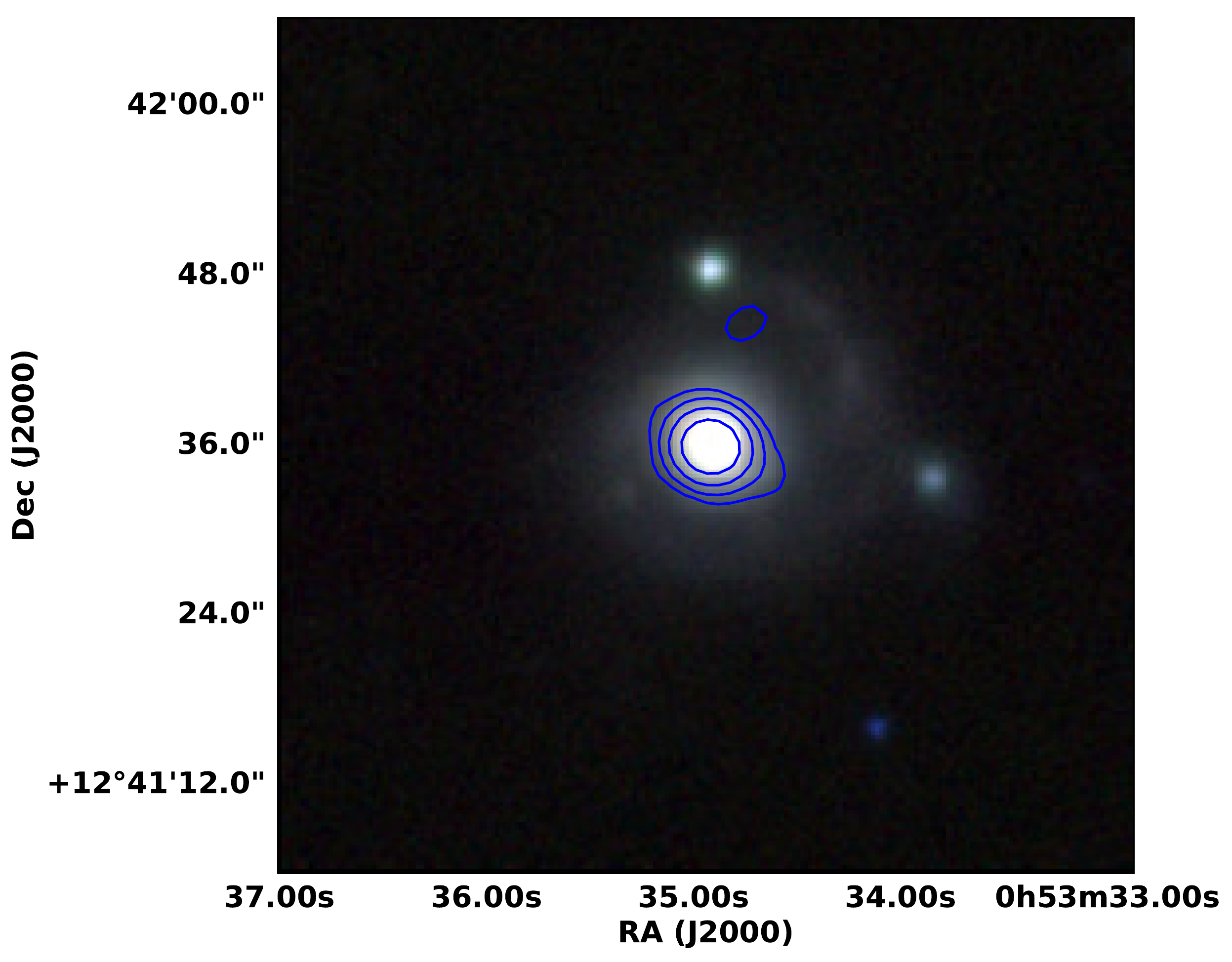}}
\caption{Left: 685 MHz total intensity contour images of PG 0003+199, PG 0049+171 and PG 0050+124 from top to bottom. The peak contour flux is {\it x} mJy beam$^{-1}$ and the contour levels are {\it y} $\times$ (-1, 1, 1.4, 2, 2.8, 4, 5.6, 8, 11.20, 16, 23, 32, 45, 64, 90, 128, 180, 256, 362, 512) mJy beam$^{-1}$, where {\it (x ; y)} for PG 0003+199, PG 0049+171 and PG 0050+124 are (144;0.23), (1055.7;0.28) and (112.6;0.30) respectively. Right: 685 MHz total intensity contours in blue superimposed on PanSTARRS $\it{griz}$-color composite optical images of PG 0003+199, PG 0049+171 and PG 0050+124 from top to bottom. The peak contour flux is {\it x} mJy beam$^{-1}$ and the contour levels are {\it y} $\times$ (2, 4, 8, 16, 32, 64, 128, 256, 512) mJy beam$^{-1}$, where {\it (x ; y)} for PG 0003+199, PG 0049+171 and PG 0050+124 are (144;0.23), (1055.7;0.28) and (112.6;0.30) respectively.}
\end{figure*}

\begin{figure*}
\centerline{
\includegraphics[width=7cm]{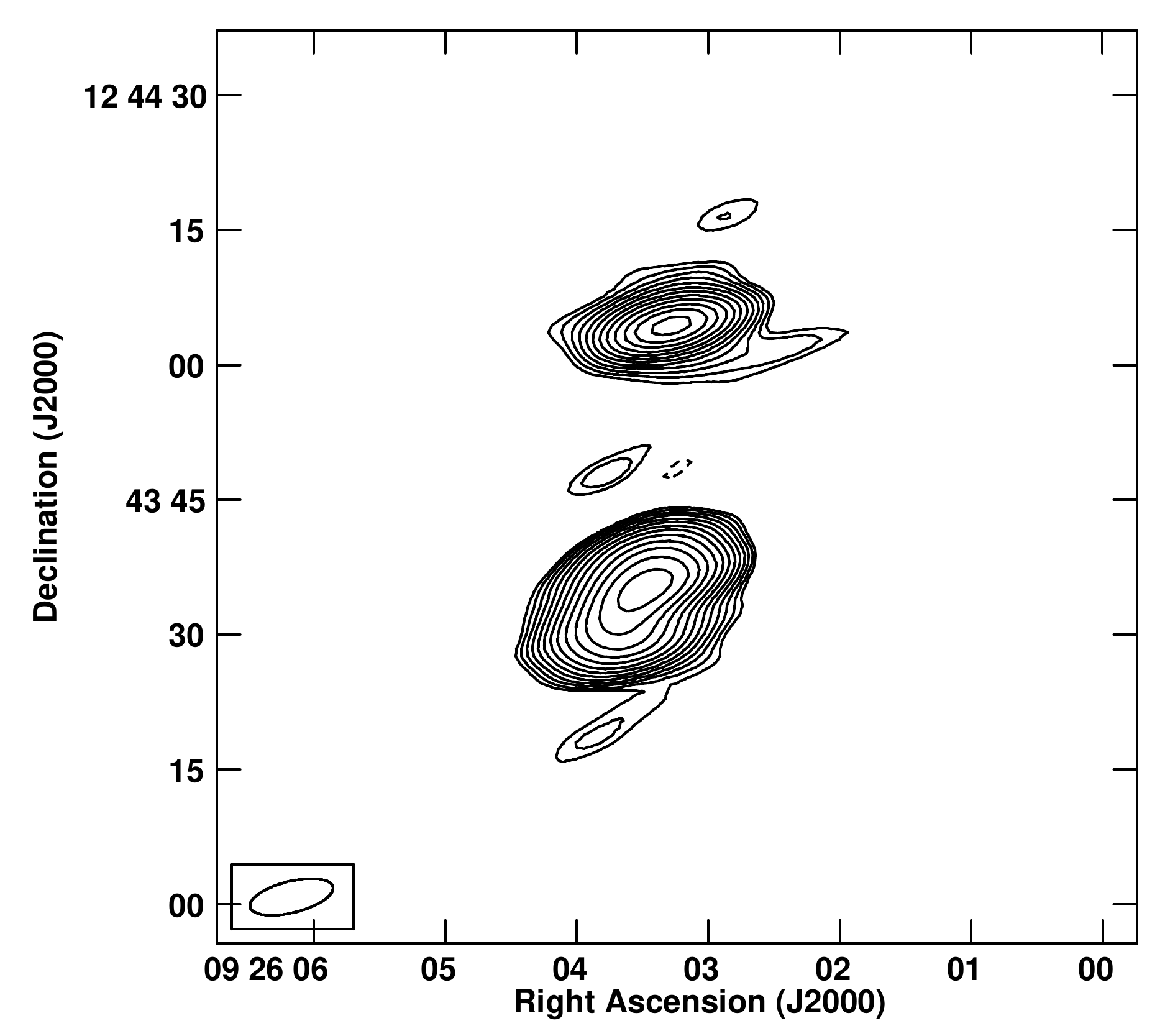}
\includegraphics[width=7cm]{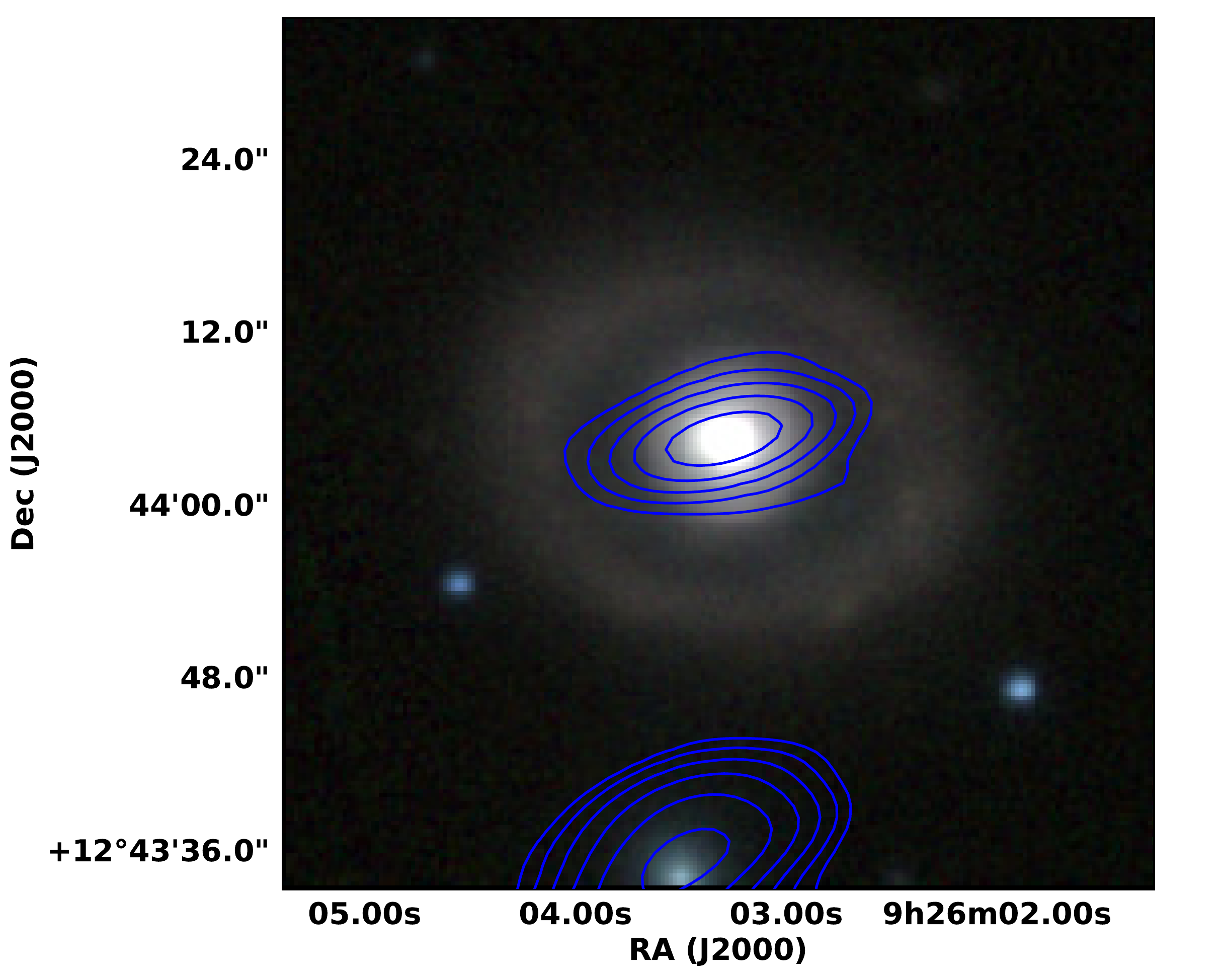}}
\end{figure*}

\begin{figure*}
\centerline{
\includegraphics[width=7cm]{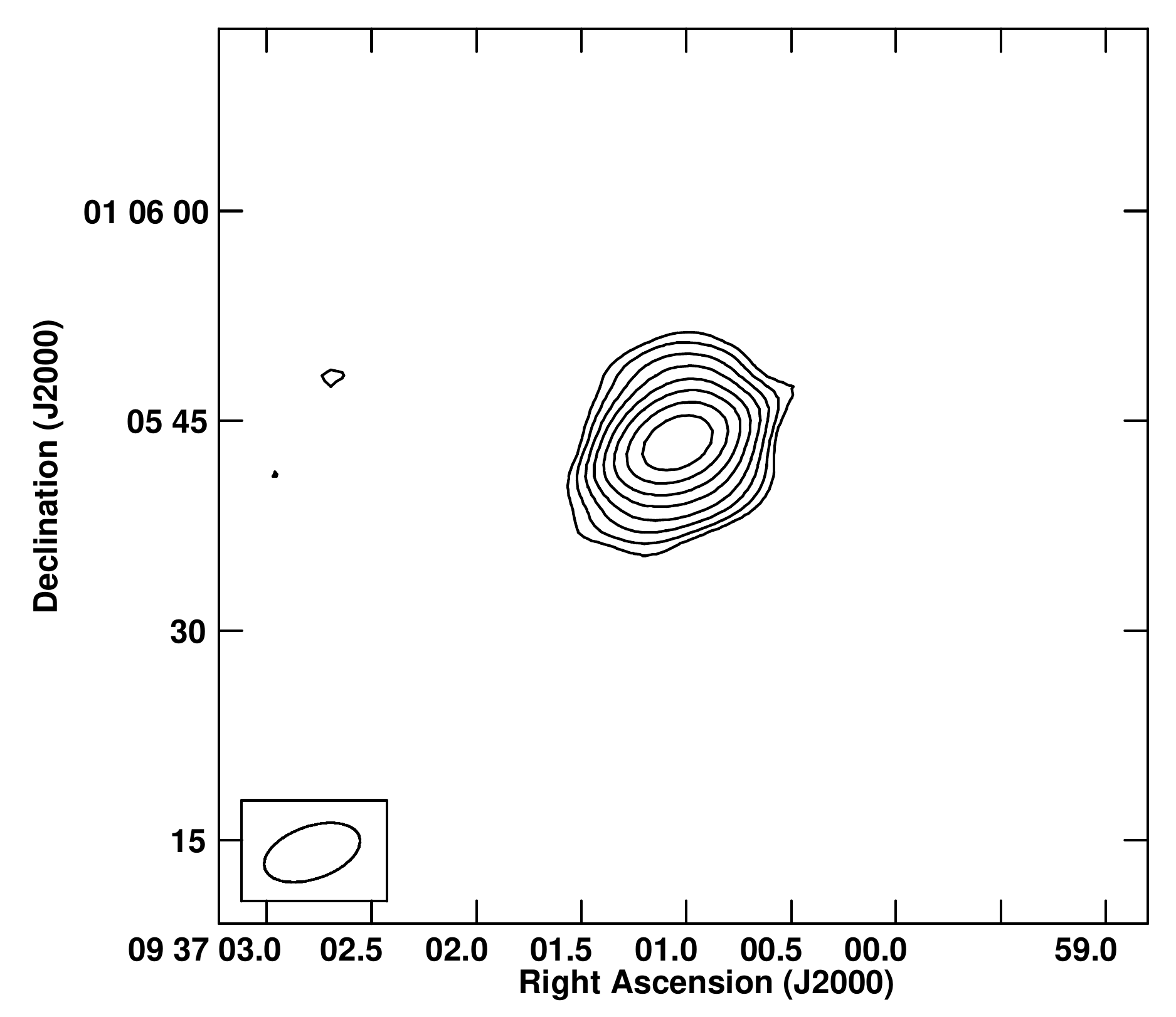}
\includegraphics[width=7cm]{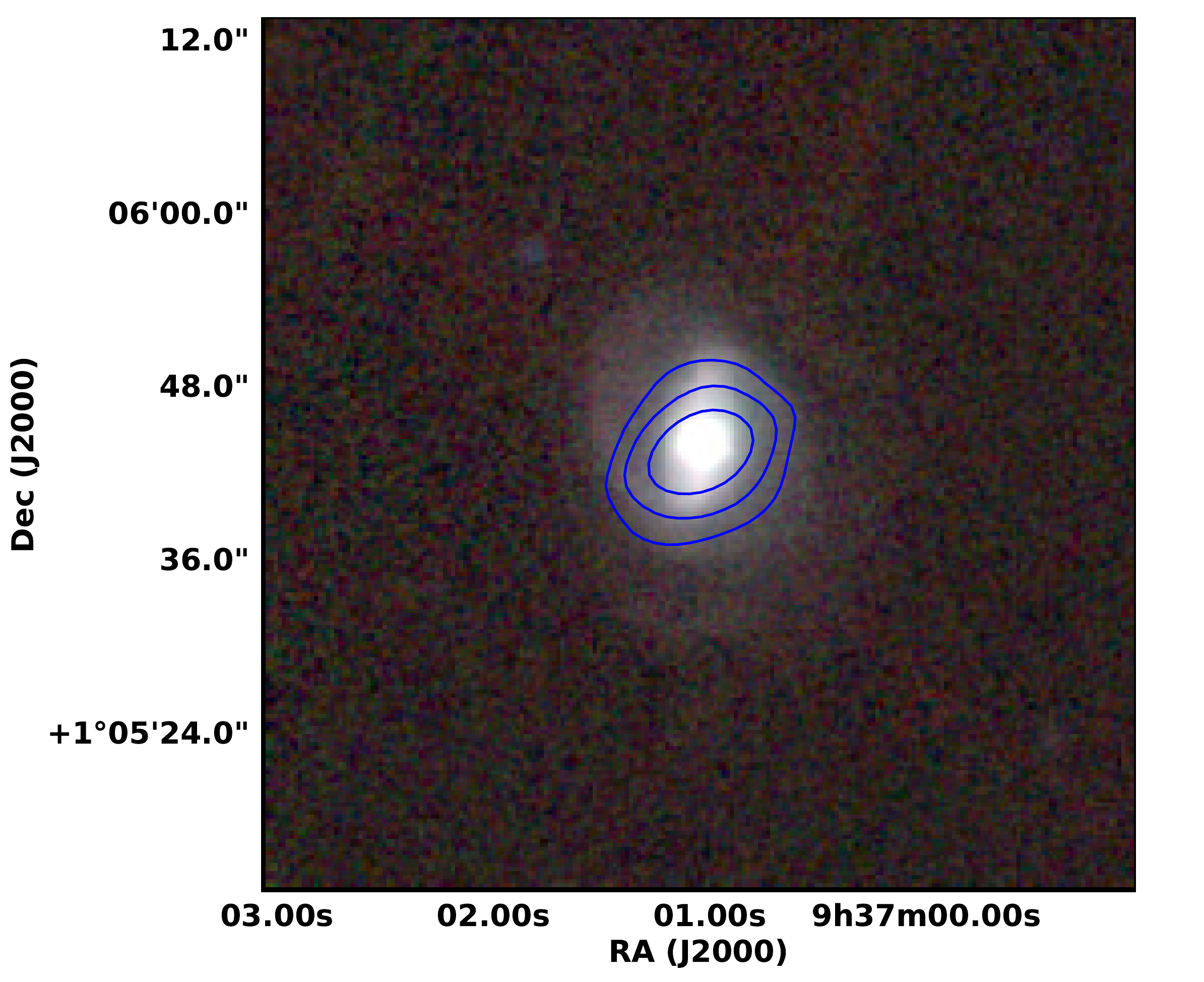}}
\end{figure*}

\begin{figure*}
\centerline{
\includegraphics[width=7cm]{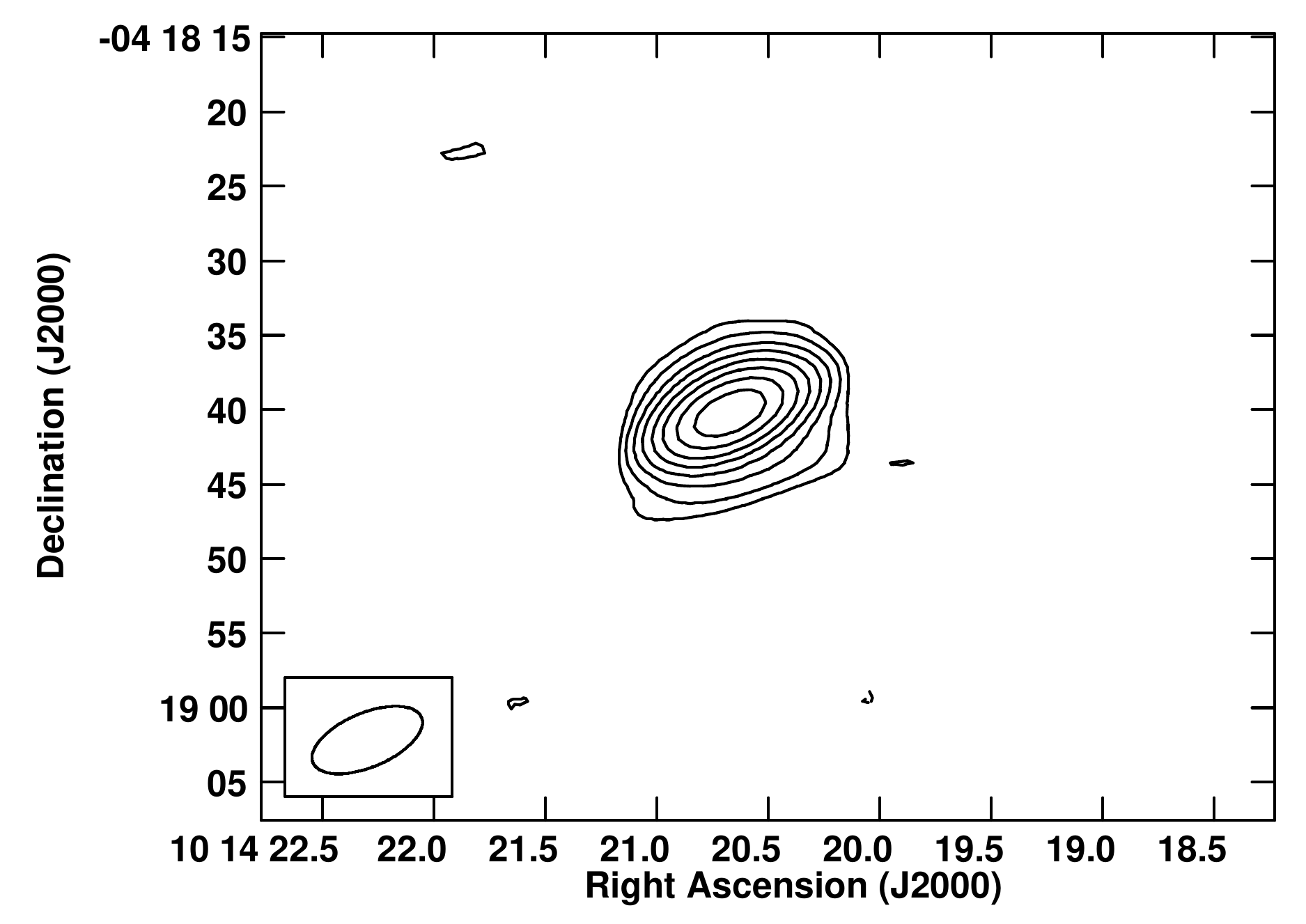}
\includegraphics[width=7cm]{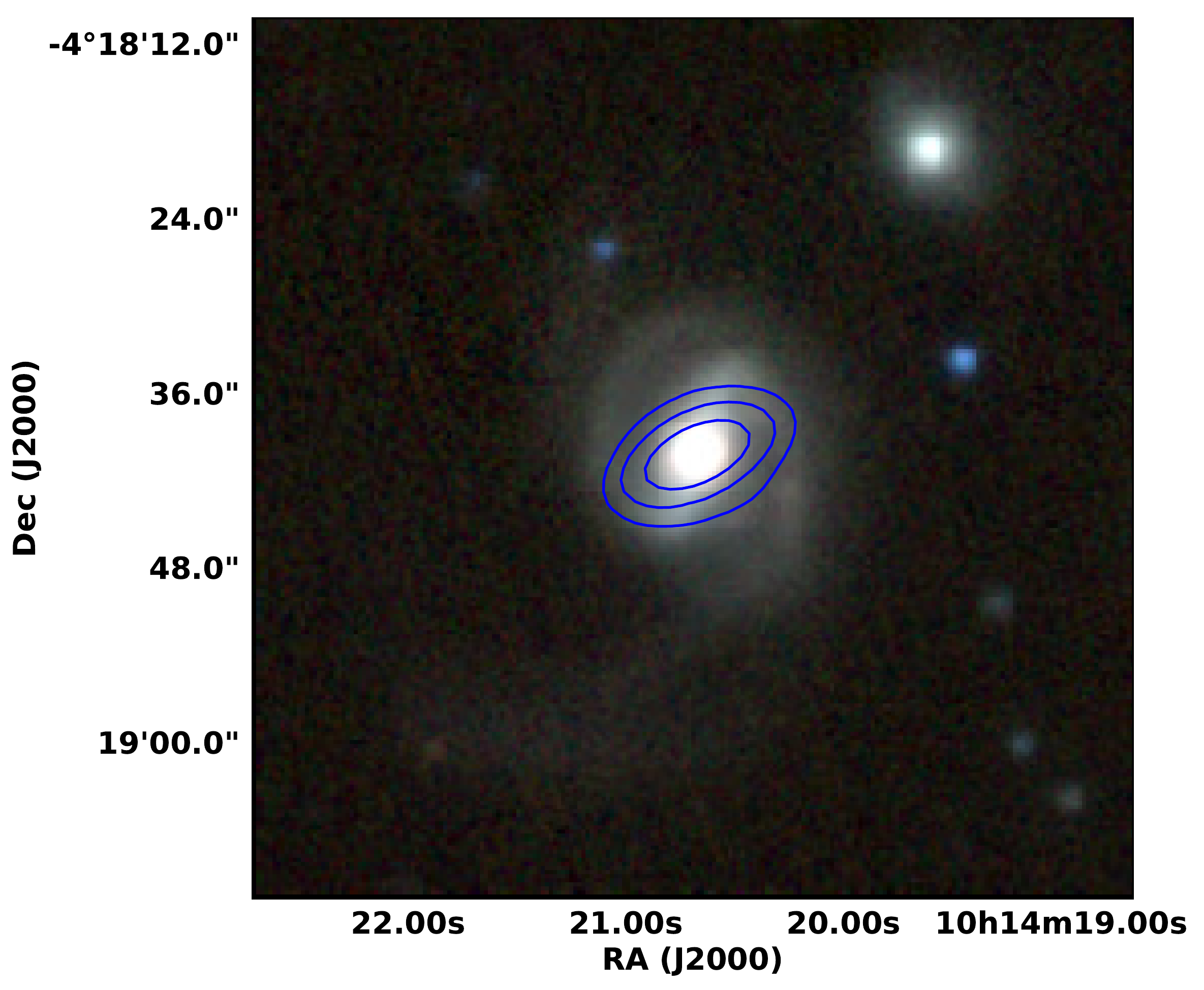}}
\caption{Left: 685 MHz total intensity contour images of PG 0923+129, PG 0934+013 and PG 1011$-$040 from top to bottom. The peak contour flux is {\it x} mJy beam$^{-1}$ and the contour levels are {\it y} $\times$ (-1, 1, 1.4, 2, 2.8, 4, 5.6, 8, 11.20, 16, 23, 32, 45, 64, 90, 128, 180, 256, 362, 512) mJy beam$^{-1}$, where {\it (x ; y)} for PG 0923+129, PG 0934+013 and PG 1011$-$040 are (30;0.22), (35.9;0.08) and (42.3;0.08) respectively. Right: 685 MHz total intensity contours in blue superimposed on PanSTARRS $\it{griz}$-color composite optical image of PG 0923+129, PG 0934+013 and PG 1011$-$040 from top to bottom. The peak contour flux is {\it x} mJy beam$^{-1}$ and the contour levels are {\it y} $\times$ (2, 4, 8, 16, 32, 64, 128, 256, 512) mJy beam$^{-1}$, where {\it (x ; y)} for PG 0923+129, PG 0934+013 and PG 1011$-$040 are (30;0.22), (35.9;0.08) and (42.3;0.08) respectively.}
\label{FigureB2}
\end{figure*}

\begin{figure*}
\centerline{
\includegraphics[width=7cm]{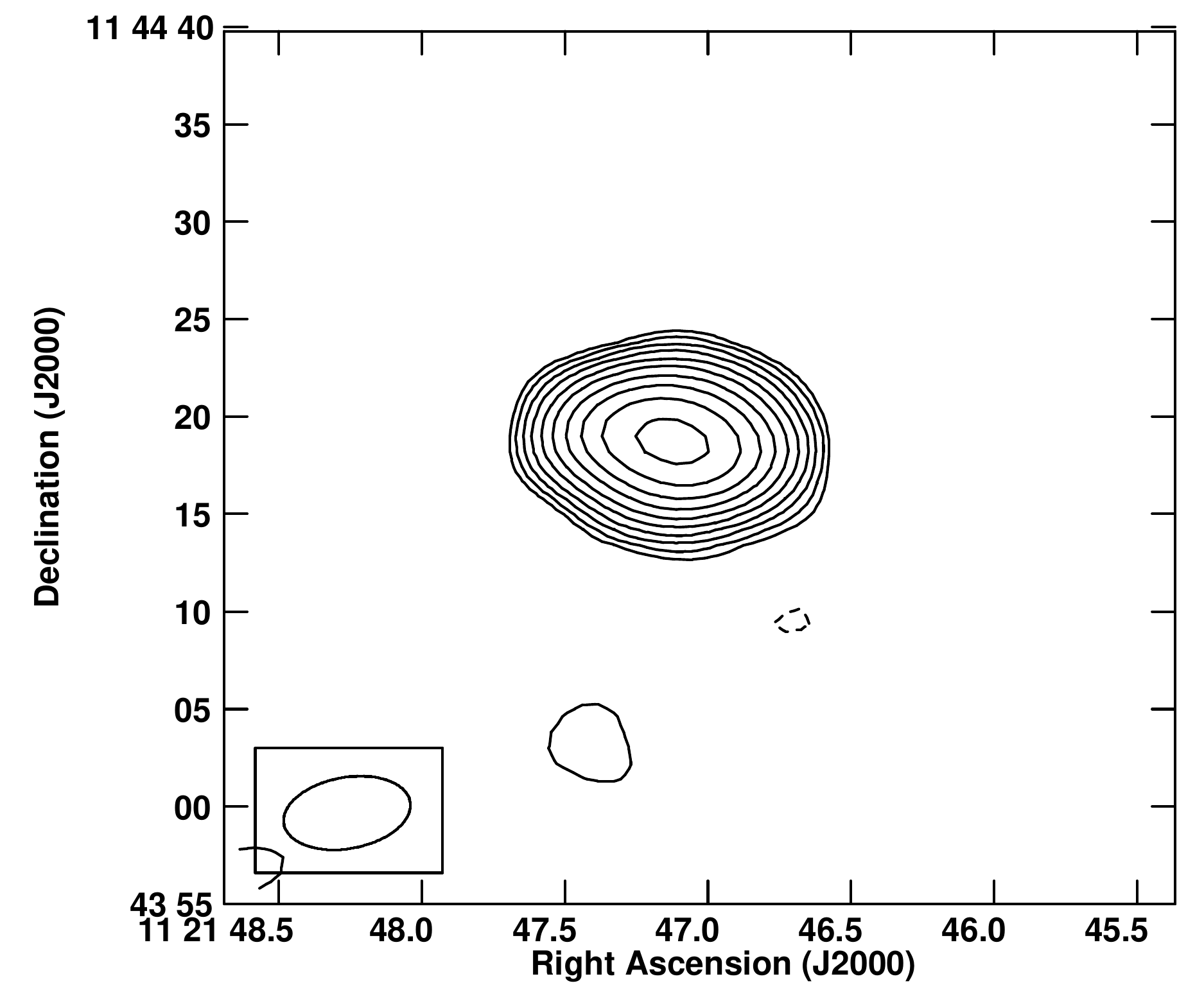}
\includegraphics[width=7cm]{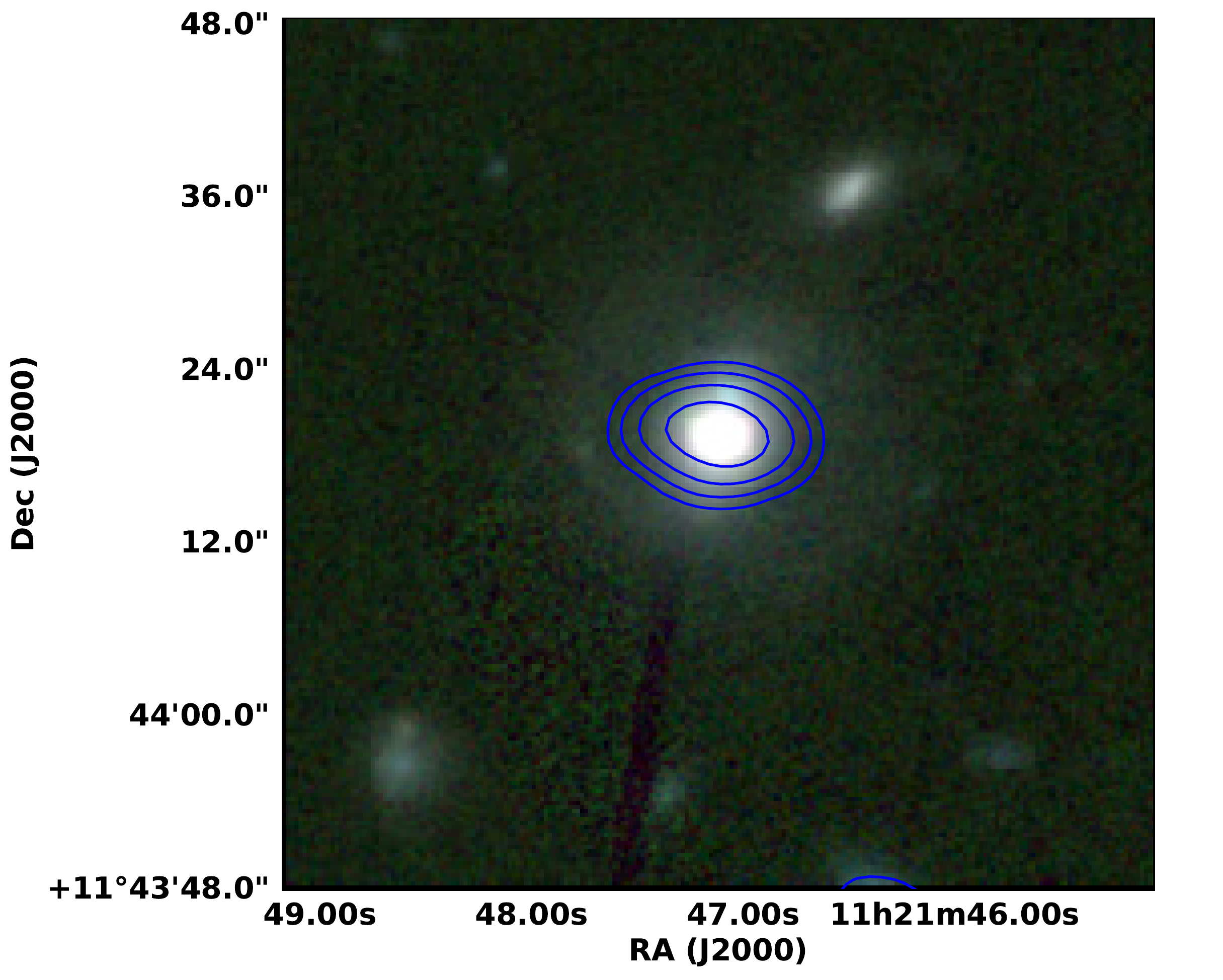}}
\end{figure*}

\begin{figure*}
\centerline{
\includegraphics[width=7cm]{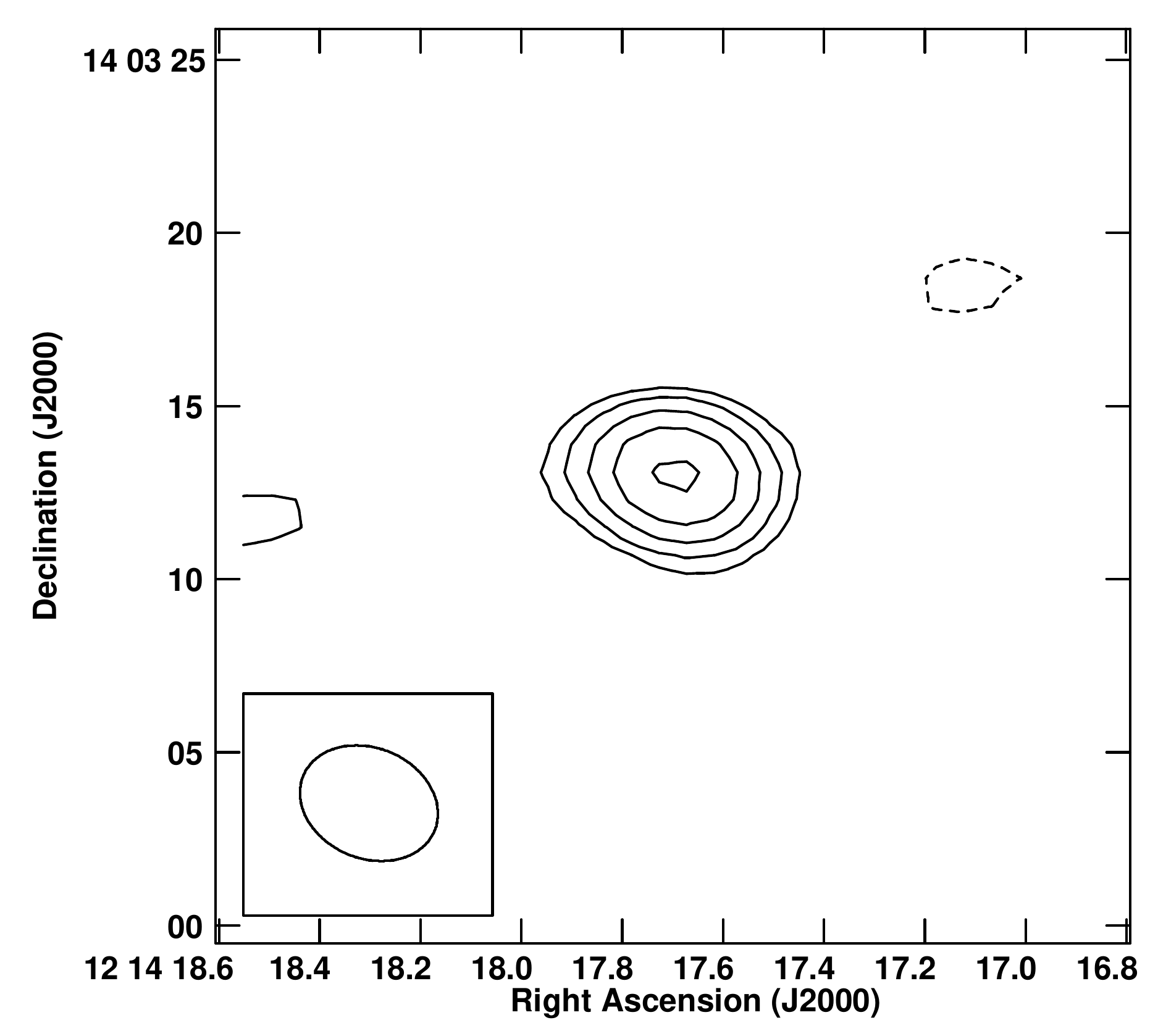}
\includegraphics[width=7cm]{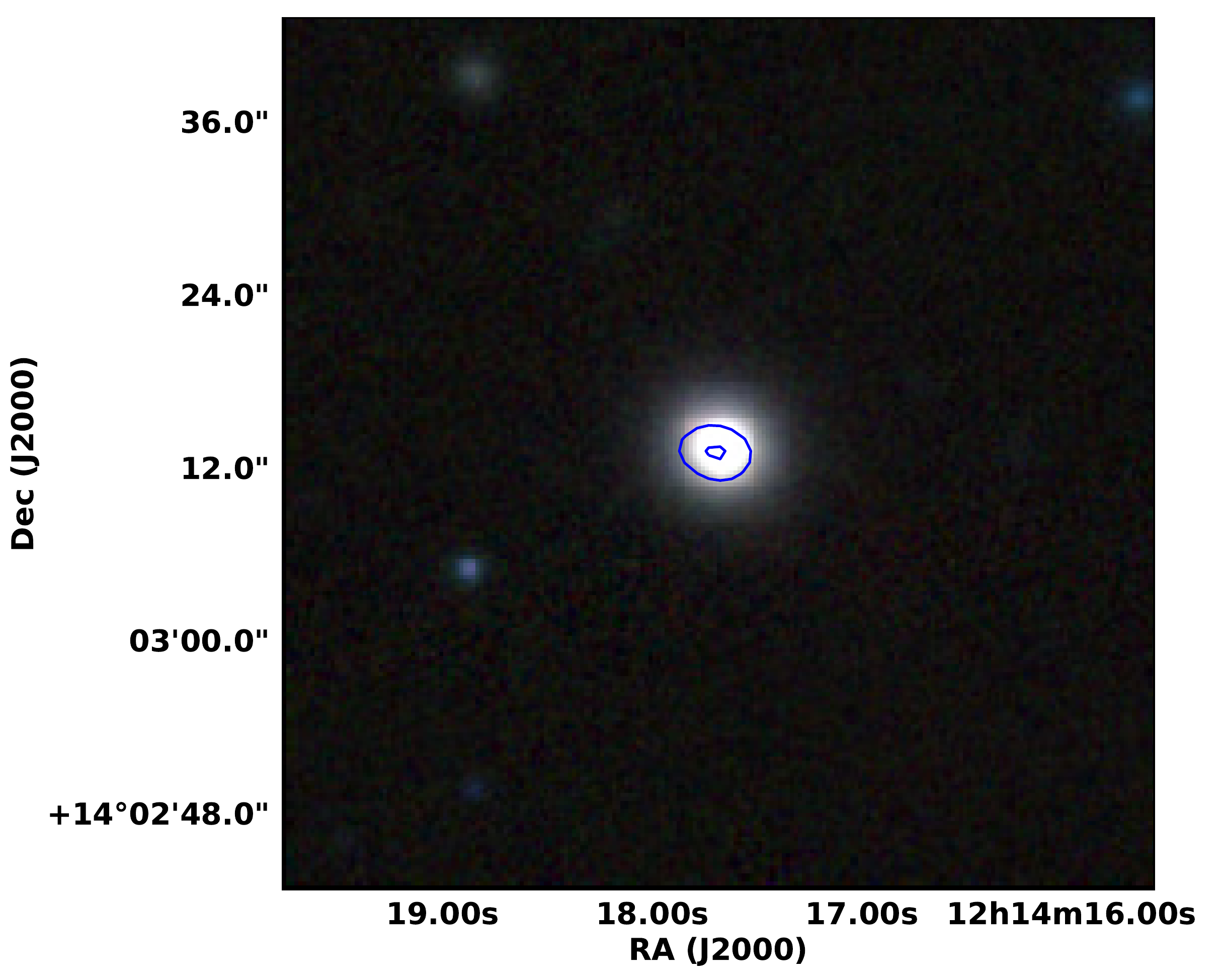}}
\end{figure*}

\begin{figure*}
\centerline{
\includegraphics[width=7cm]{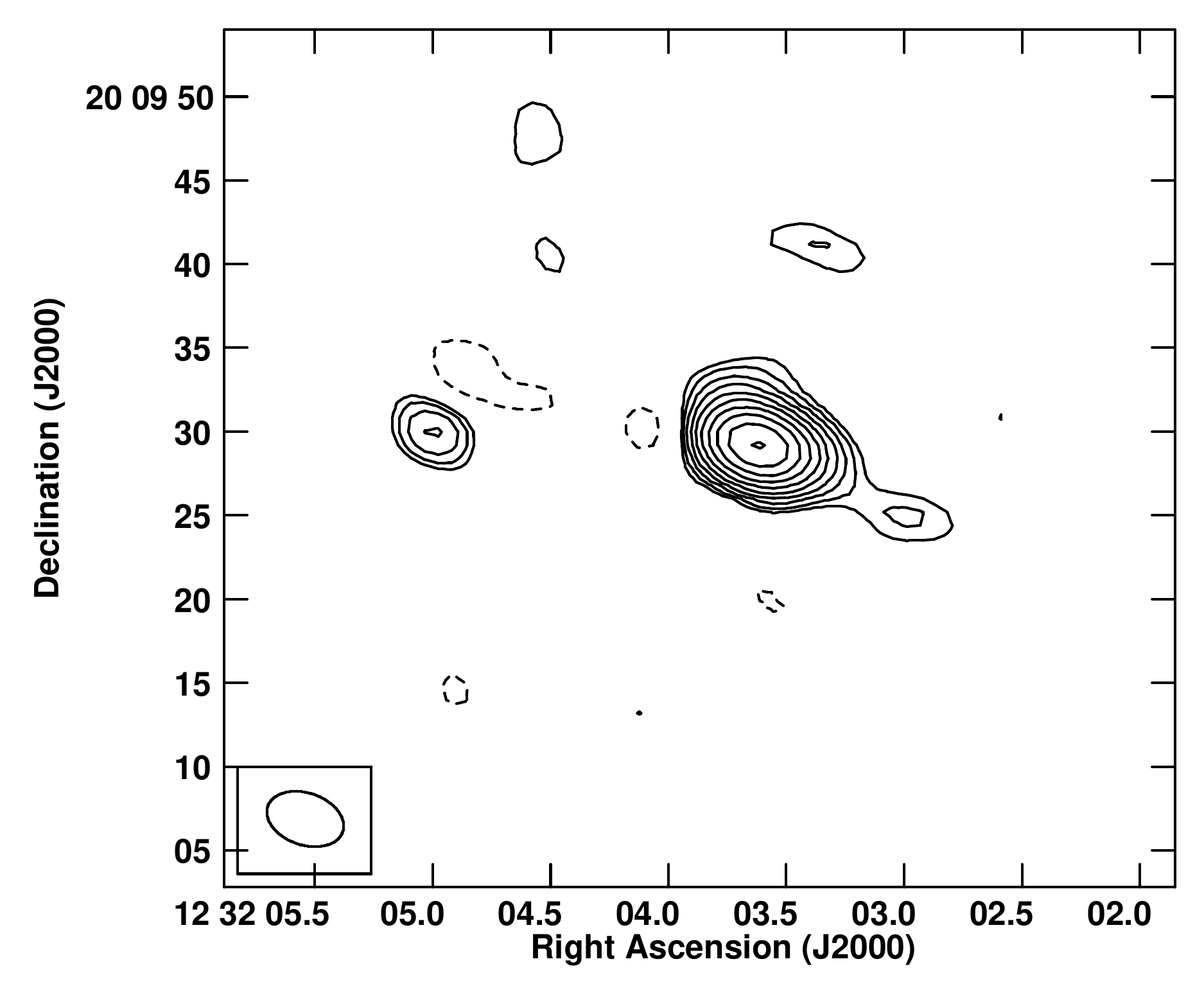}
\includegraphics[width=7cm]{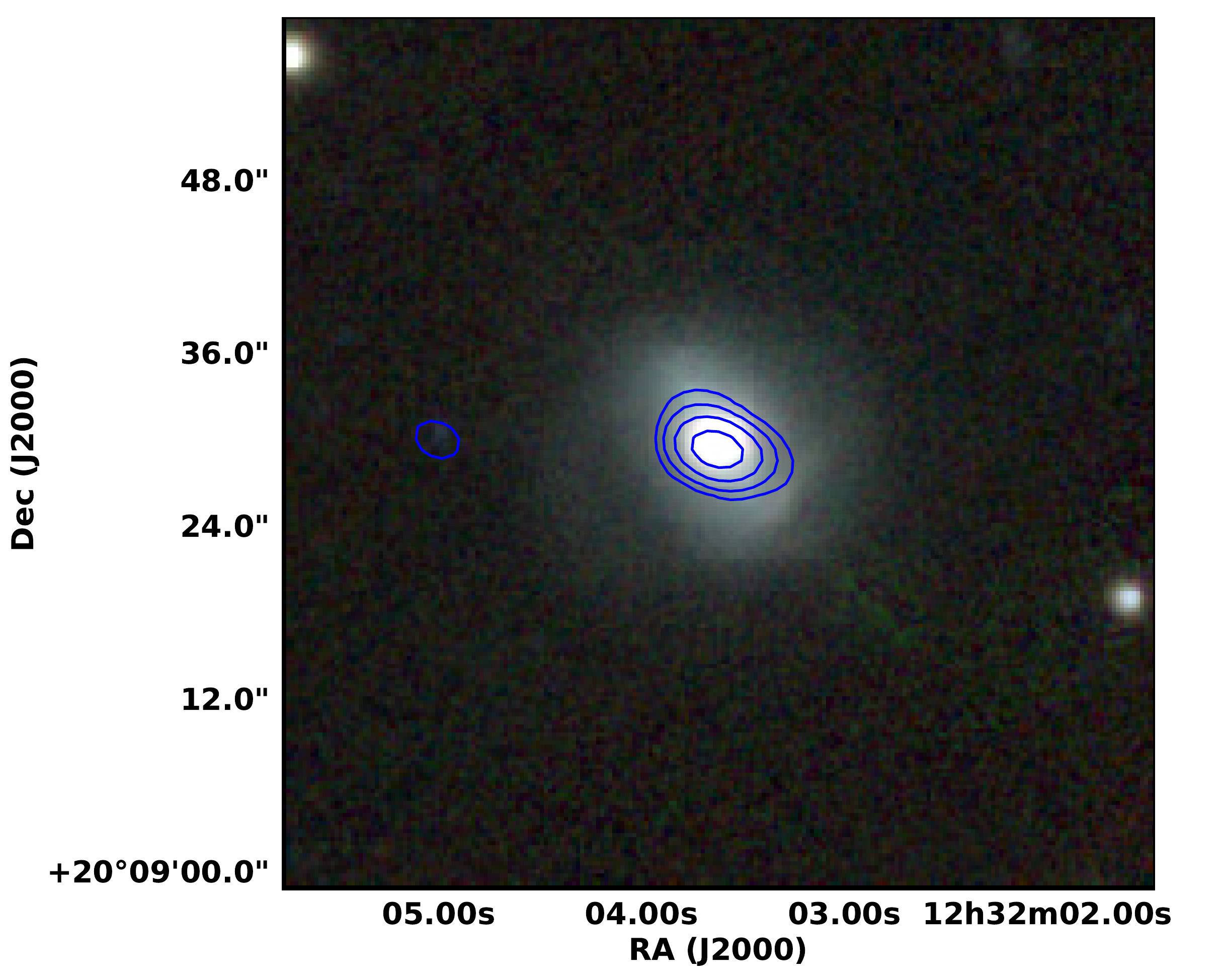}}
\caption{Left: 685 MHz total intensity contour images of PG 1119+120, PG 1211+143 and PG 1229+204 from top to bottom. The peak contour flux is {\it x} mJy beam$^{-1}$ and the contour levels are {\it y} $\times$ (-1, 1, 1.4, 2, 2.8, 4, 5.6, 8, 11.20, 16, 23, 32, 45, 64, 90, 128, 180, 256, 362, 512) mJy beam$^{-1}$, where {\it (x ; y)} for PG 1119+120, PG 1211+143 and PG 1229+204 are (66.7;0.12), (1211.4;1.1) and (160.6;0.08) respectively. Right: 685 MHz total intensity contours in blue superimposed on PanSTARRS $\it{griz}$-color composite optical image of PG 1119+120, PG 1211+143 and PG 1229+204 from top to bottom. The peak contour flux is {\it x} mJy beam$^{-1}$ and the contour levels are {\it y} $\times$ (2, 4, 8, 16, 32, 64, 128, 256, 512) mJy beam$^{-1}$, where {\it (x ; y)} for PG 1119+120, PG 1211+143 and PG 1229+204 are (66.7;0.12), (1211.4;1.1) and (160.6;0.08) respectively.}
\end{figure*}

\begin{figure*}
\centerline{
\includegraphics[width=7cm]{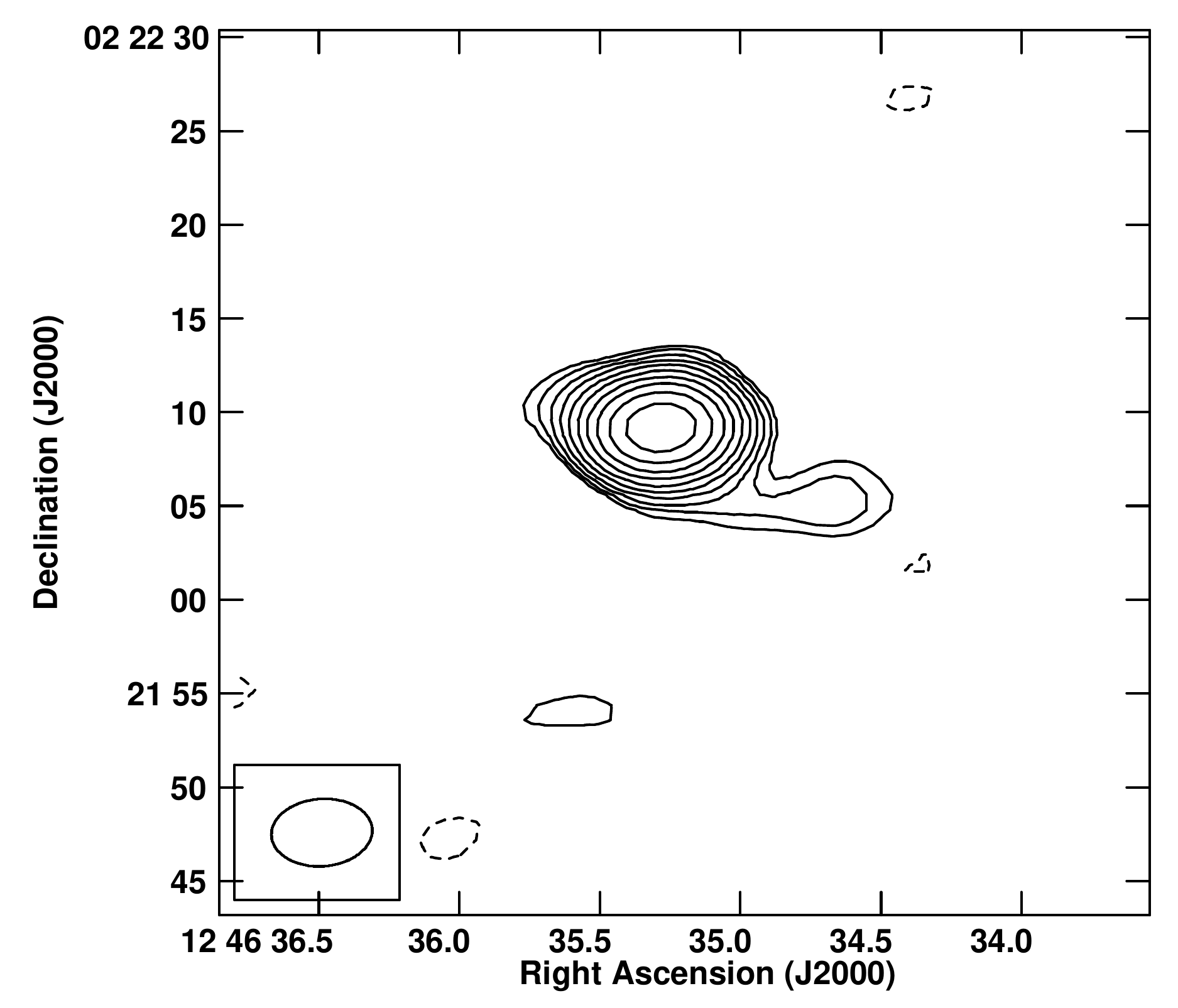}
\includegraphics[width=7cm]{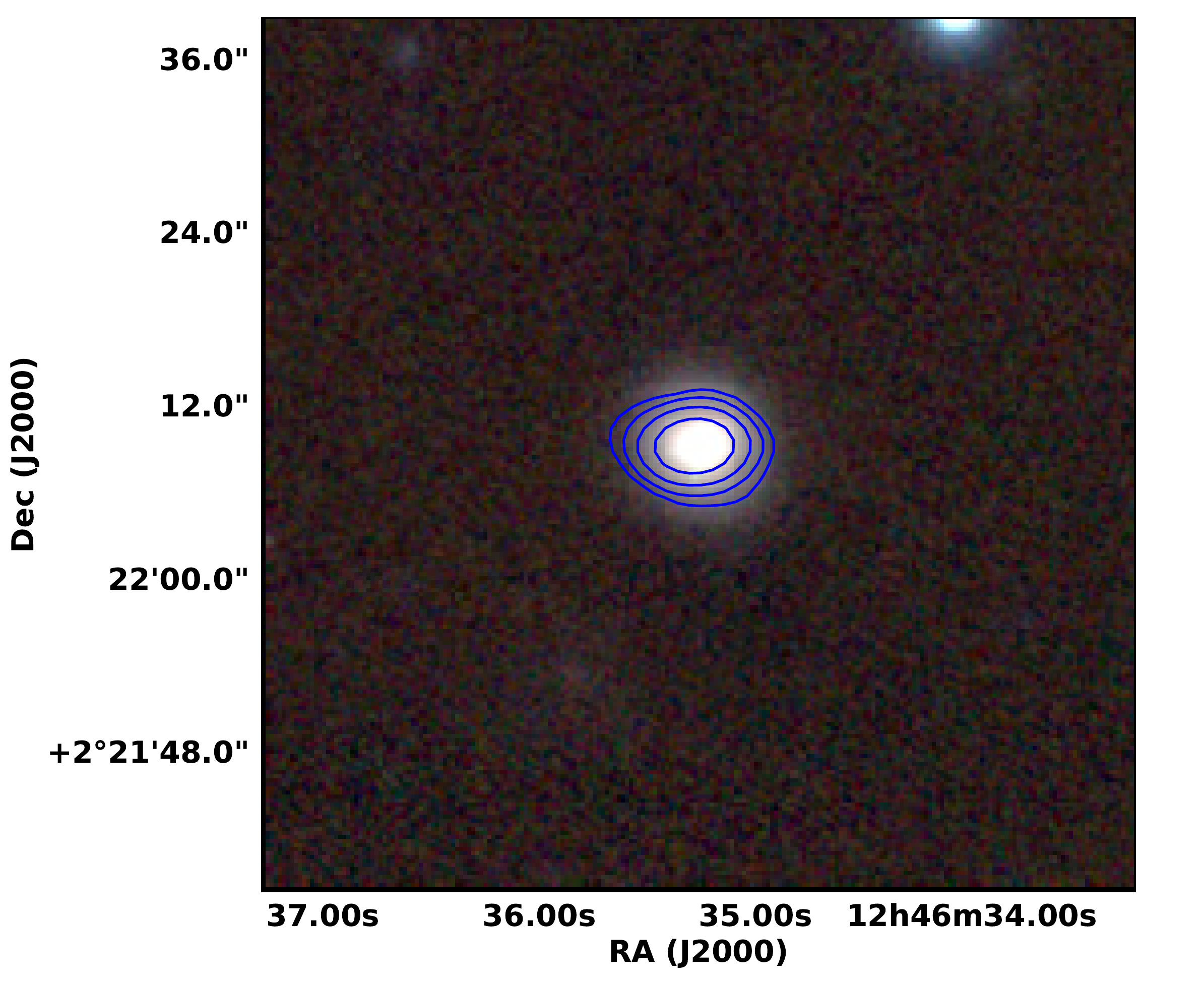}}
\end{figure*}

\begin{figure*}
\centerline{
\includegraphics[width=7cm]{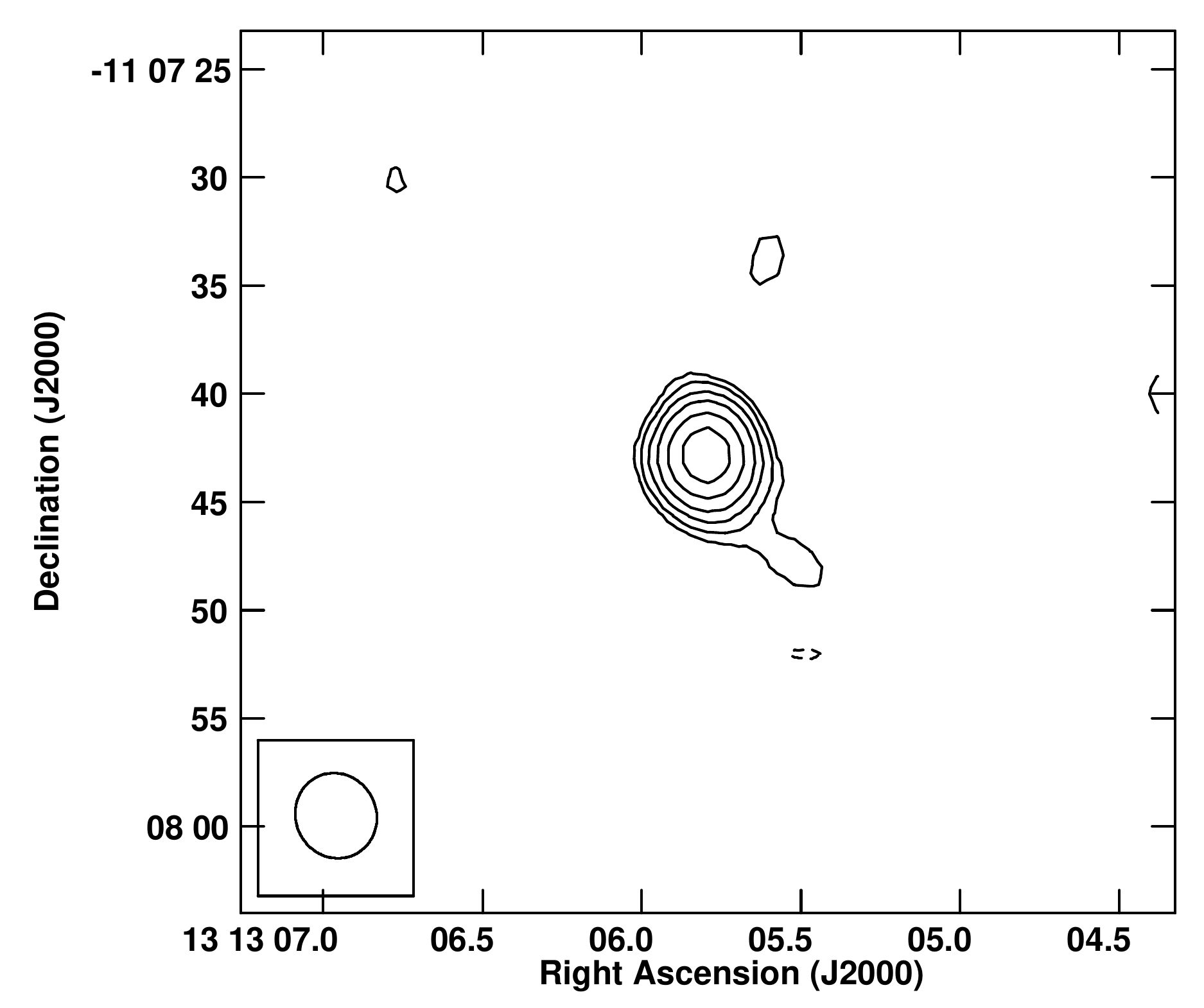}
\includegraphics[width=7cm]{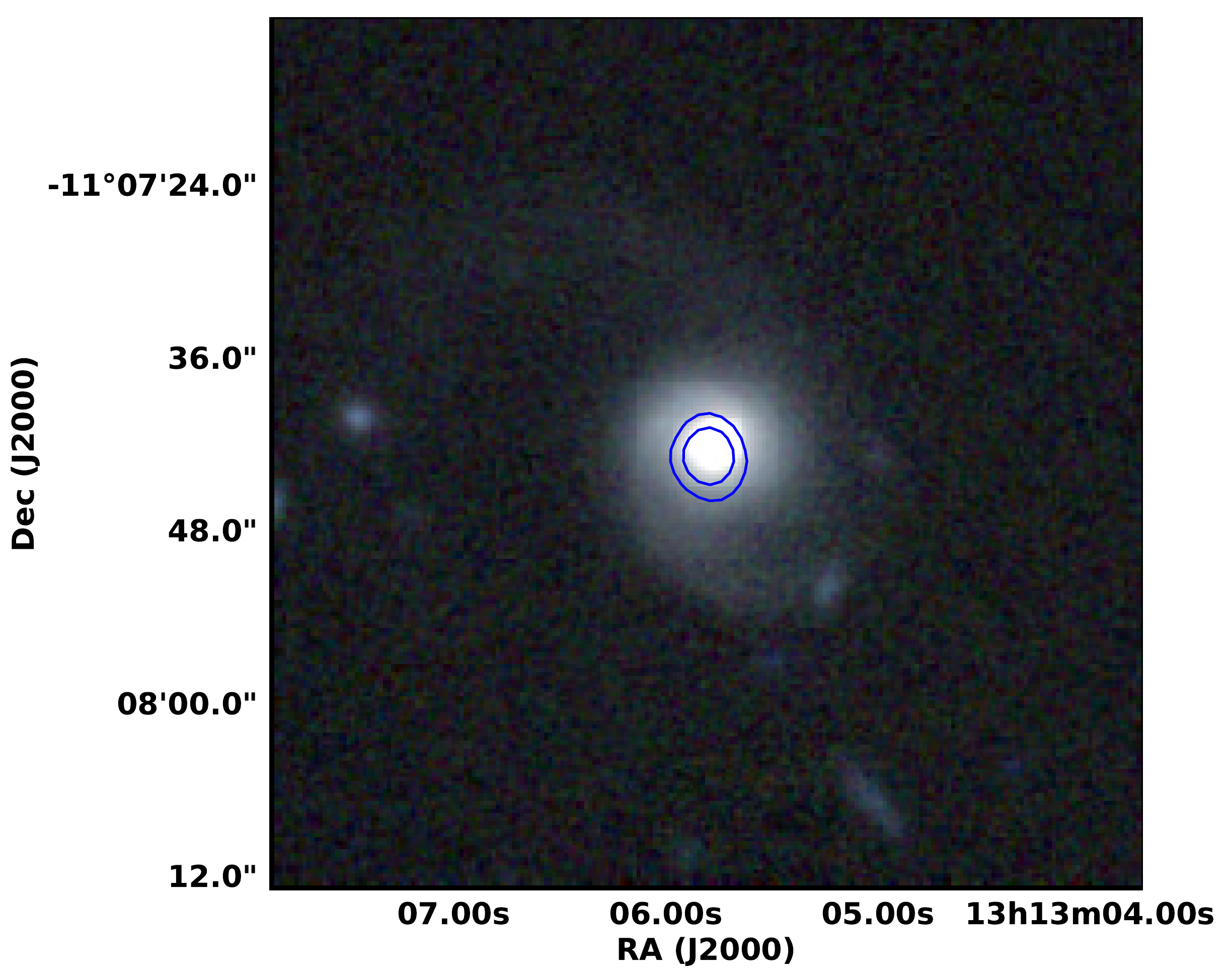}}
\end{figure*}

\begin{figure*}
\centerline{
\includegraphics[width=7cm]{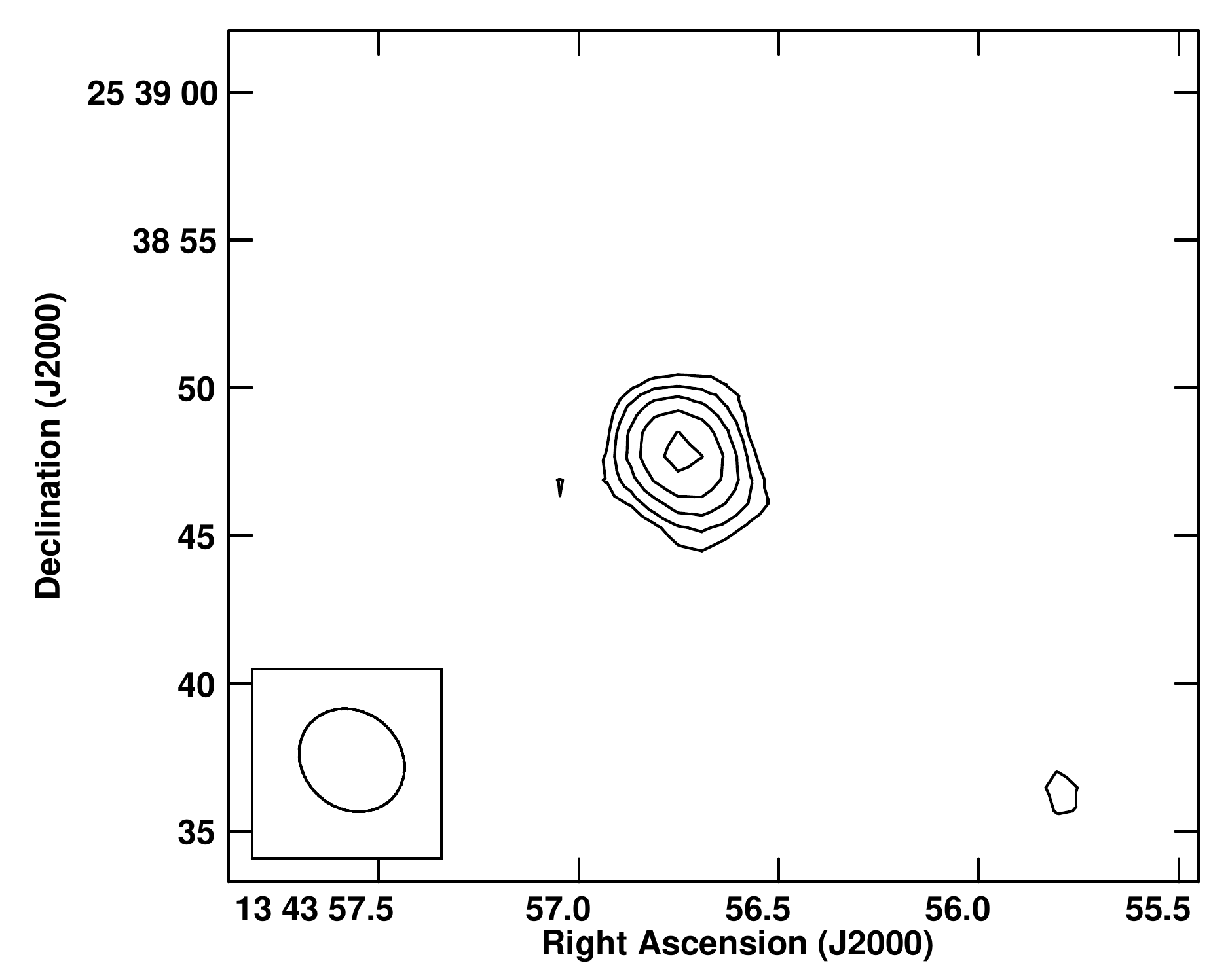}
\includegraphics[width=7cm]{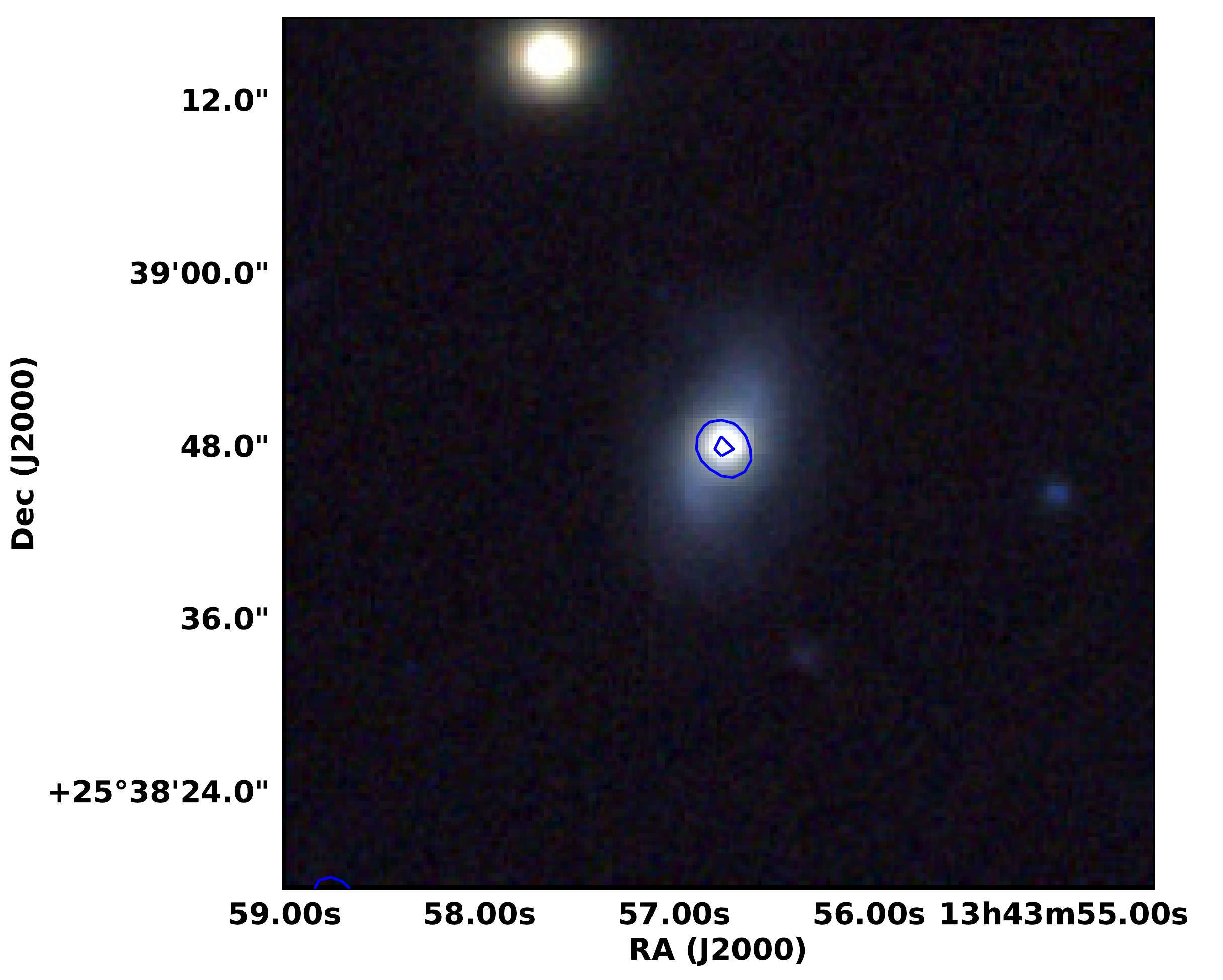}}
\caption{Left: 685 MHz total intensity contour images of PG 1244+026, PG 1310$-$108 and PG 1341+258 from top to bottom. The peak contour flux is {\it x} mJy beam$^{-1}$ and the contour levels are {\it y} $\times$ (-1, 1, 1.4, 2, 2.8, 4, 5.6, 8, 11.20, 16, 23, 32, 45, 64, 90, 128, 180, 256, 362, 512) mJy beam$^{-1}$, where {\it (x ; y)} for PG 1244+026, PG 1310$-$108 and PG 1341+258 are (41.9;0.09), (51.67;0.10) and (95.69;0.07) respectively. Right: 685 MHz total intensity contours in blue superimposed on PanSTARRS $\it{griz}$-color composite optical image of PG 1244+026, PG 1310$-$108 and PG 1341+258 from top to bottom. The peak contour flux is {\it x} mJy beam$^{-1}$ and the contour levels are {\it y} $\times$ (2, 4, 8, 16, 32, 64, 128, 256, 512) mJy beam$^{-1}$, where {\it (x ; y)} for PG 1244+026, PG 1310$-$108 and PG 1341+258 are (41.9;0.09), (51.67;0.10) and (95.69;0.07) respectively.}
\end{figure*}

\begin{figure*}
\centerline{
\includegraphics[width=7cm]{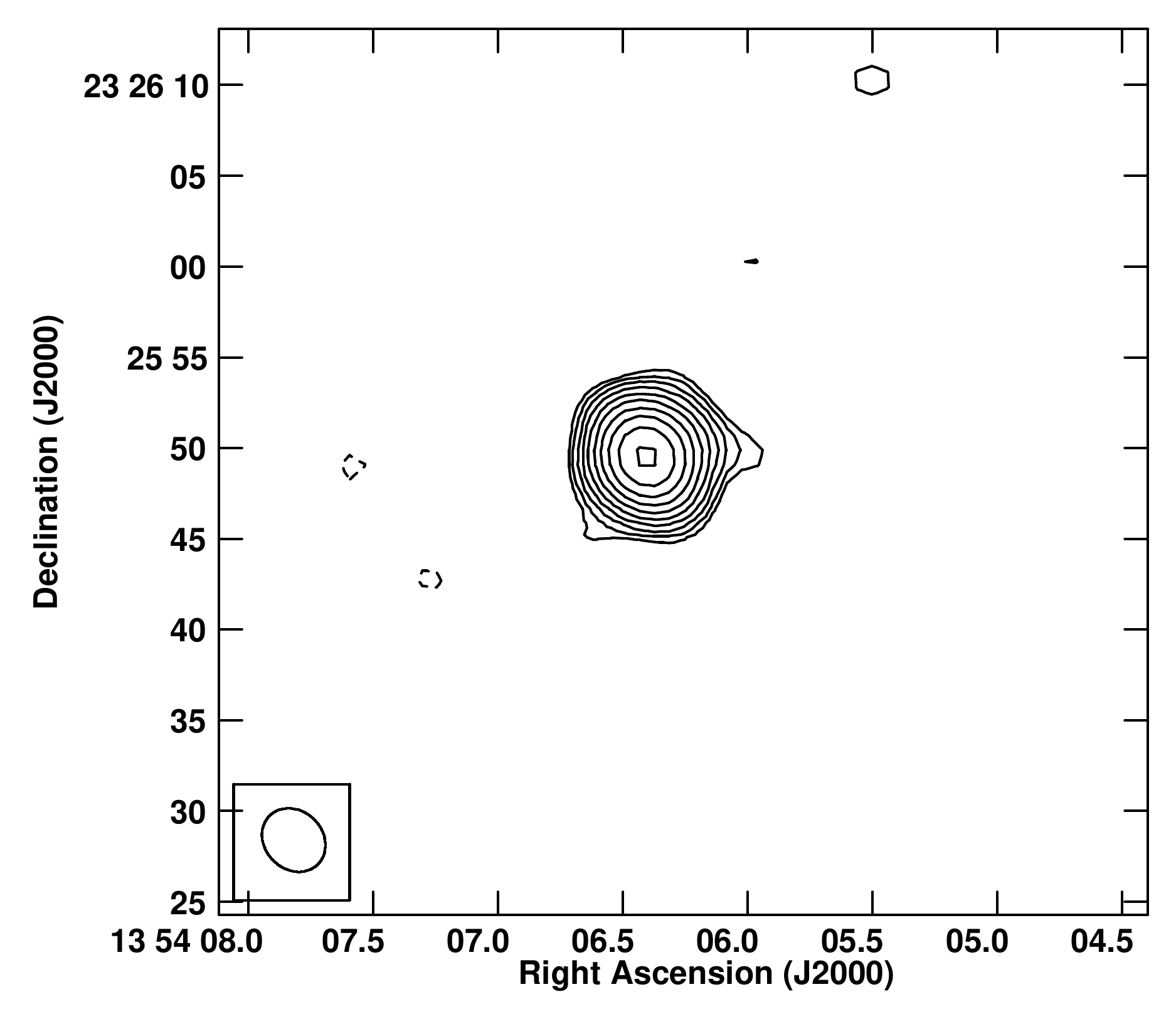}
\includegraphics[width=7cm]{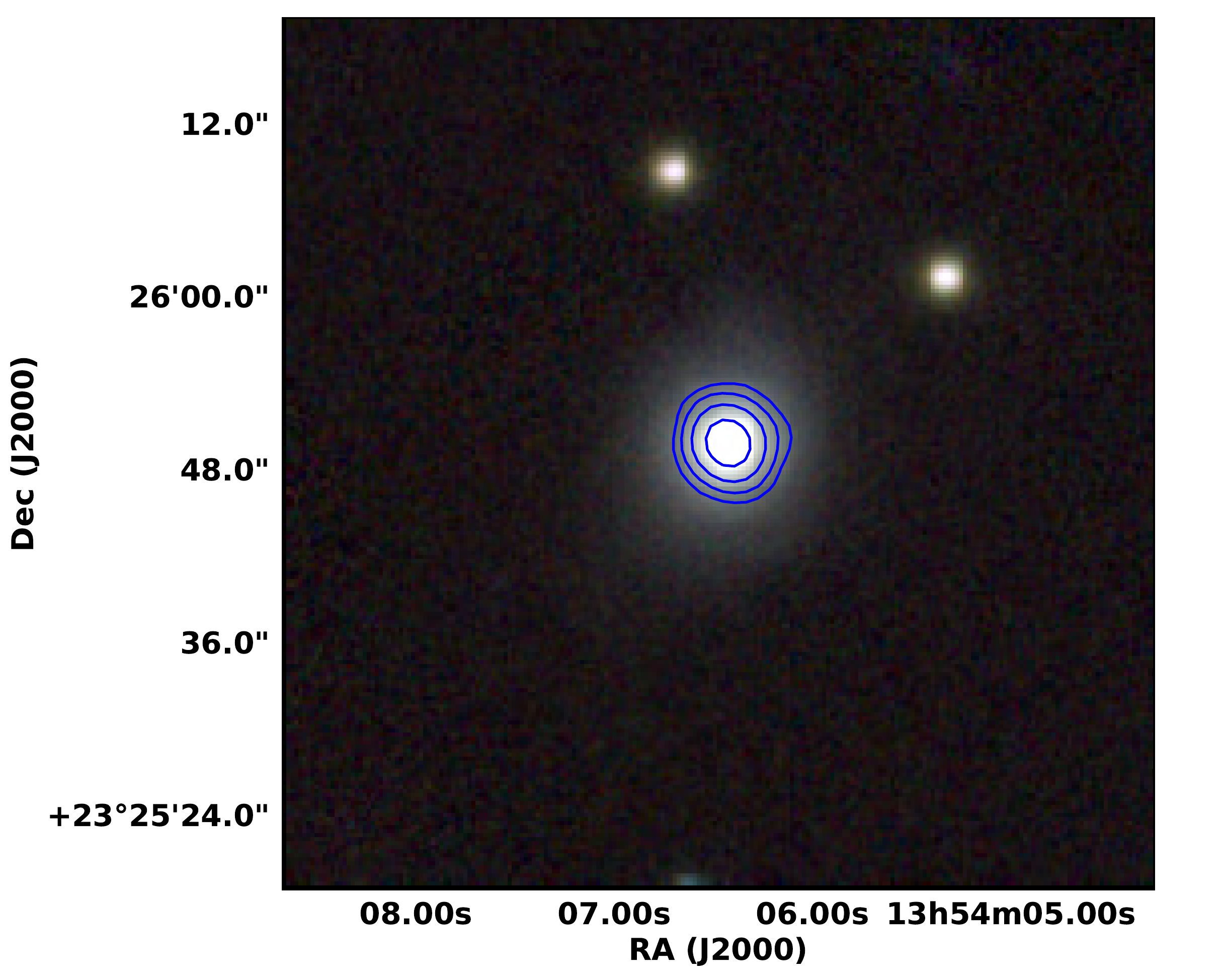}}
\end{figure*}

\begin{figure*}
\centerline{
\includegraphics[width=7cm]{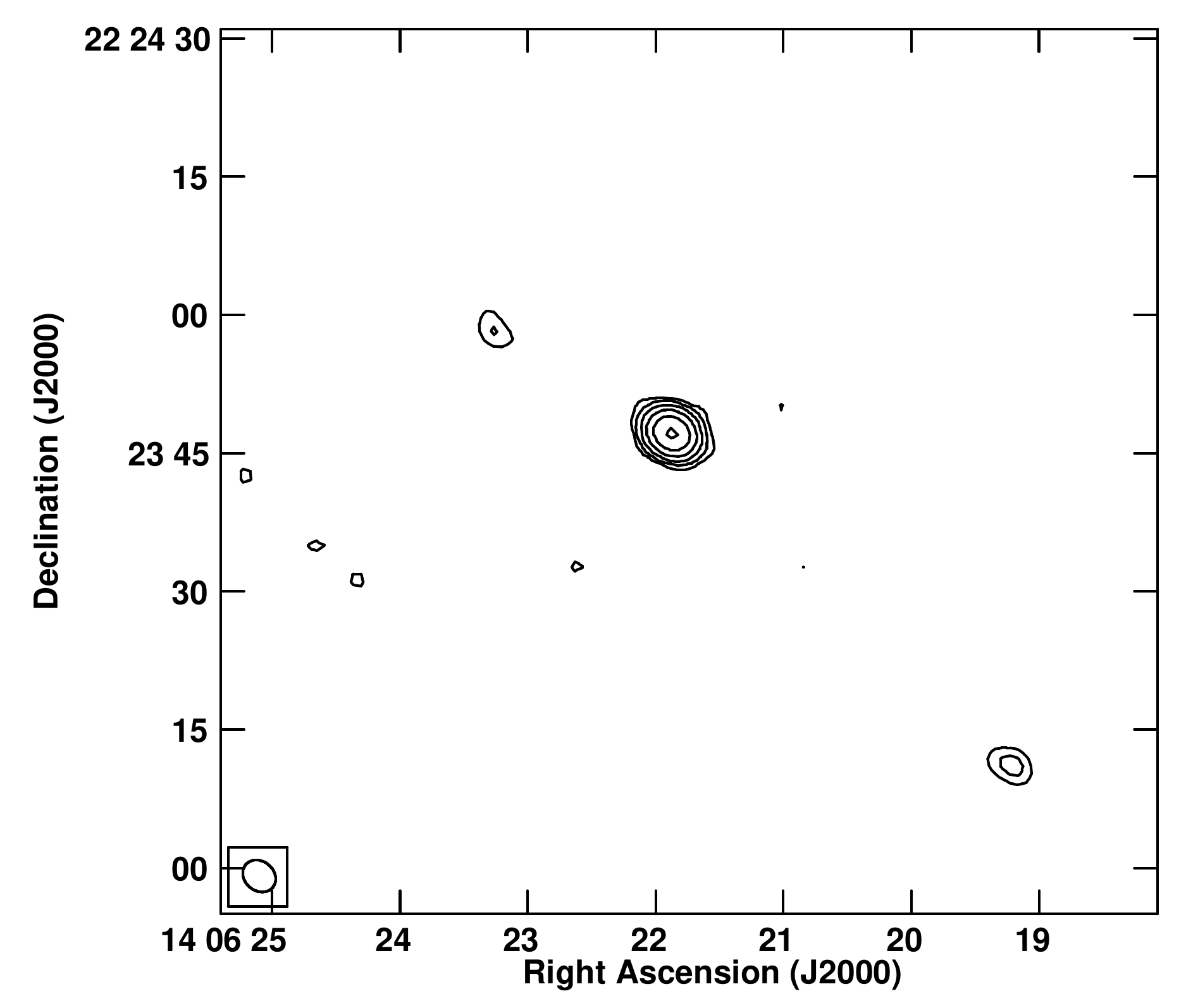}
\includegraphics[width=7cm]{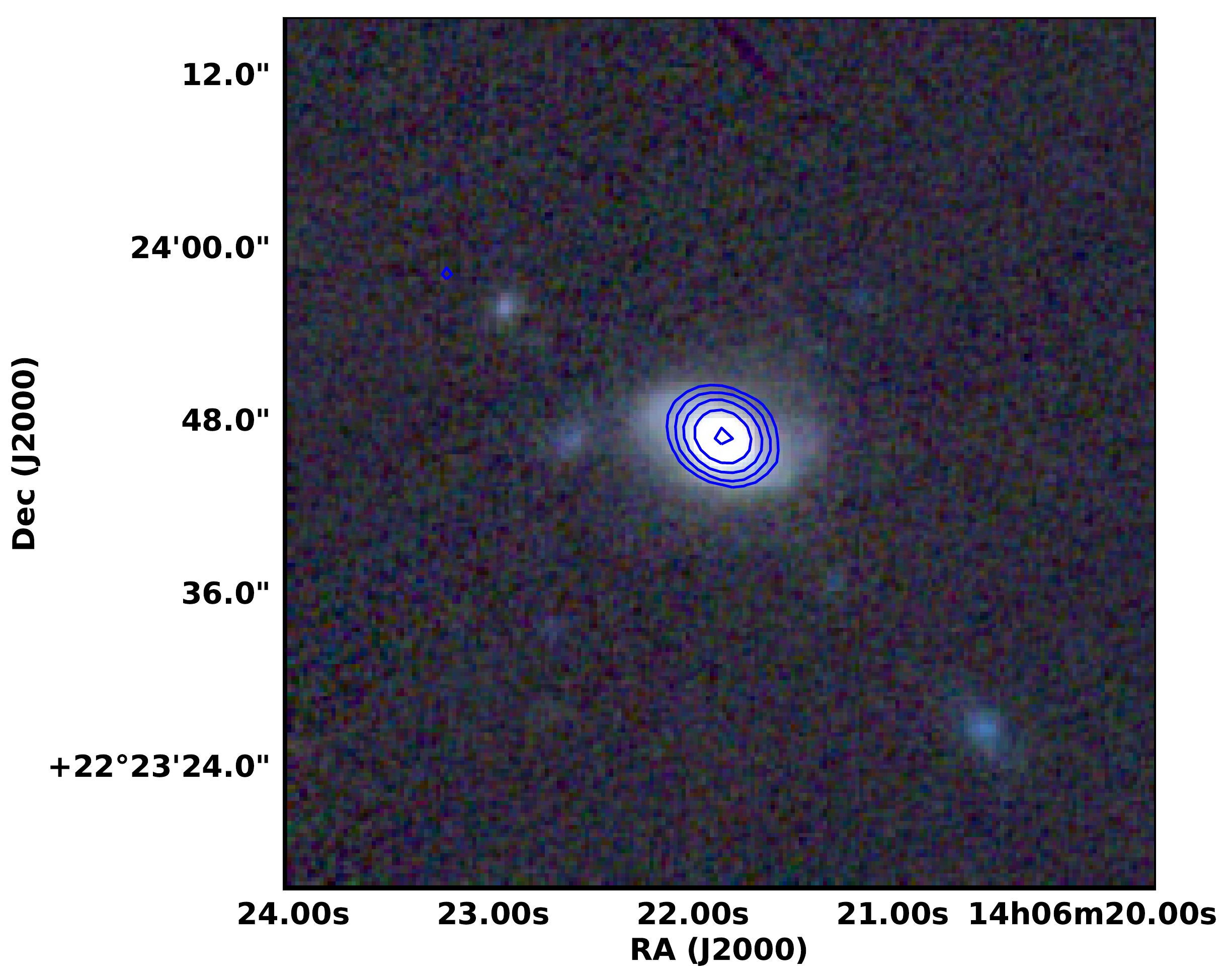}}
\end{figure*}

\begin{figure*}
\centerline{
\includegraphics[width=7cm]{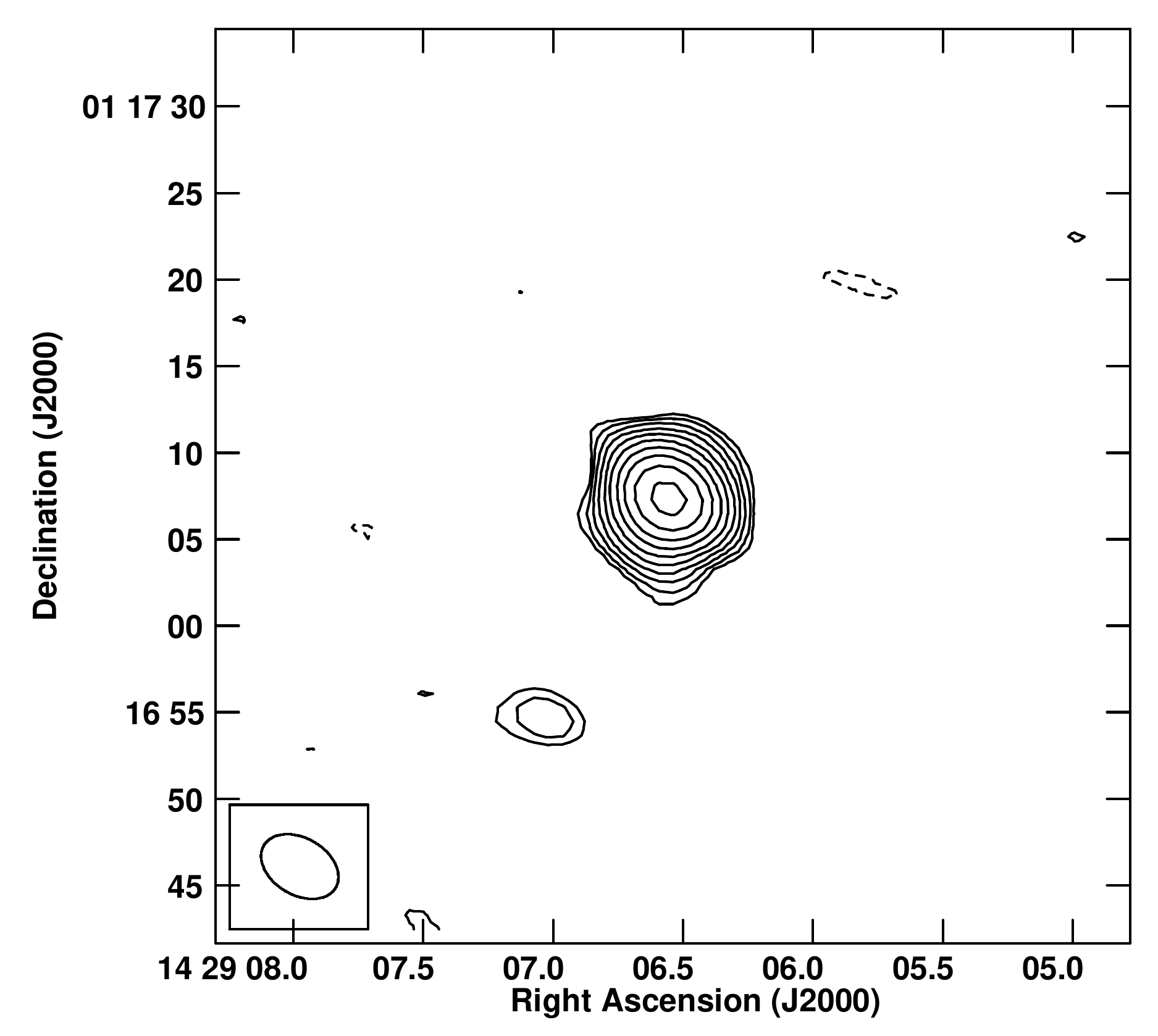}
\includegraphics[width=7cm]{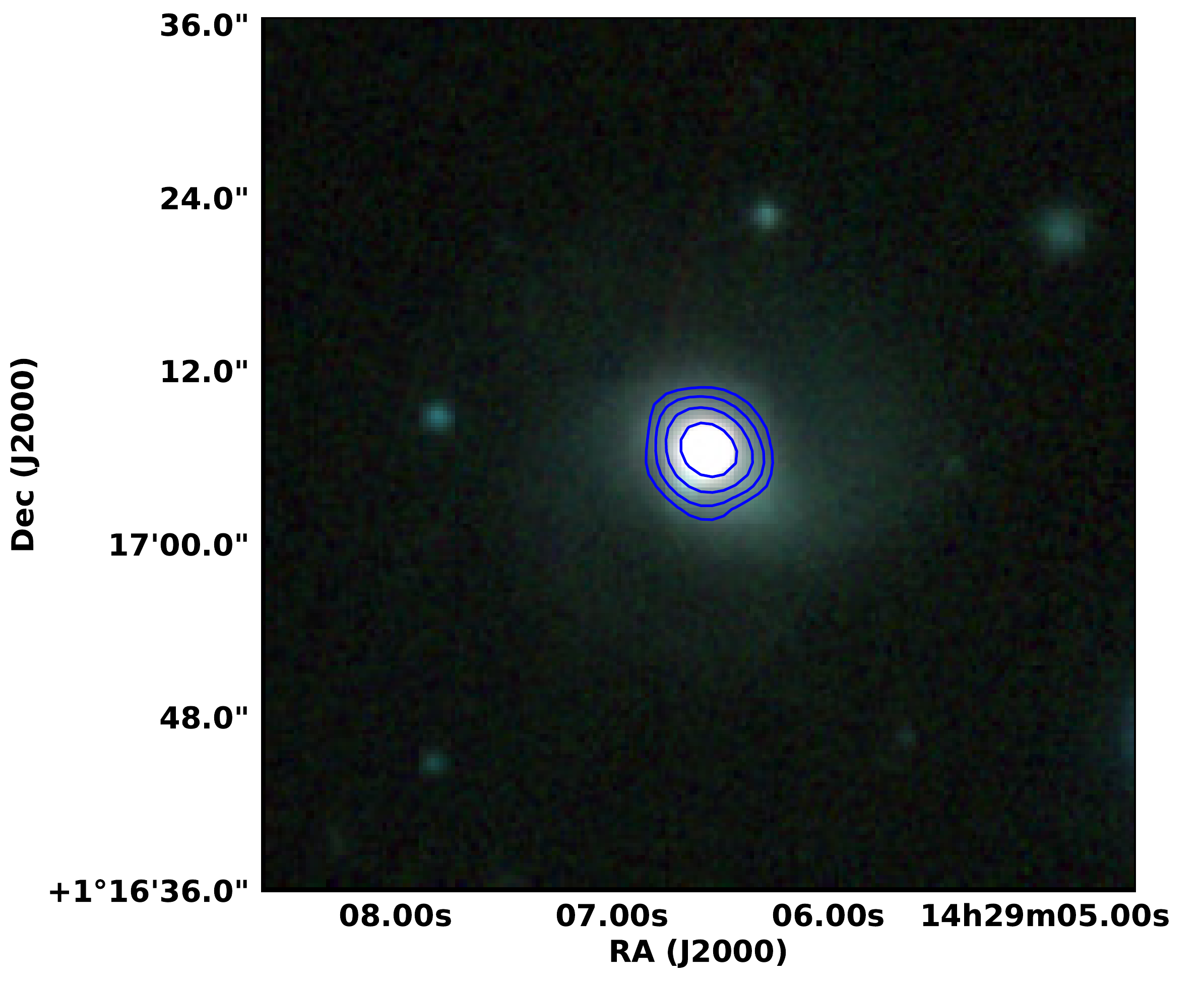}}
\caption{Left: 685 MHz total intensity contour images of PG 1351+236, PG 1404+226 and PG 1426+015 from top to bottom. The peak contour flux is {\it x} mJy beam$^{-1}$ and the contour levels are {\it y} $\times$ (-1, 1, 1.4, 2, 2.8, 4, 5.6, 8, 11.20, 16, 23, 32, 45, 64, 90, 128, 180, 256, 362, 512) mJy beam$^{-1}$, where {\it (x ; y)} for PG 1351+236 and PG 1426+015 are (34.2;0.08) and (48.8;0.08) respectively. The peak contour flux for PG 1404+226 is 143.3 mJy beam$^{-1}$ and the contour levels are 0.08 $\times$ (-1, 1, 2, 4, 8, 16, 32, 64, 128, 256, 512) mJy beam$^{-1}$. Right: 685 MHz total intensity contours in blue superimposed on PanSTARRS $\it{griz}$-color composite optical image of PG 1351+236, PG 1404+226 and PG 1426+015 from top to bottom. The peak contour flux is {\it x} mJy beam$^{-1}$ and the contour levels are {\it y} $\times$ (2, 4, 8, 16, 32, 64, 128, 256, 512) mJy beam$^{-1}$, where {\it (x ; y)} for PG 1351+236, PG 1404+226 and PG 1426+015 are (34.2;0.08), (143.3;0.08) and (48.8;0.08) respectively.}
\end{figure*}

\begin{figure*}
\centerline{
\includegraphics[width=7cm]{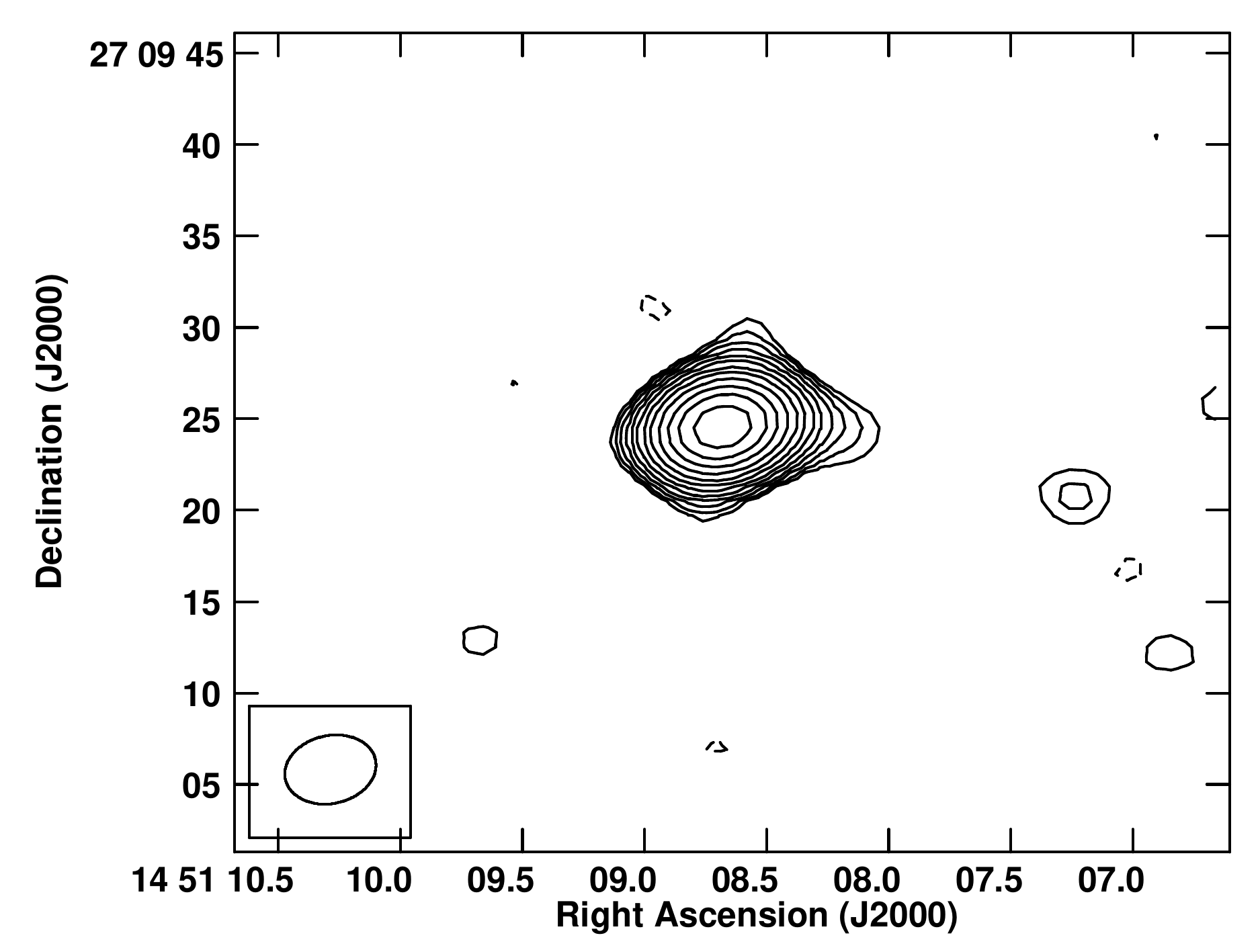}
\includegraphics[width=7cm]{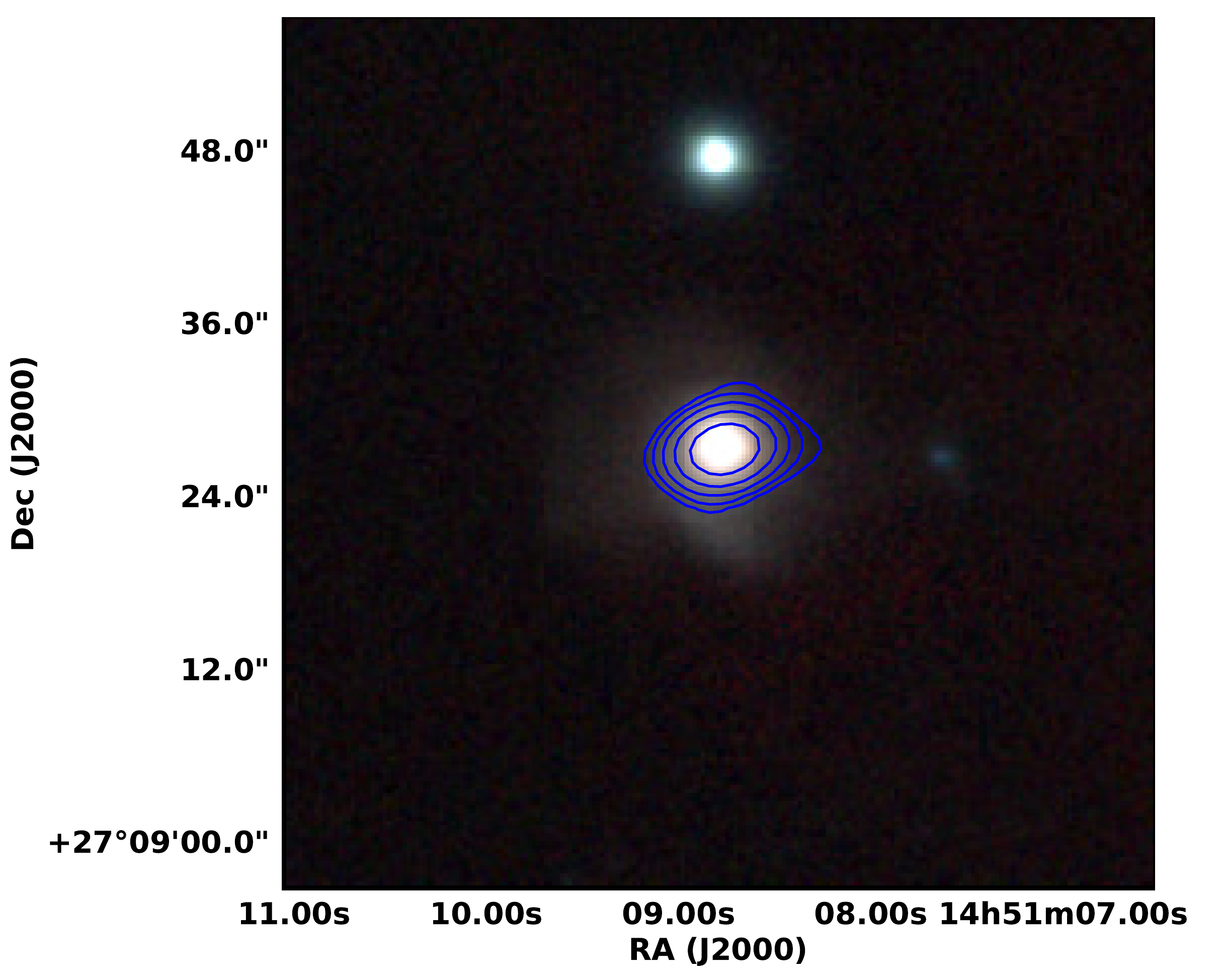}}
\end{figure*}

\begin{figure*}
\centerline{
\includegraphics[width=7cm]{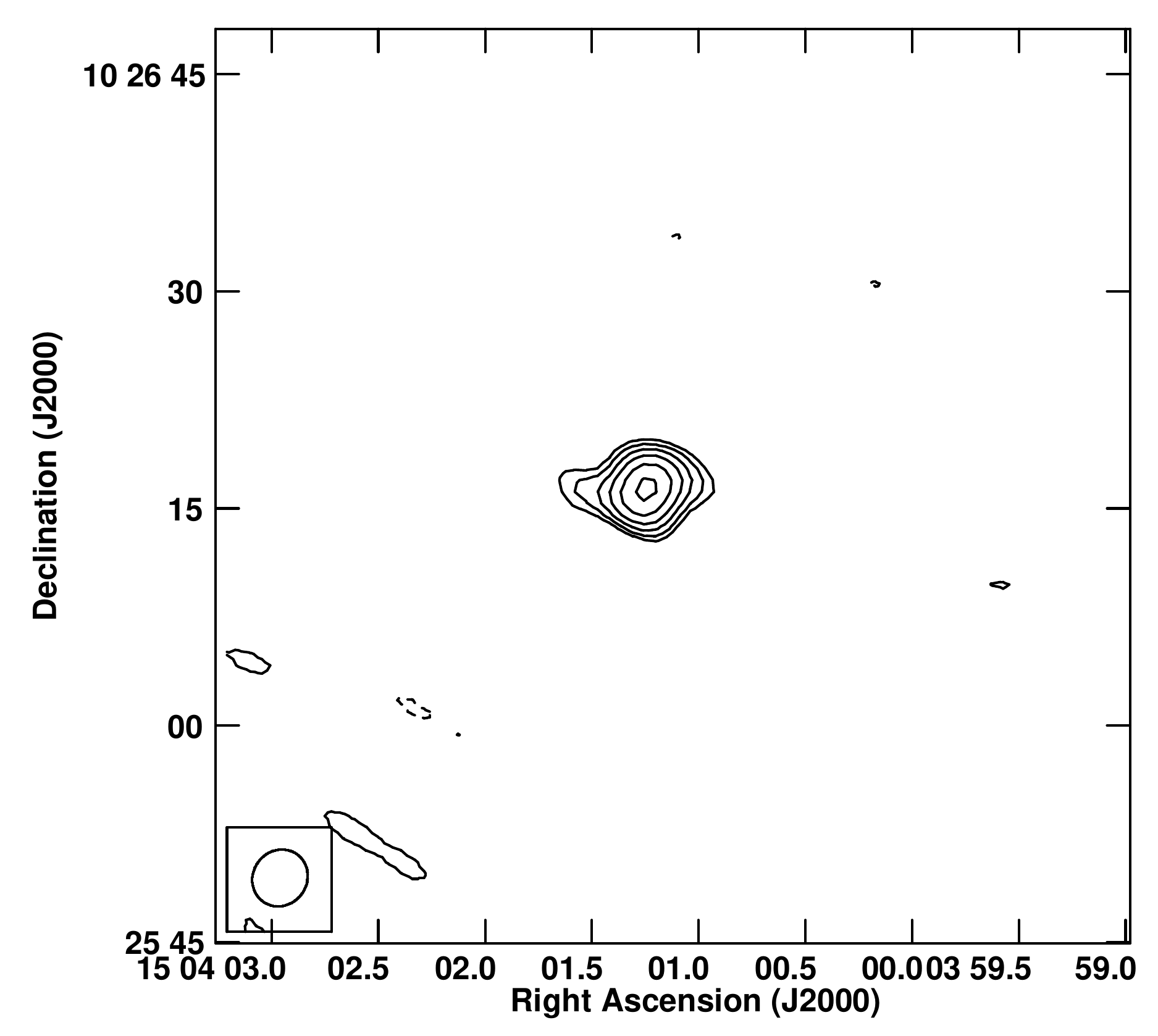}
\includegraphics[width=7cm]{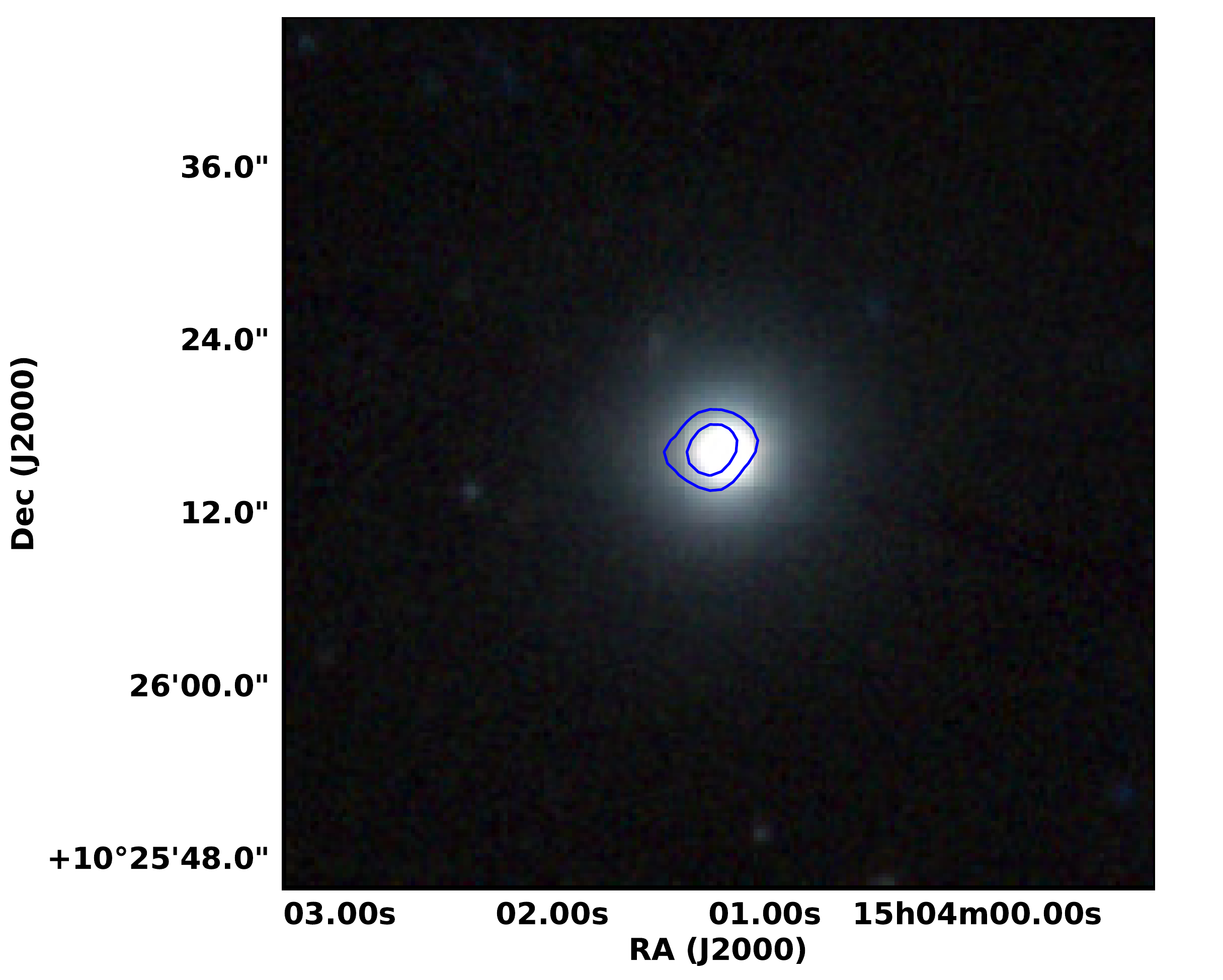}}
\end{figure*}

\begin{figure*}
\centerline{
\includegraphics[width=7cm]{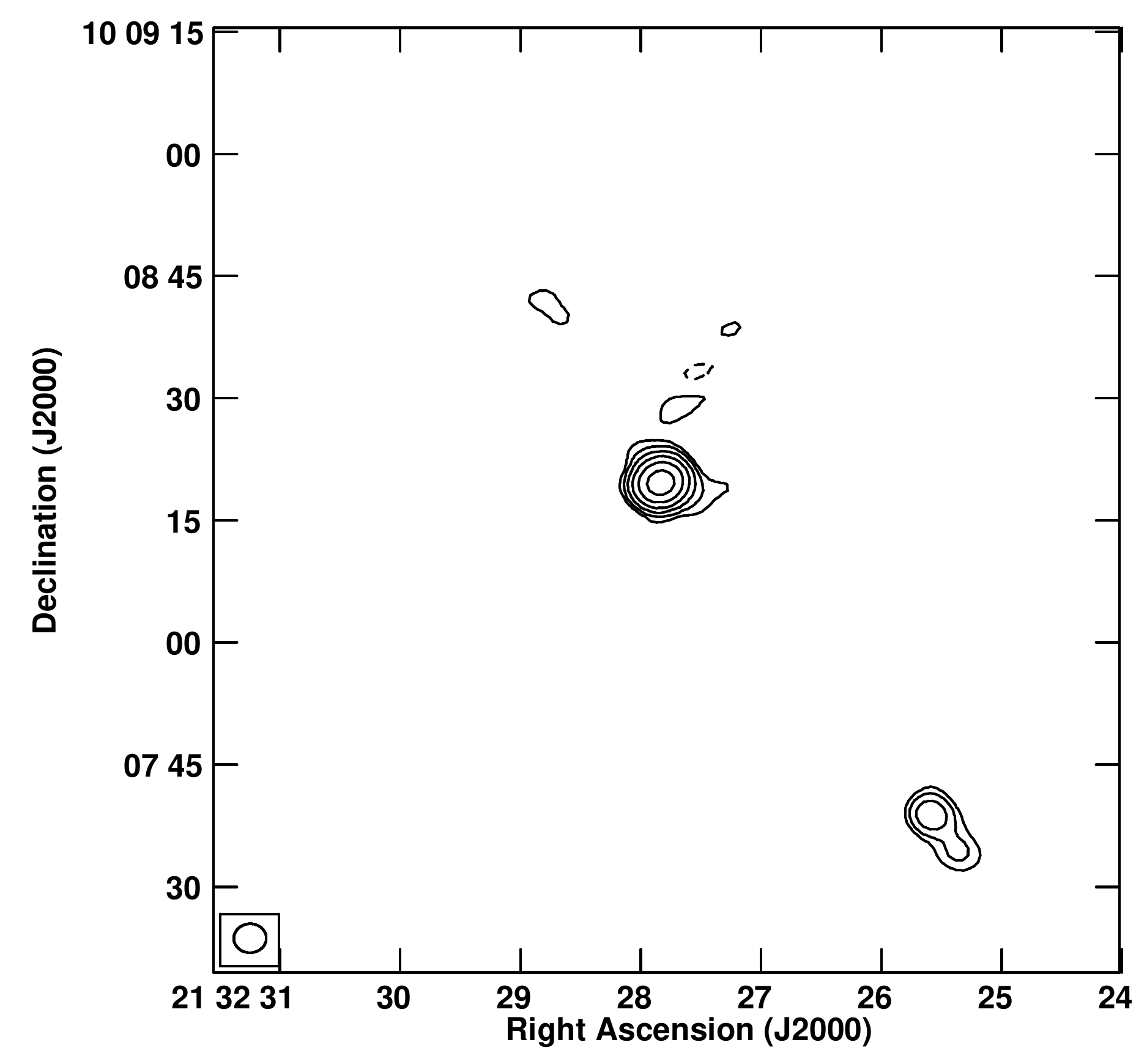}
\includegraphics[width=7cm]{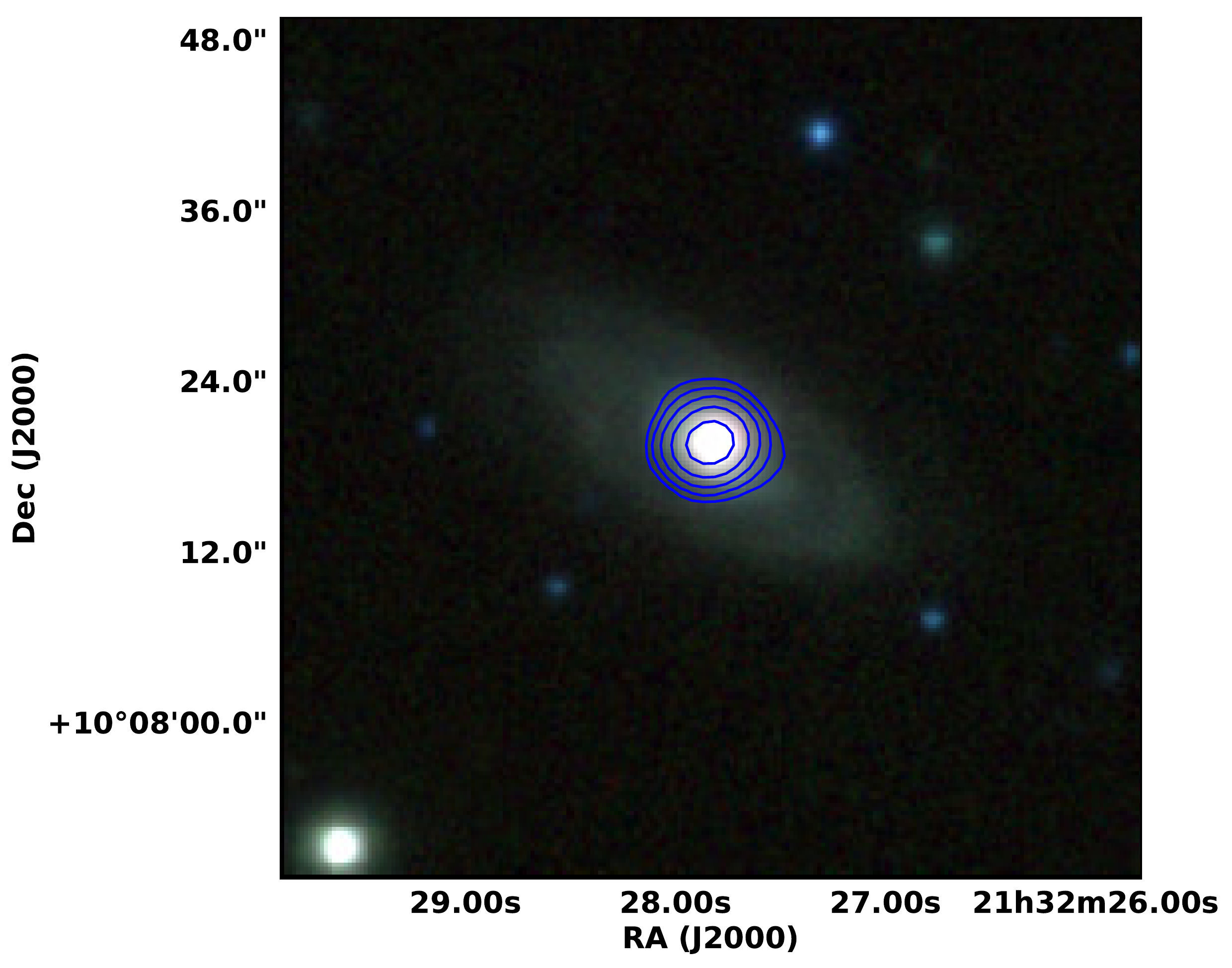}}
\caption{Left: 685 MHz total intensity contour images of PG 1448+273, PG 1501+106 and PG 2130+099 from top to bottom. The peak contour flux is {\it x} mJy beam$^{-1}$ and the contour levels are {\it y} $\times$ (-1, 1, 1.4, 2, 2.8, 4, 5.6, 8, 11.20, 16, 23, 32, 45, 64, 90, 128, 180, 256, 362, 512) mJy beam$^{-1}$, where {\it (x ; y)} for PG 1448+273 and PG 1501+106 are (18.4;0.09) and (1469.5;0.6) respectively. The peak contour flux for PG 2130+099 is 67.2 mJy beam$^{-1}$ and the contour levels are 0.15 $\times$ (-1, 1, 2, 4, 8, 16, 32, 64, 128, 256, 512) mJy beam$^{-1}$. Right: 685 MHz total intensity contours in blue superimposed on PanSTARRS $\it{griz}$-color composite optical image of PG 1448+273, PG 1501+106 and PG 2130+099 from top to bottom. The peak contour flux is {\it x} mJy beam$^{-1}$ and the contour levels are {\it y} $\times$ (2, 4, 8, 16, 32, 64, 128, 256, 512) mJy beam$^{-1}$, where {\it (x ; y)} for PG 1448+273, PG 1501+106 and PG 2130+099 are (18.4;0.09), (1469.5;0.6) and (67.2;0.15) respectively.}
\end{figure*}

\begin{figure*}
\centerline{
\includegraphics[width=7cm]{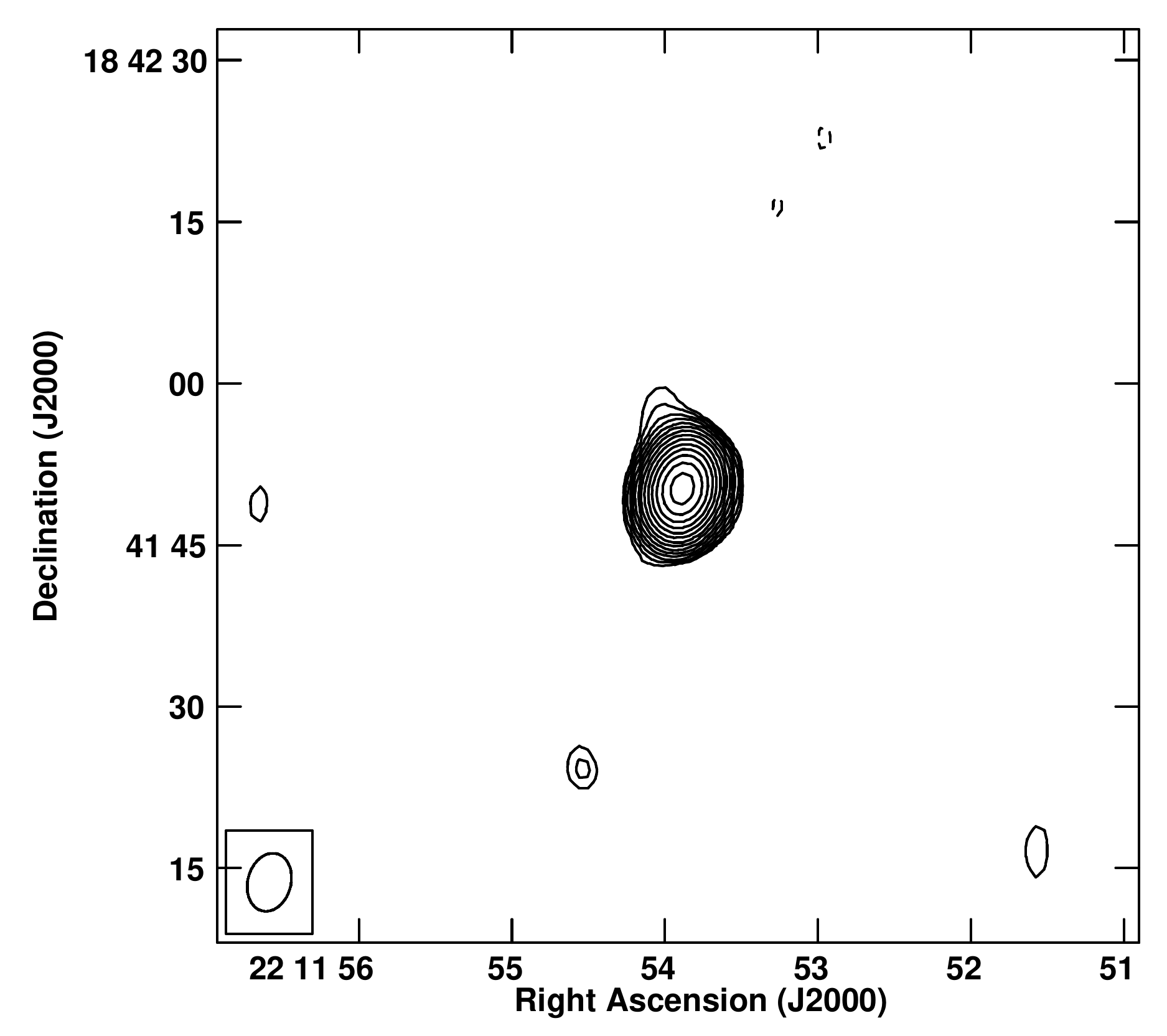}
\includegraphics[width=7cm]{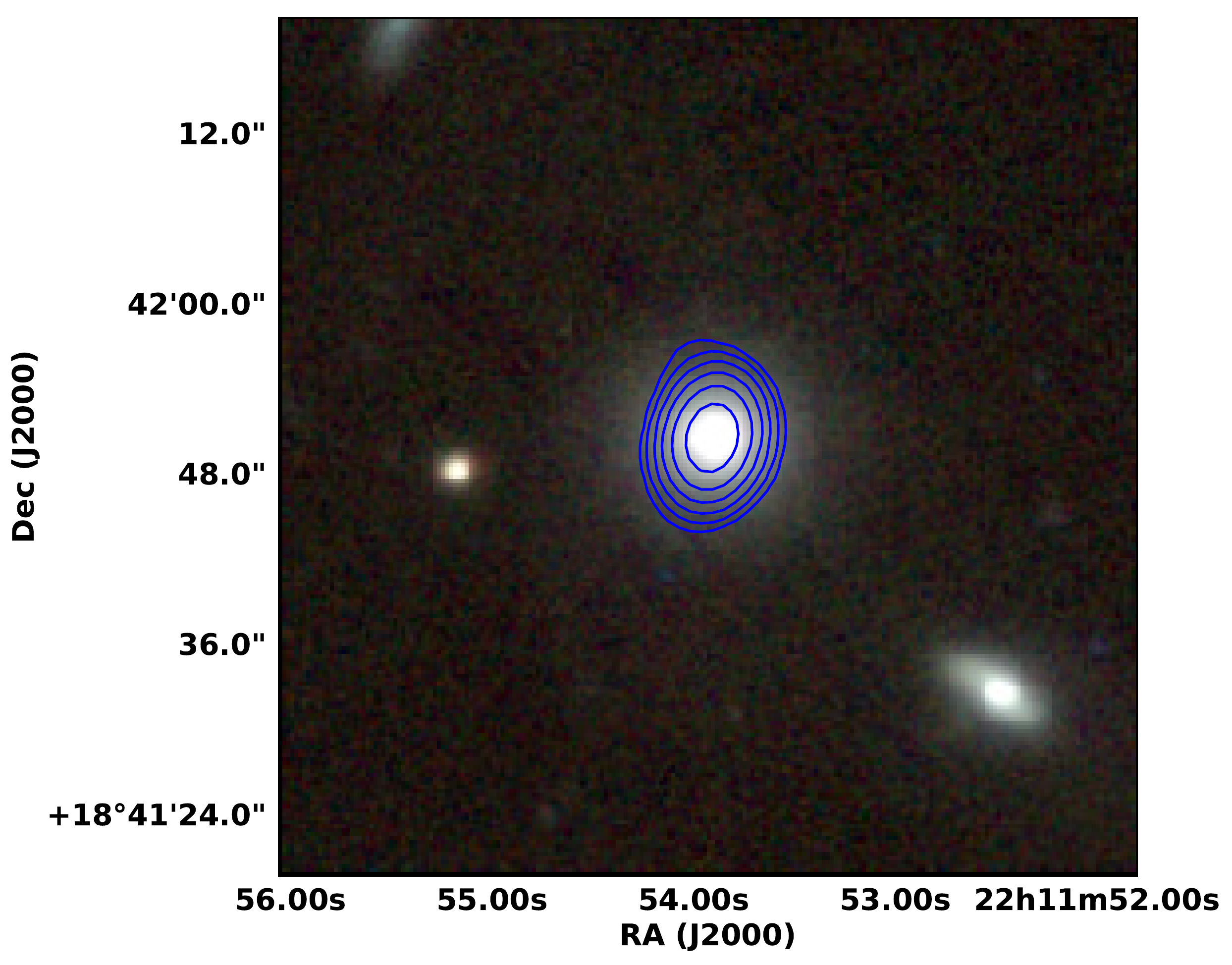}}
\end{figure*}

\begin{figure*}
\centerline{
\includegraphics[width=7cm]{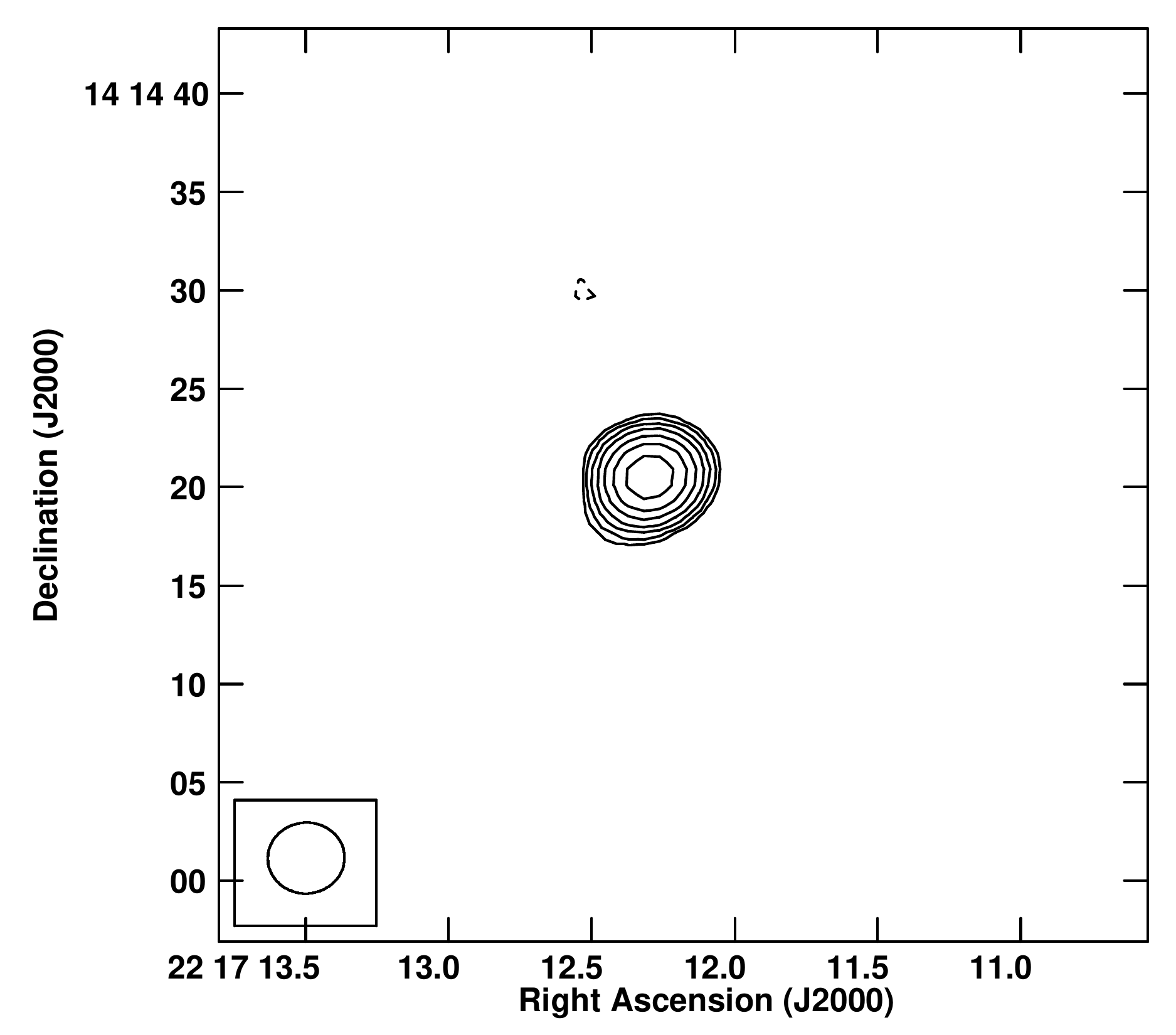}
\includegraphics[width=7cm]{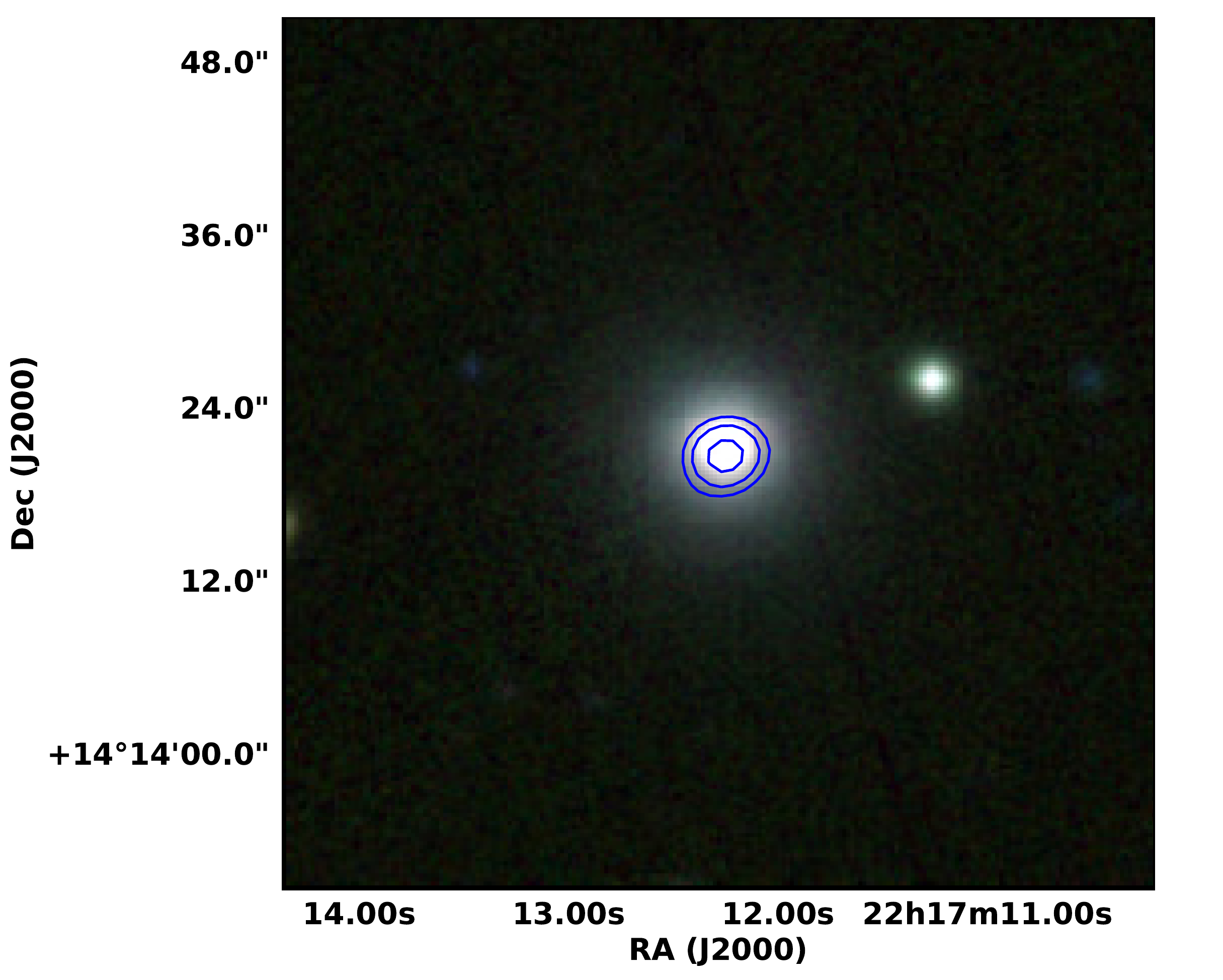}}
\end{figure*}

\begin{figure*}
\centerline{
\includegraphics[width=7cm]{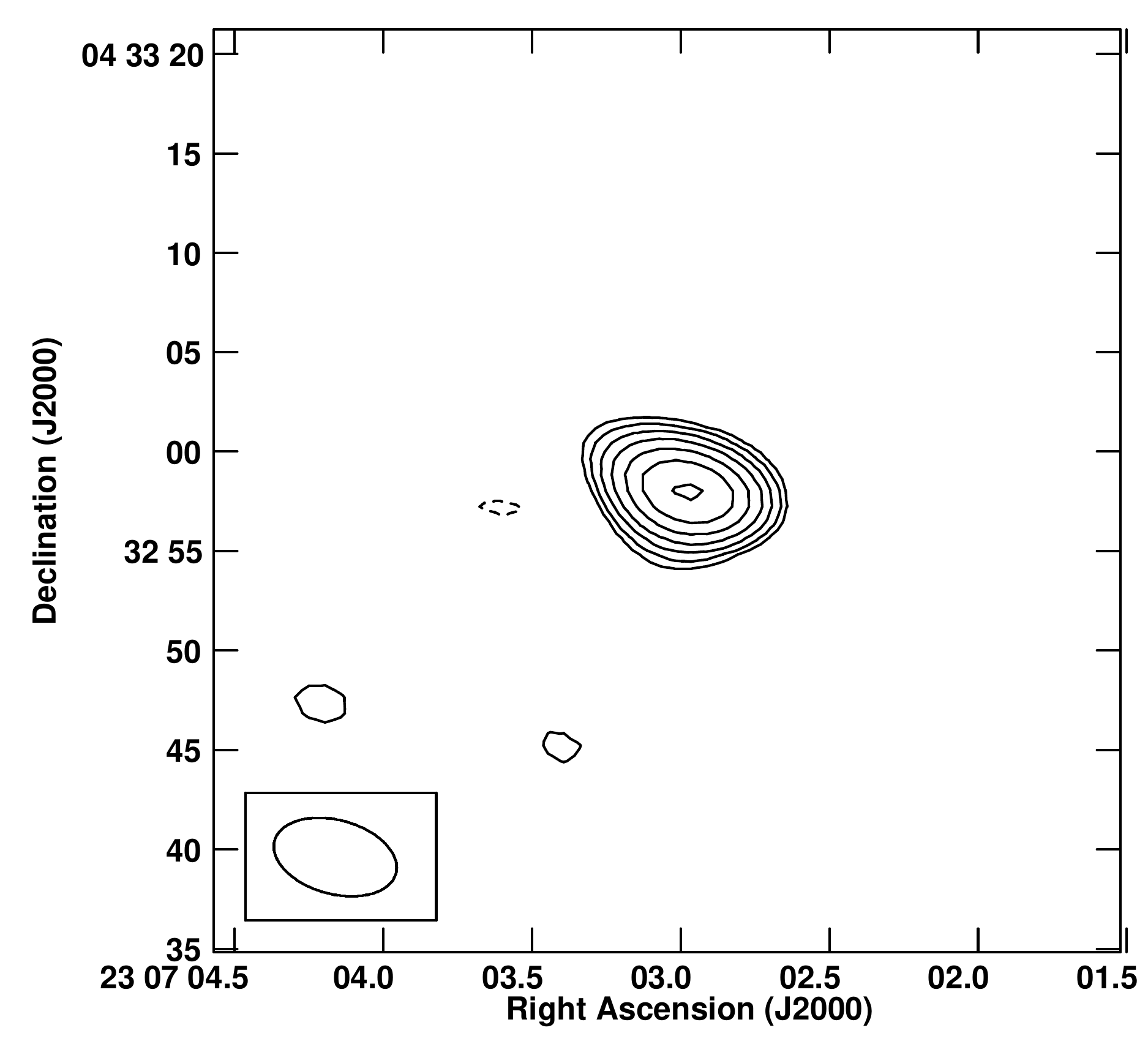}
\includegraphics[width=7cm]{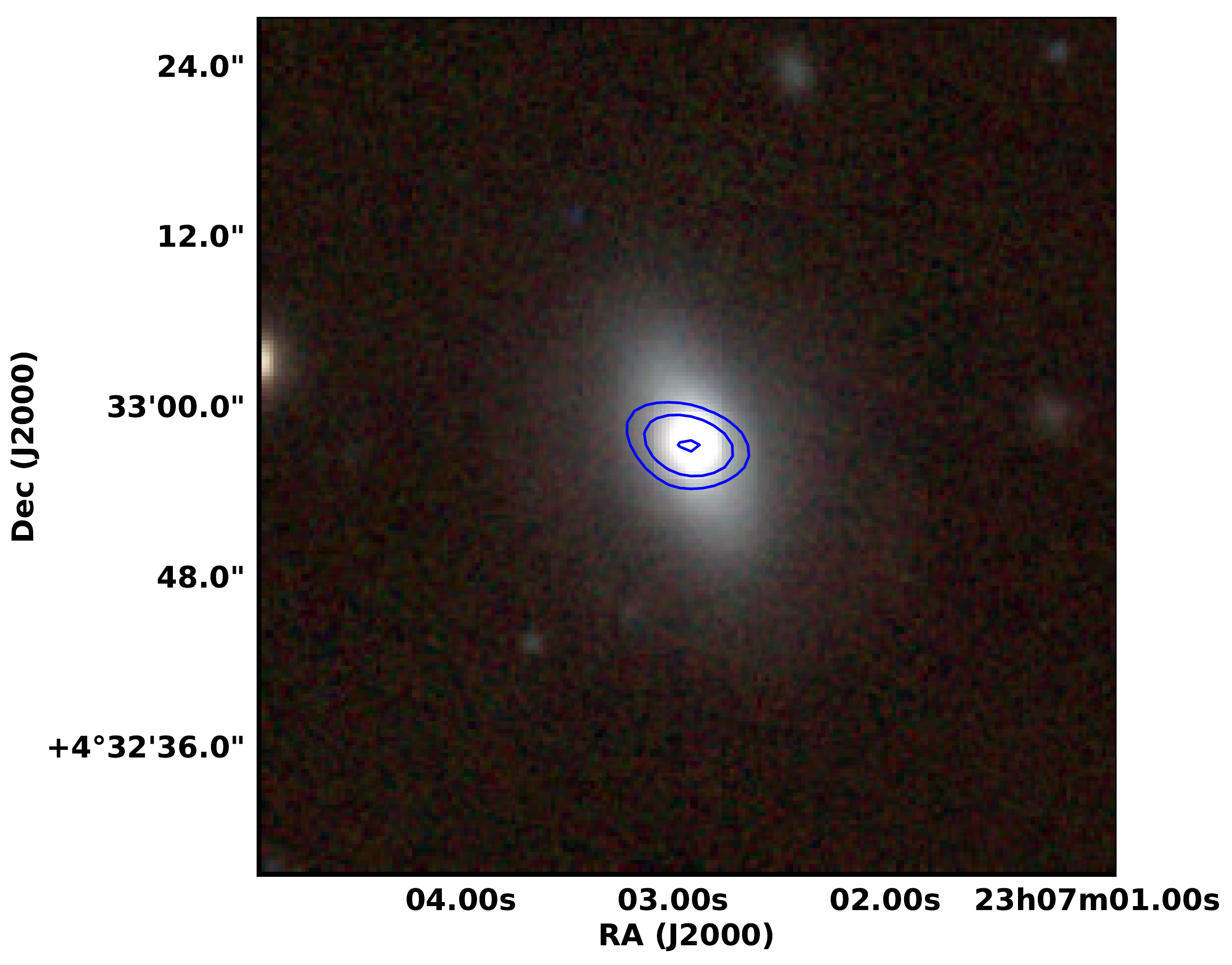}}
\caption{ Left: 685 MHz total intensity contour images of PG 2209+184, PG 2214+139 and PG 2304+042 from top to bottom. The peak contour flux is {\it x} mJy beam$^{-1}$ and the contour levels are {\it y} $\times$ (-1, 1, 1.4, 2, 2.8, 4, 5.6, 8, 11.20, 16, 23, 32, 45, 64, 90, 128, 180, 256, 362, 512) mJy beam$^{-1}$, where {\it (x ; y)} for PG 2209+184, PG 2214+139 and PG 2304+042 are (333;0.85), (27.34;0.09) and (321.8;0.12) respectively. Right: 685 MHz total intensity contours in blue superimposed on PanSTARRS $\it{griz}$-color composite optical image of PG 2209+184, PG 2214+139 and PG 2304+042 from top to bottom. The peak contour flux is {\it x} mJy beam$^{-1}$ and the contour levels are {\it y} $\times$ (2, 4, 8, 16, 32, 64, 128, 256, 512) mJy beam$^{-1}$, where {\it (x ; y)} for PG 2209+184, PG 2214+139 and PG 2304+042 are (333;0.85), (27.34;0.09) and (321.8;0.12) respectively.}
\end{figure*}

\begin{figure*}
\centerline{
\includegraphics[width=6cm]{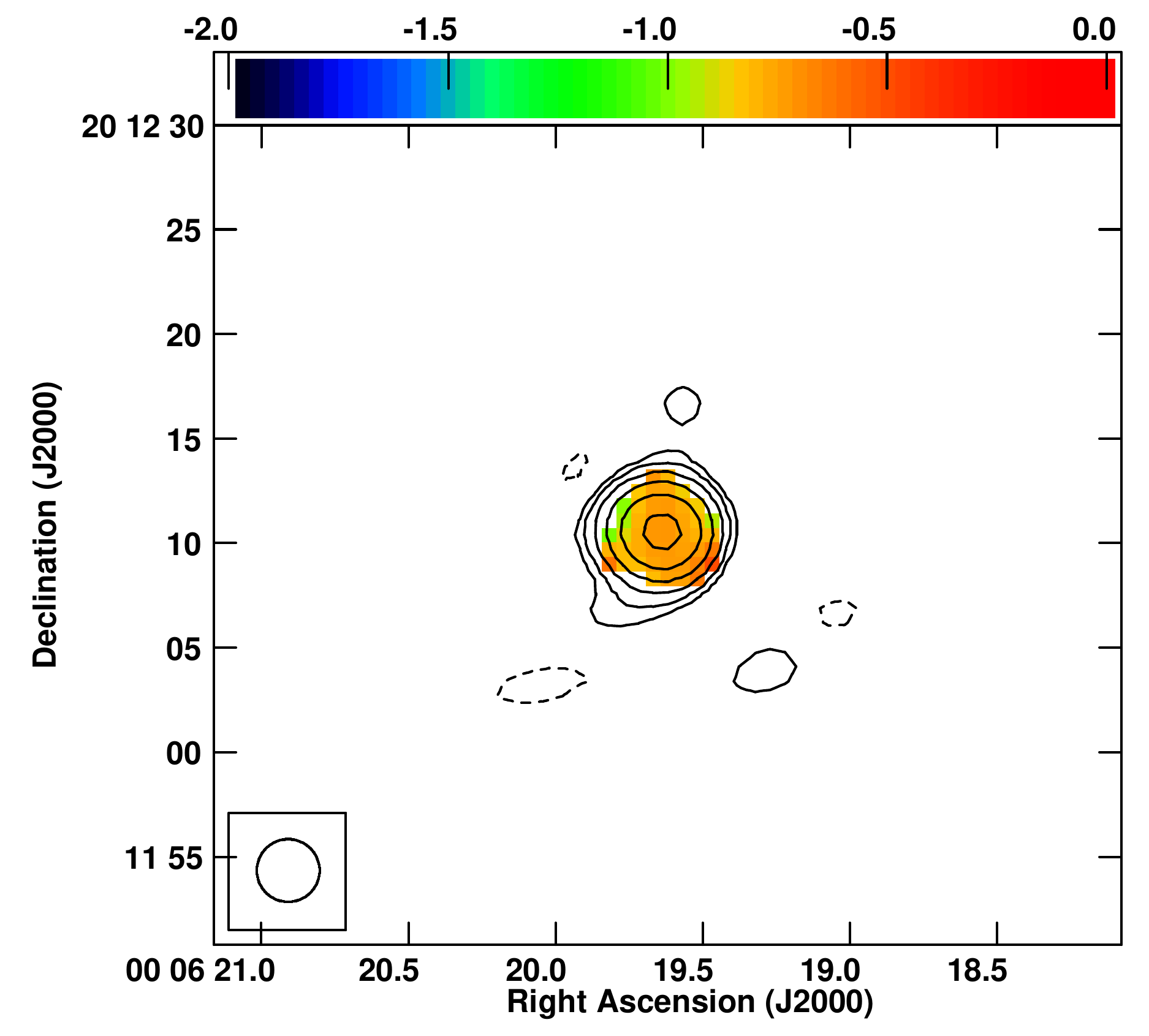}
\includegraphics[width=6cm]{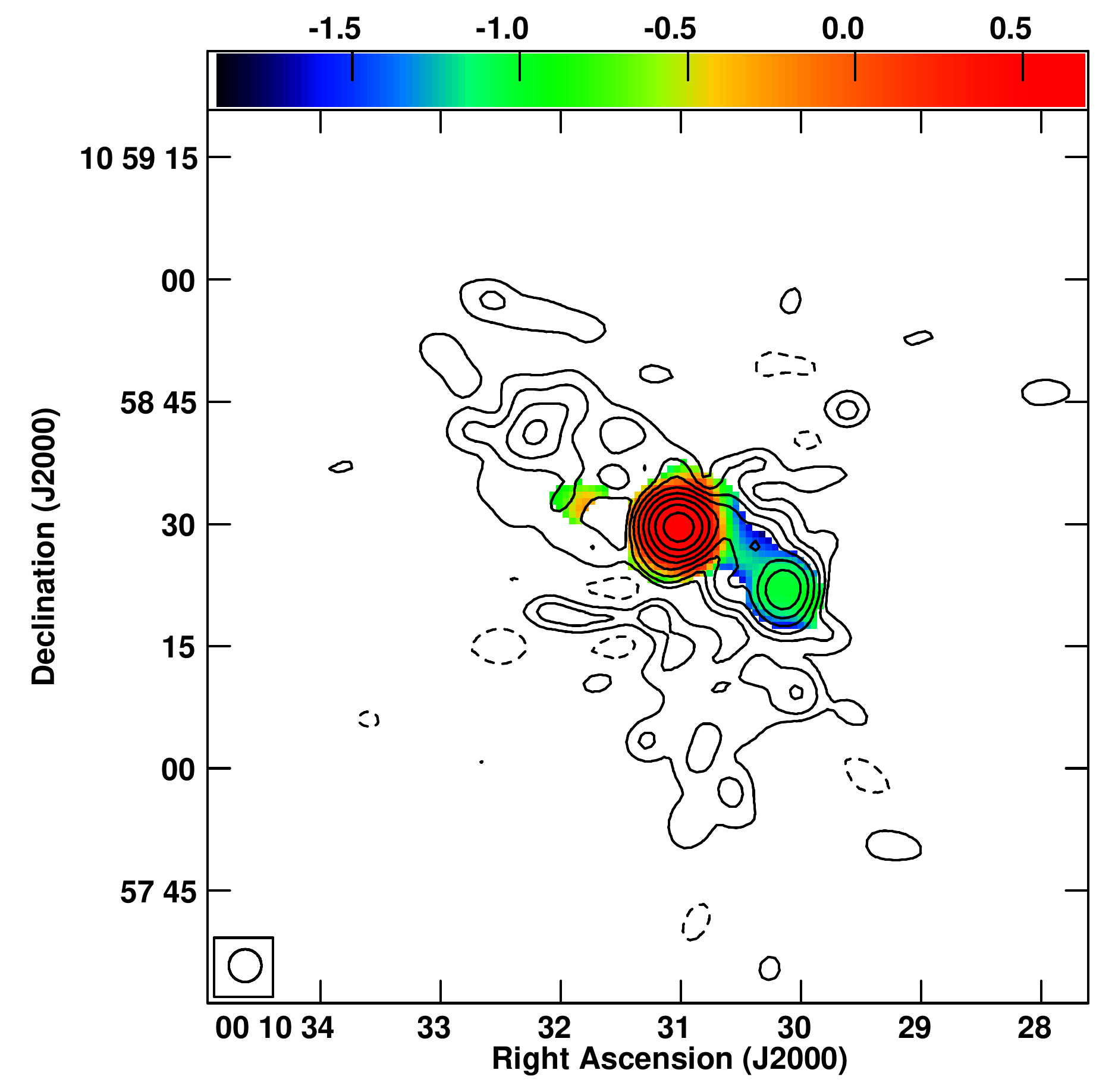}}
\end{figure*}

\begin{figure*}
\centerline{
\includegraphics[width=6cm]{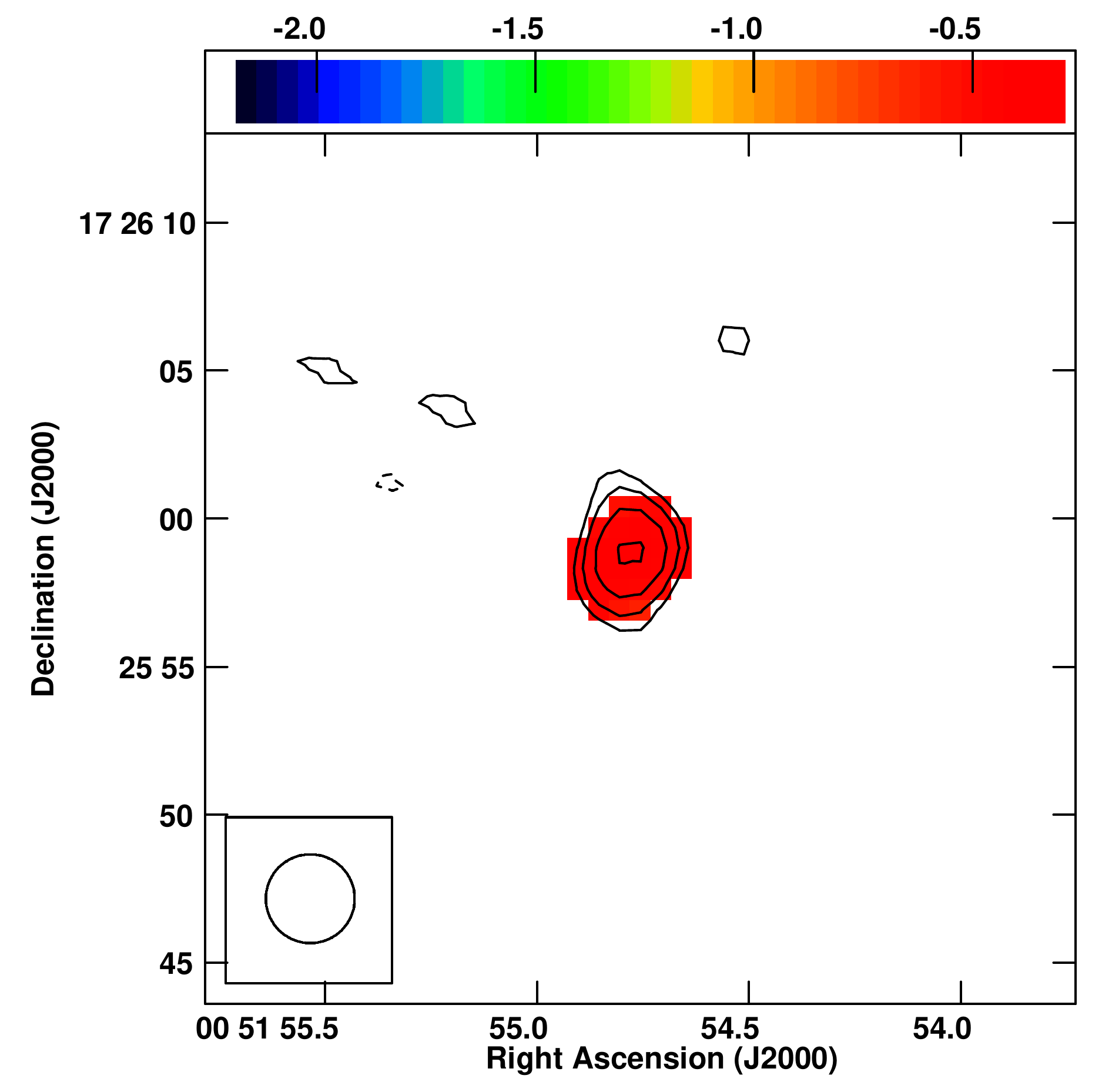}
\includegraphics[width=6cm]{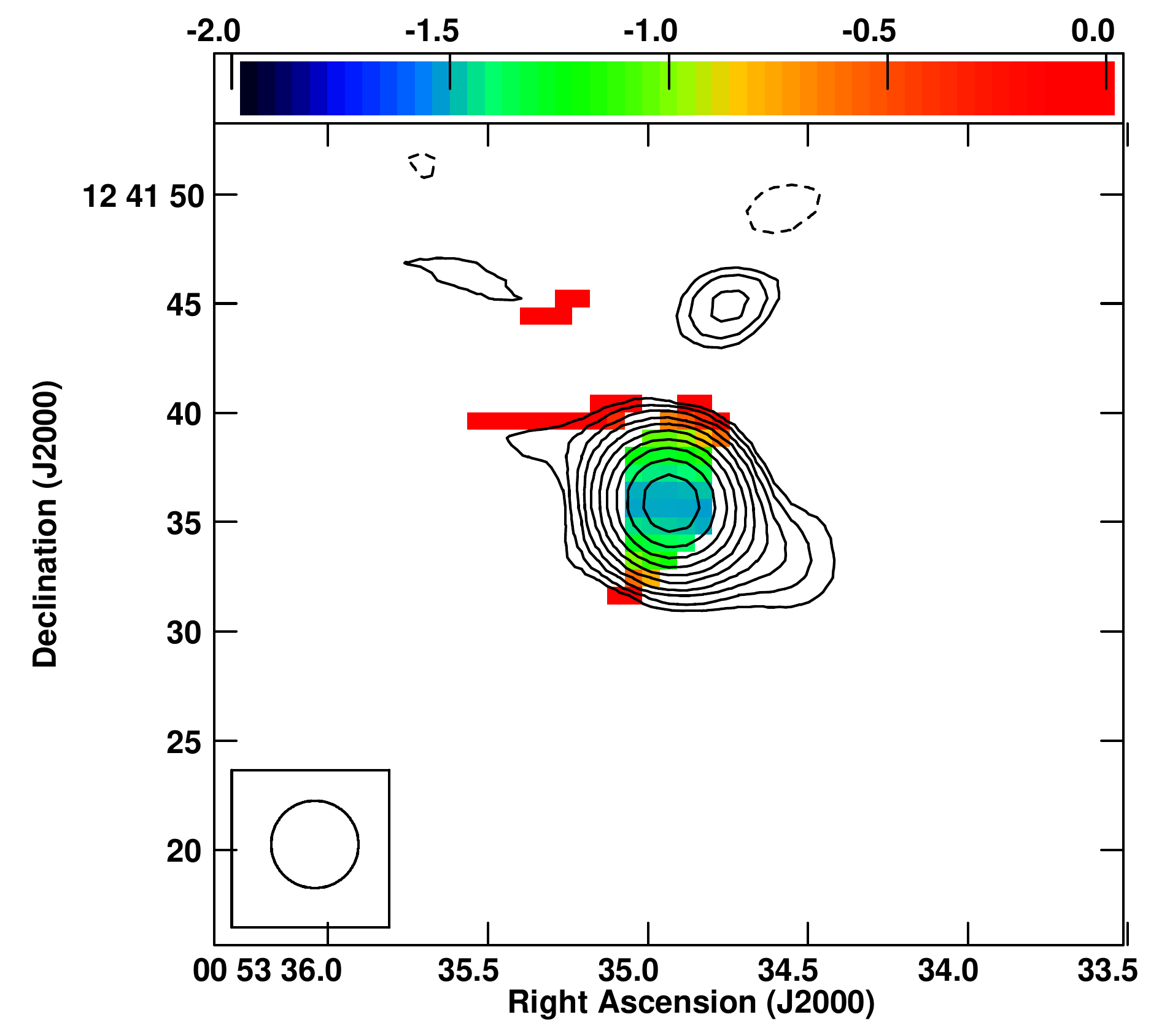}}
\end{figure*}

\begin{figure*}
\centerline{
\includegraphics[width=6cm]{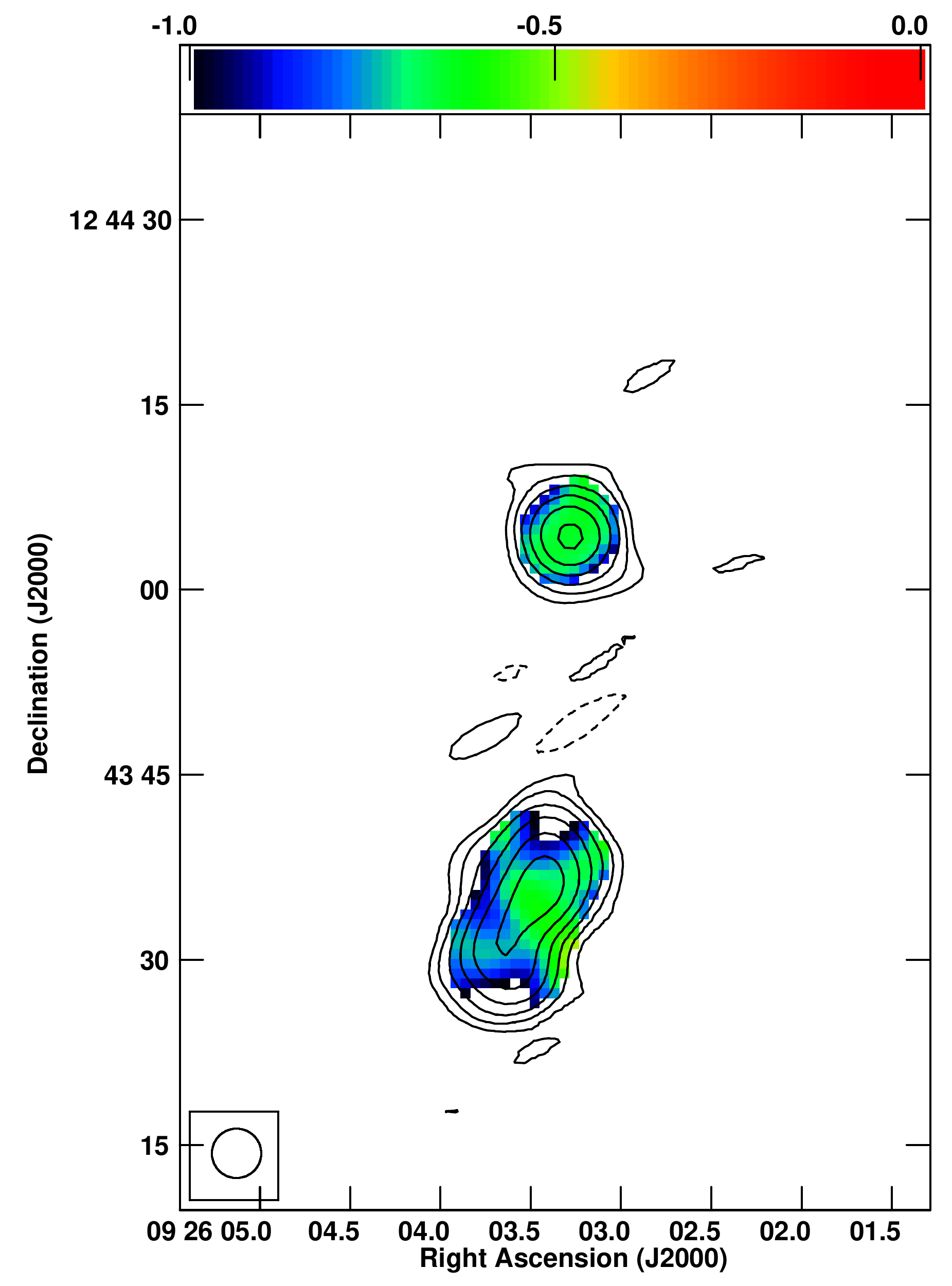}
\includegraphics[width=6cm]{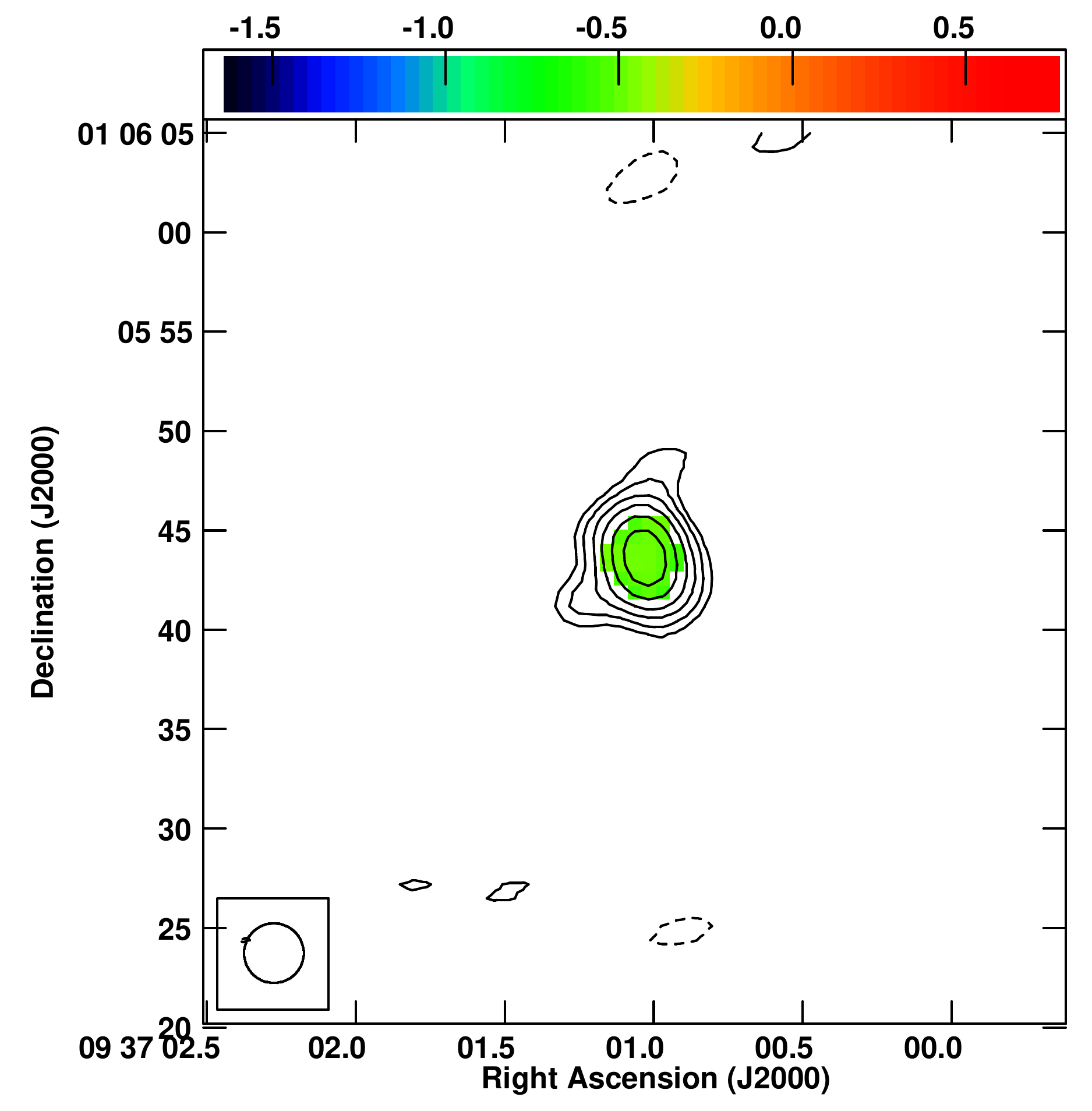}}
\caption{685 MHz total intensity contours in black superimposed on spectral index image in color for Top left: PG 0003+199; Top right: PG 0007+106; Middle left: PG 0049+171; Middle right: PG 0050+124; Bottom left: PG 0923+129; Bottom right: PG 0934+013. The peak contour flux is {\it x} mJy beam$^{-1}$ and the contour levels are {\it y} $\times$ (-1, 1, 2, 4, 8, 16, 32, 64, 128, 256, 512) mJy beam$^{-1}$, where {\it (x ; y)} for PG 0003+199, PG 0007+106 and PG 0923+129 are (135;0.26), (54;0.25) and (27;0.30) respectively. The peak contour flux is {\it x*} mJy beam$^{-1}$ and the contour levels are {\it y*} $\times$ (-1, 1, 1.4, 2, 2.8, 4, 5.6, 8, 11.20, 16, 23, 32, 45, 64, 90, 128, 180, 256, 362, 512) mJy beam$^{-1}$, where {\it (x* ; y*)} for PG 0049+171, PG 0050+124 and PG 0934+013 are  (989.4;0.59), (17.9;0.32) and (24.9;0.12) respectively.}
\end{figure*}

\begin{figure*}
\centerline{
\includegraphics[width=6cm]{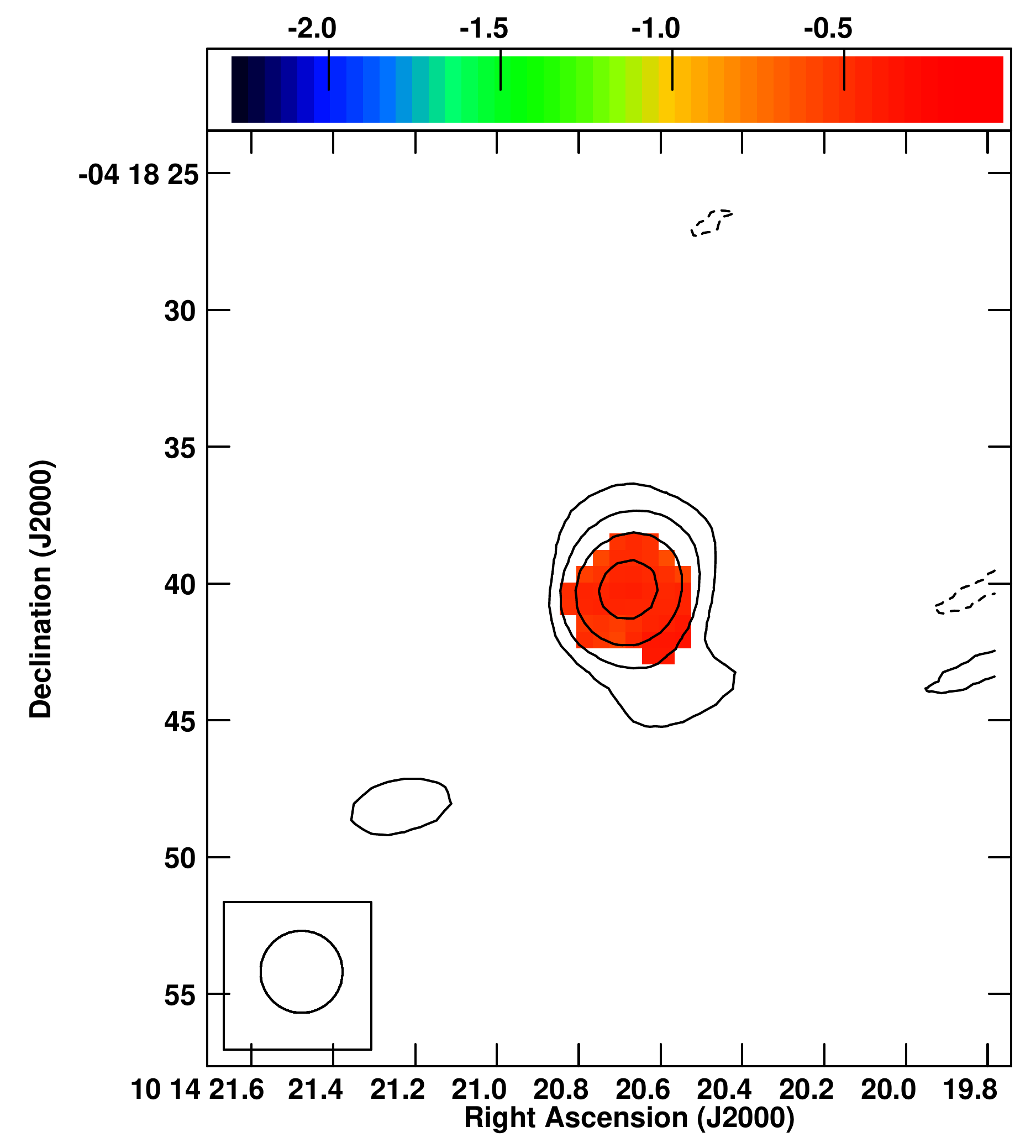}
\includegraphics[width=6cm]{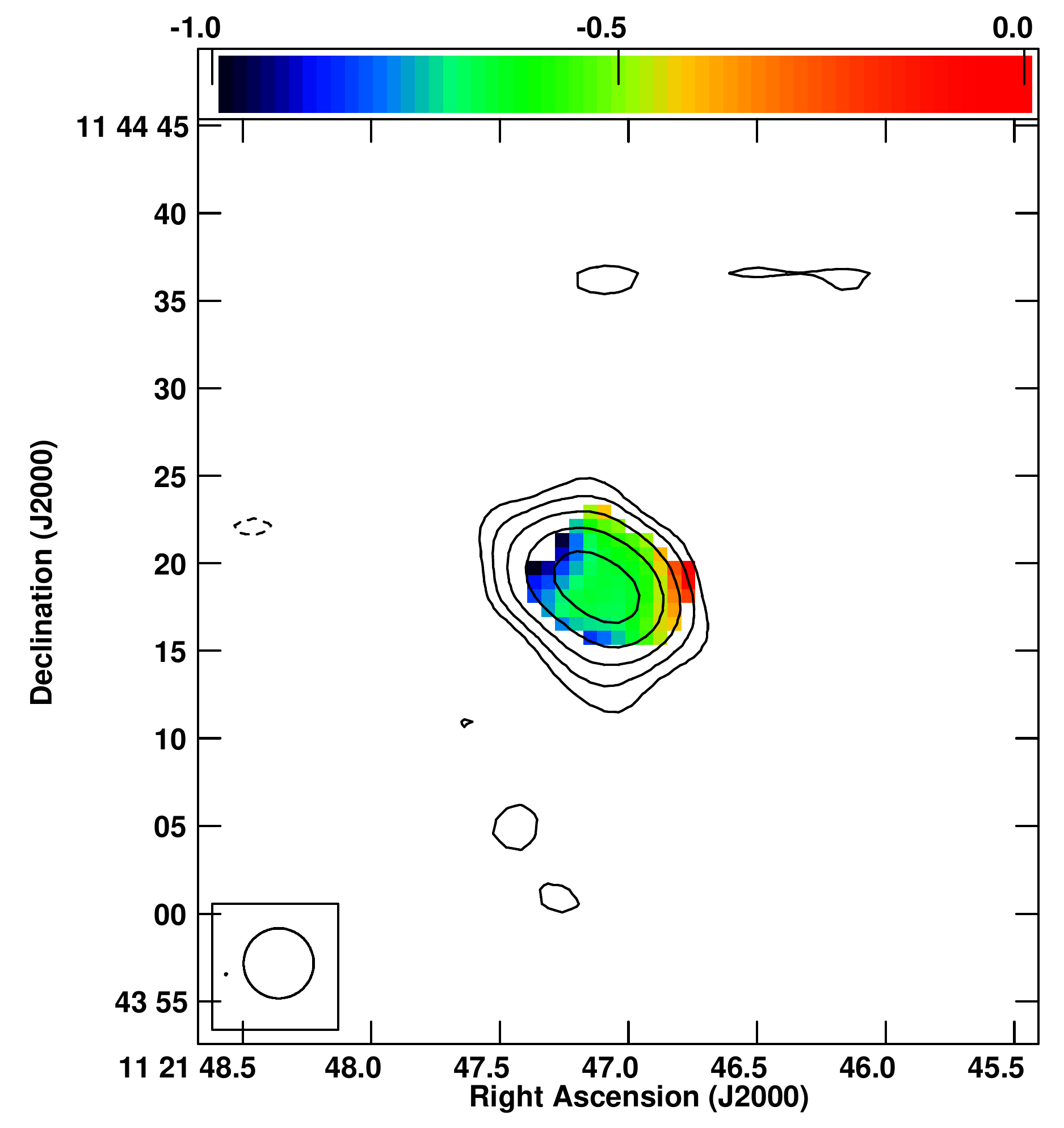}}
\end{figure*}

\begin{figure*}
\centerline{
\includegraphics[width=6cm]{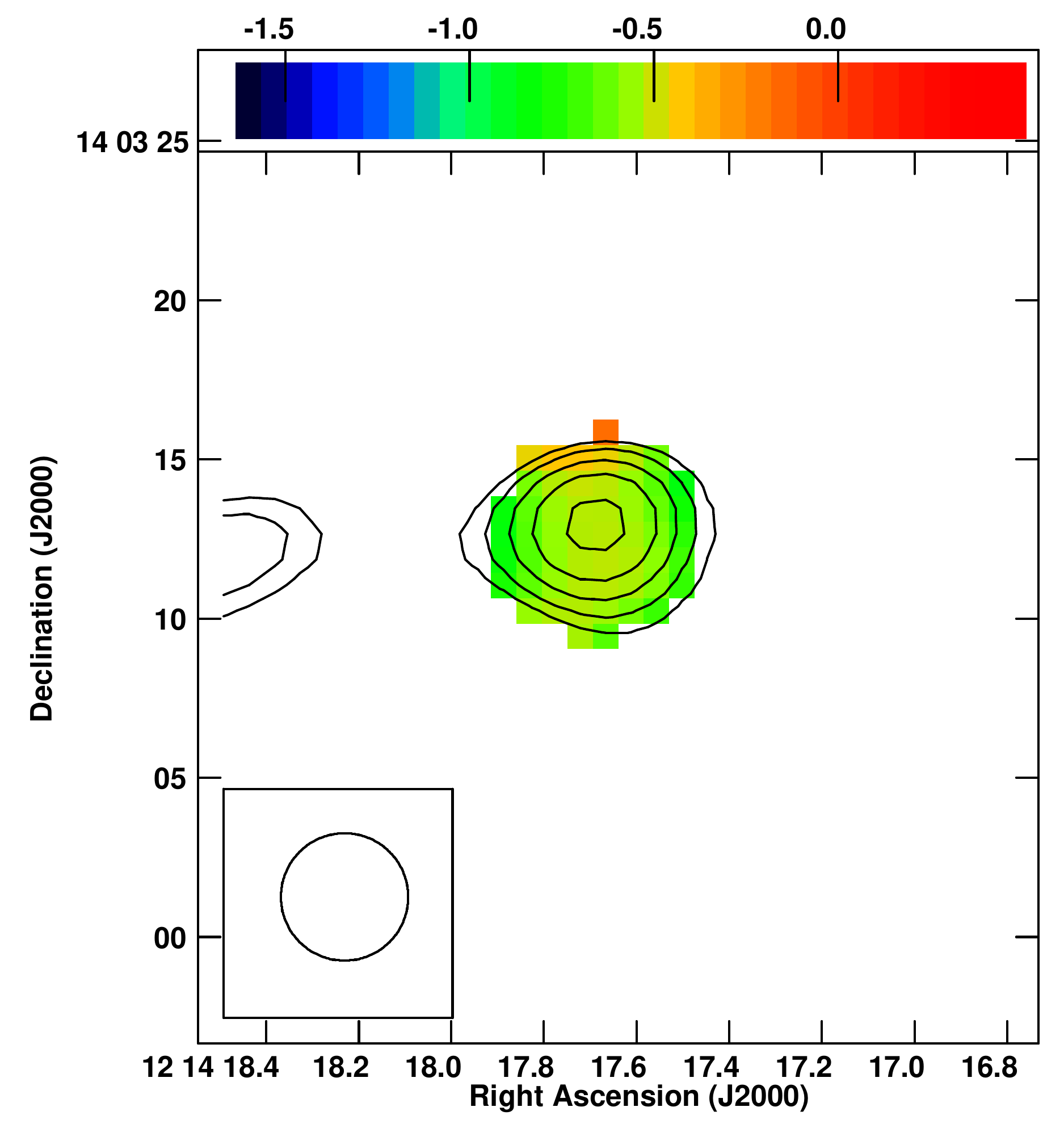}
\includegraphics[width=6cm]{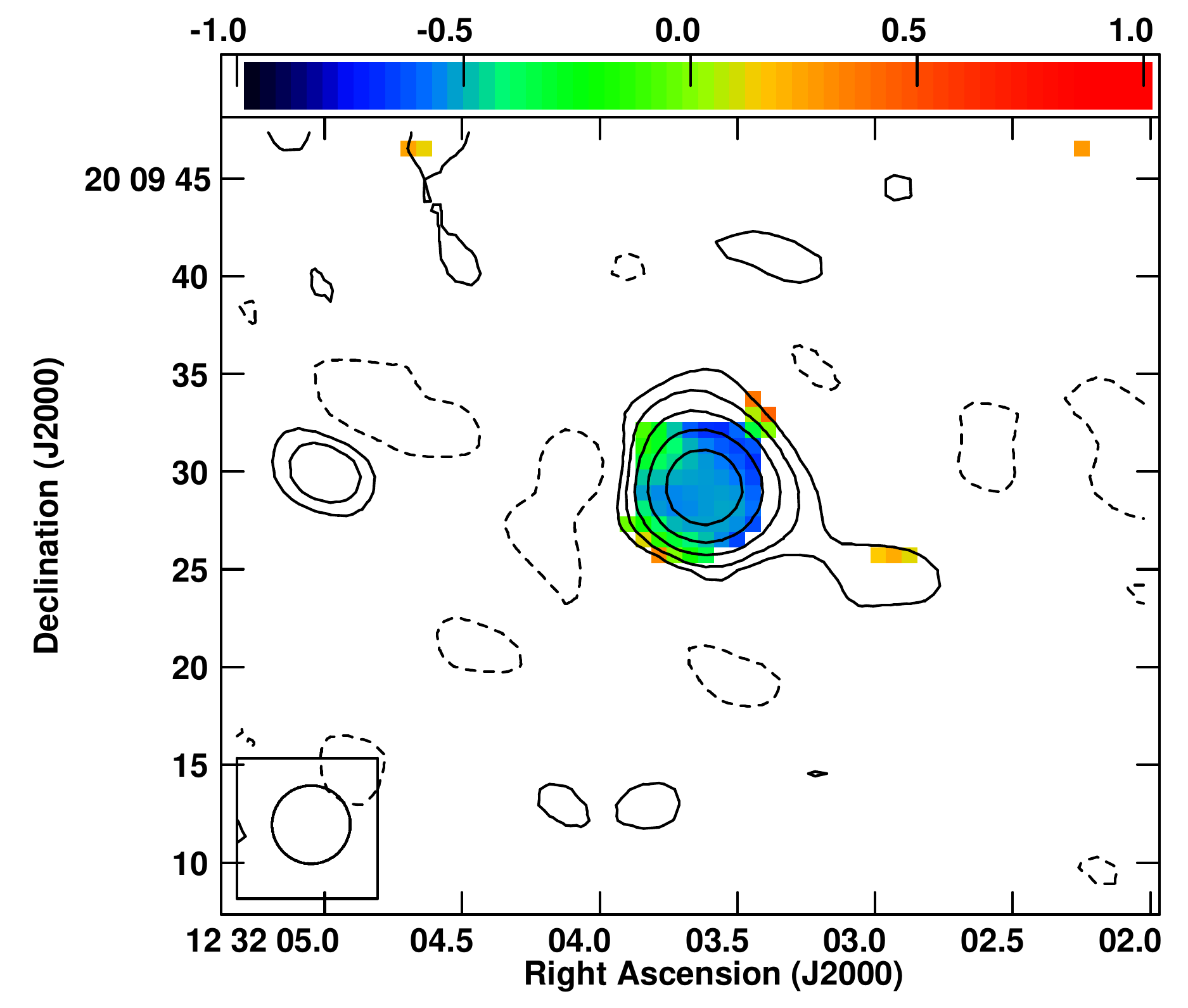}}
\end{figure*}

\begin{figure*}
\centerline{
\includegraphics[width=6cm]{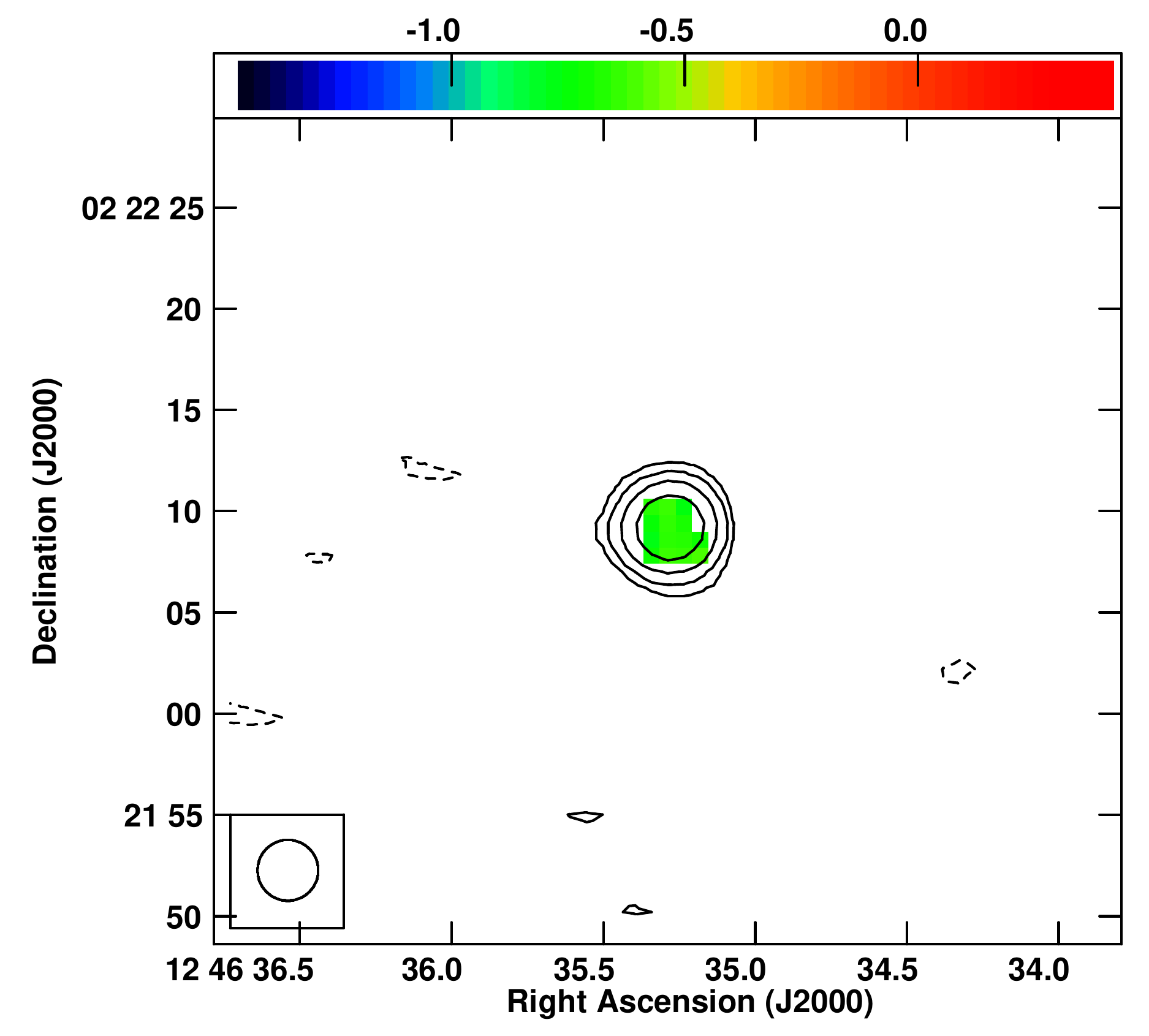}
\includegraphics[width=6cm]{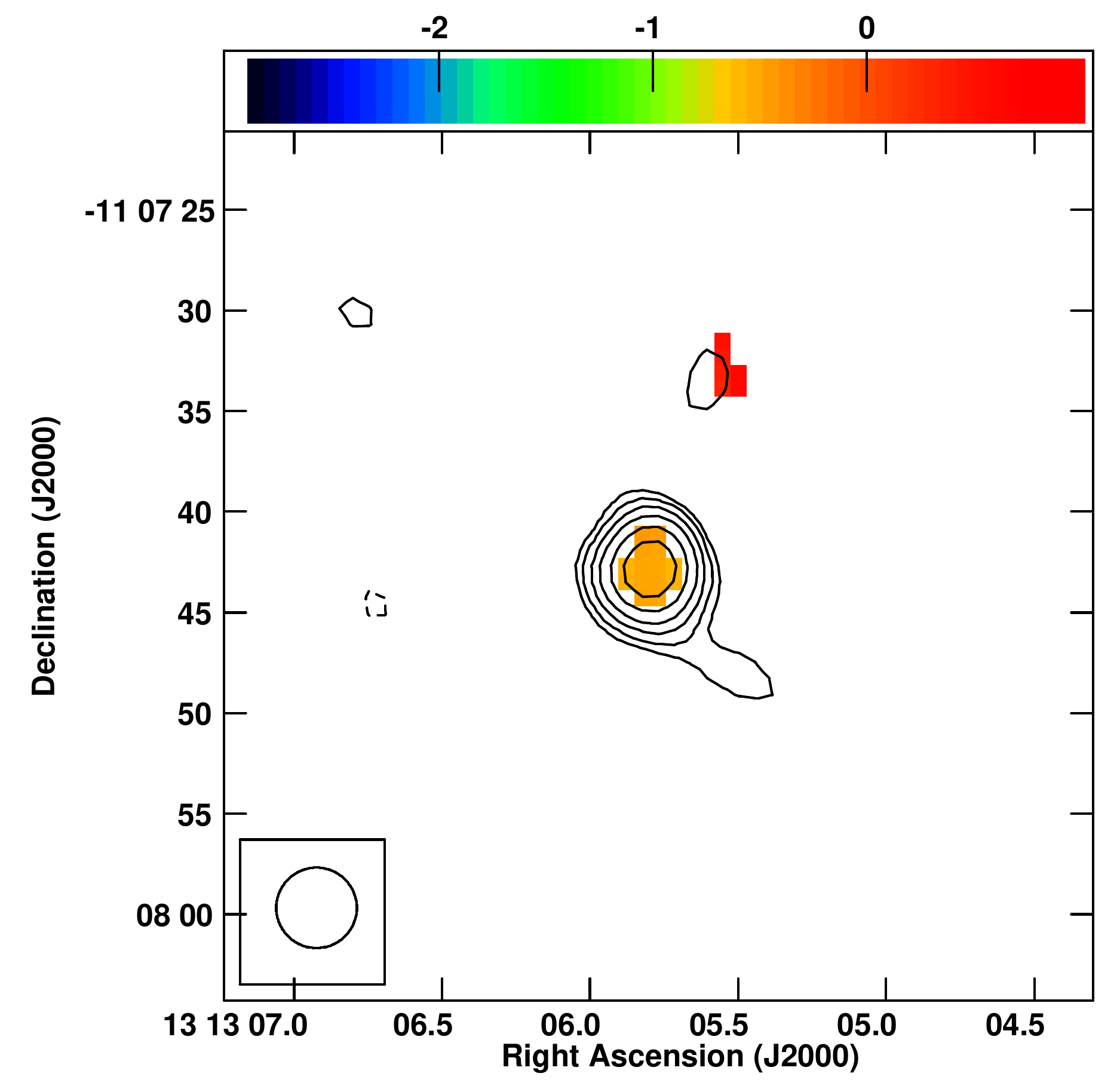}}
\caption{685 MHz total intensity contours in black superimposed on spectral index image in color for Top left: PG 1011$-$040; Top right: PG 1119+120; Middle left: PG 1211+143; Middle right: PG 1229+204; Bottom left: PG 1244+026; Bottom right: PG 1310$-$108. The peak contour flux is {\it x} mJy beam$^{-1}$ and the contour levels are {\it y} $\times$ (-1, 1, 2, 4, 8, 16, 32, 64, 128, 256, 512) mJy beam$^{-1}$, where {\it (x ; y)} for PG 1011$-$040, PG 1119+120, PG 1229+204 and PG 1244+026 are (40.8;0.10), (64.6;0.12), (166.1;0.06) and (39.1;0.16) respectively. The peak contour flux is {\it x*} mJy beam$^{-1}$ and the contour levels are {\it y*} $\times$ (-1, 1, 1.4, 2, 2.8, 4, 5.6, 8, 11.20, 16, 23, 32, 45, 64, 90, 128, 180, 256, 362, 512) mJy beam$^{-1}$, where {\it (x* ; y*)} for PG 1211+143 and PG 1310$-$108 are (1200.3;1.0) and (53.76;0.10) respectively.}
\end{figure*}

\begin{figure*}
\centerline{
\includegraphics[width=6cm]{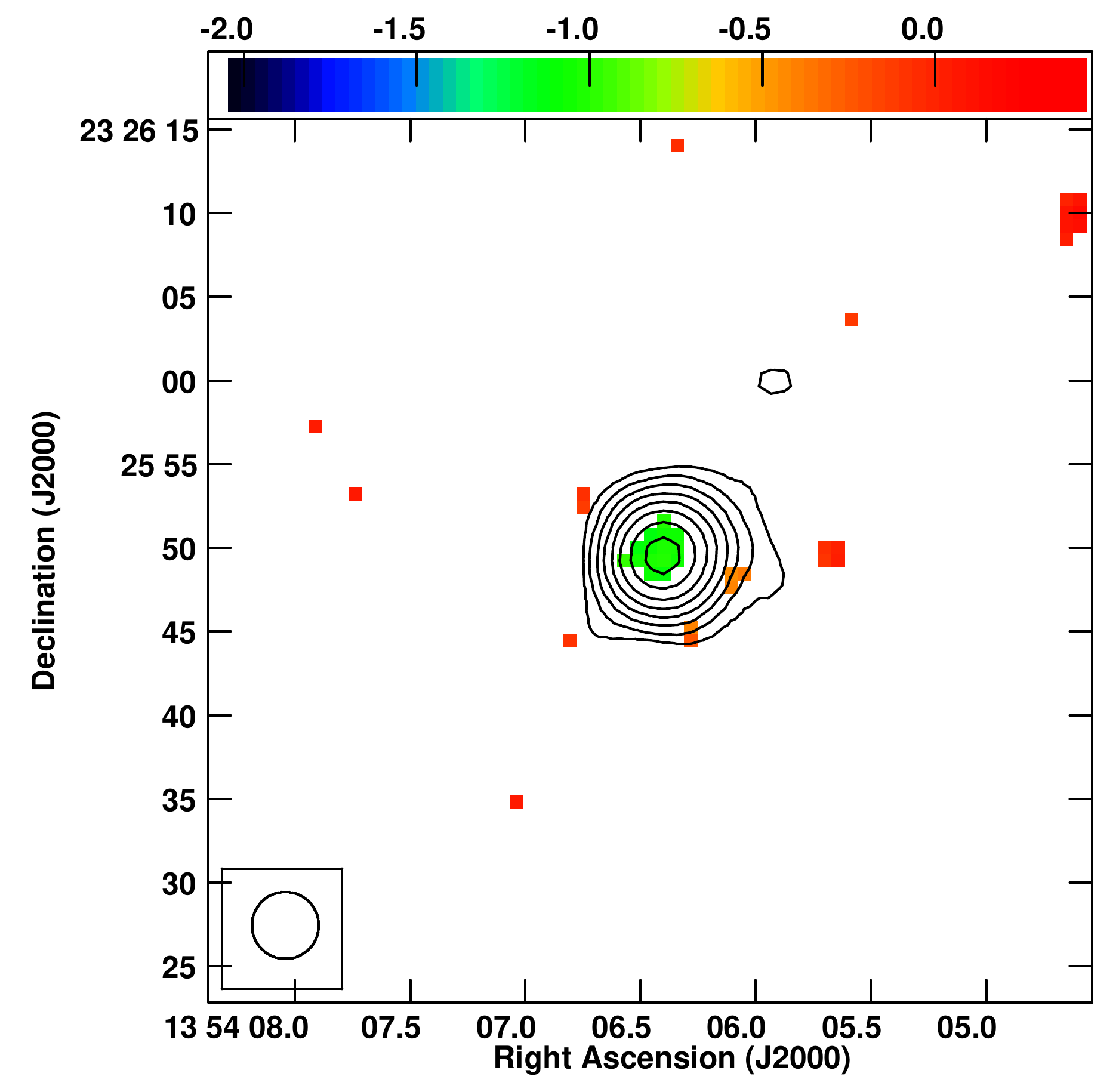}
\includegraphics[width=6cm]{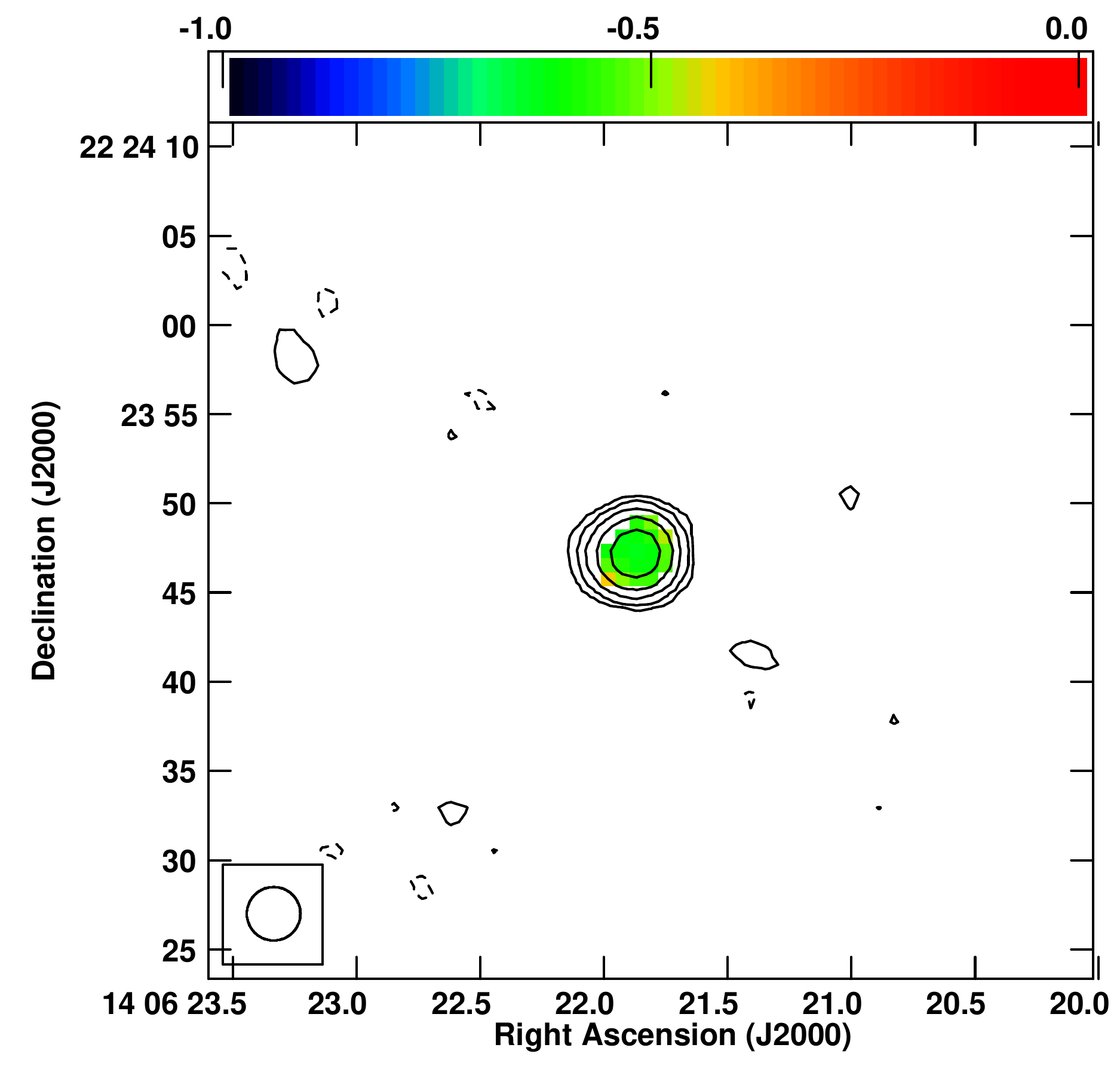}}
\end{figure*}

\begin{figure*}
\centerline{
\includegraphics[width=6cm]{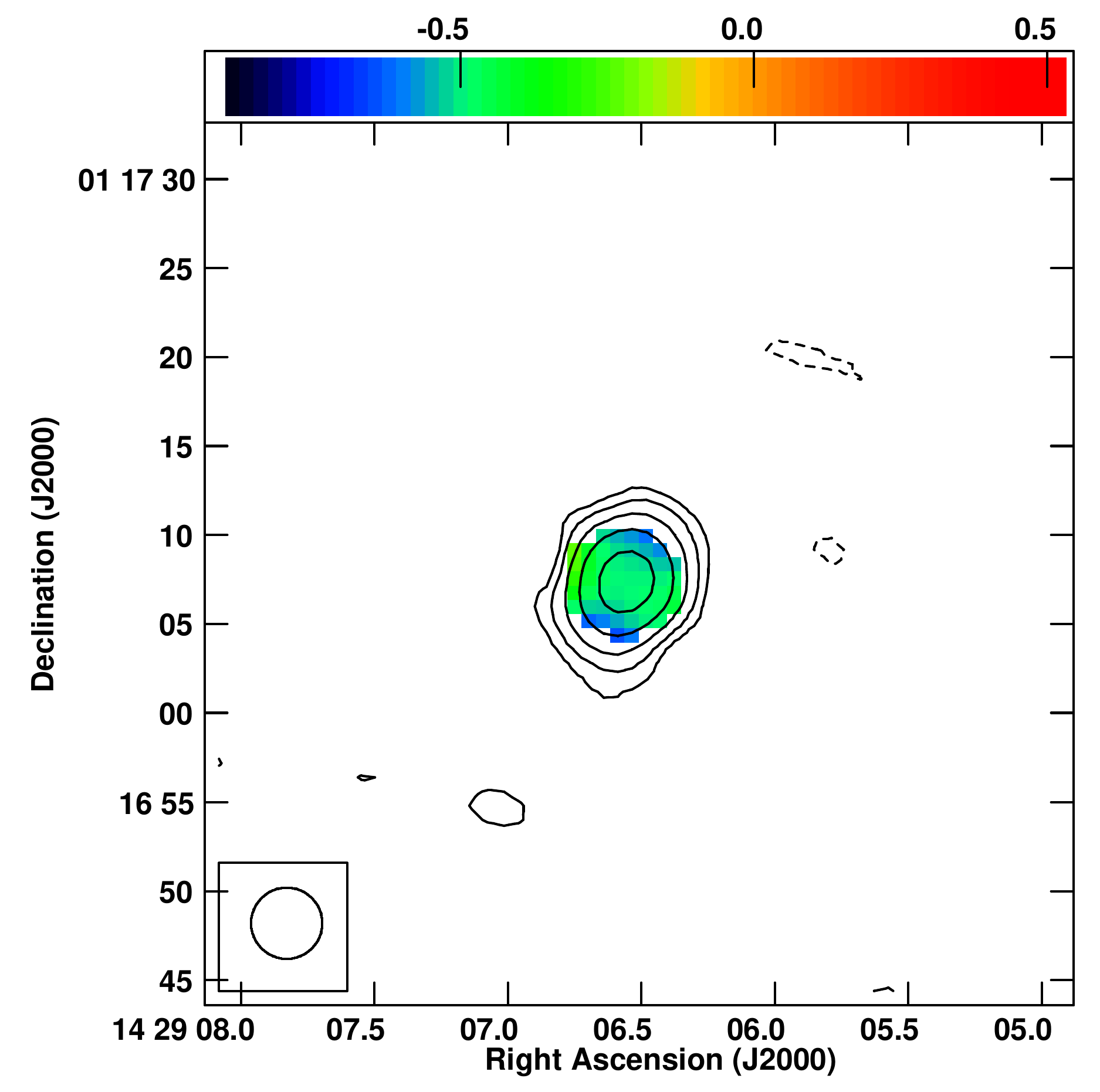}
\includegraphics[width=6cm]{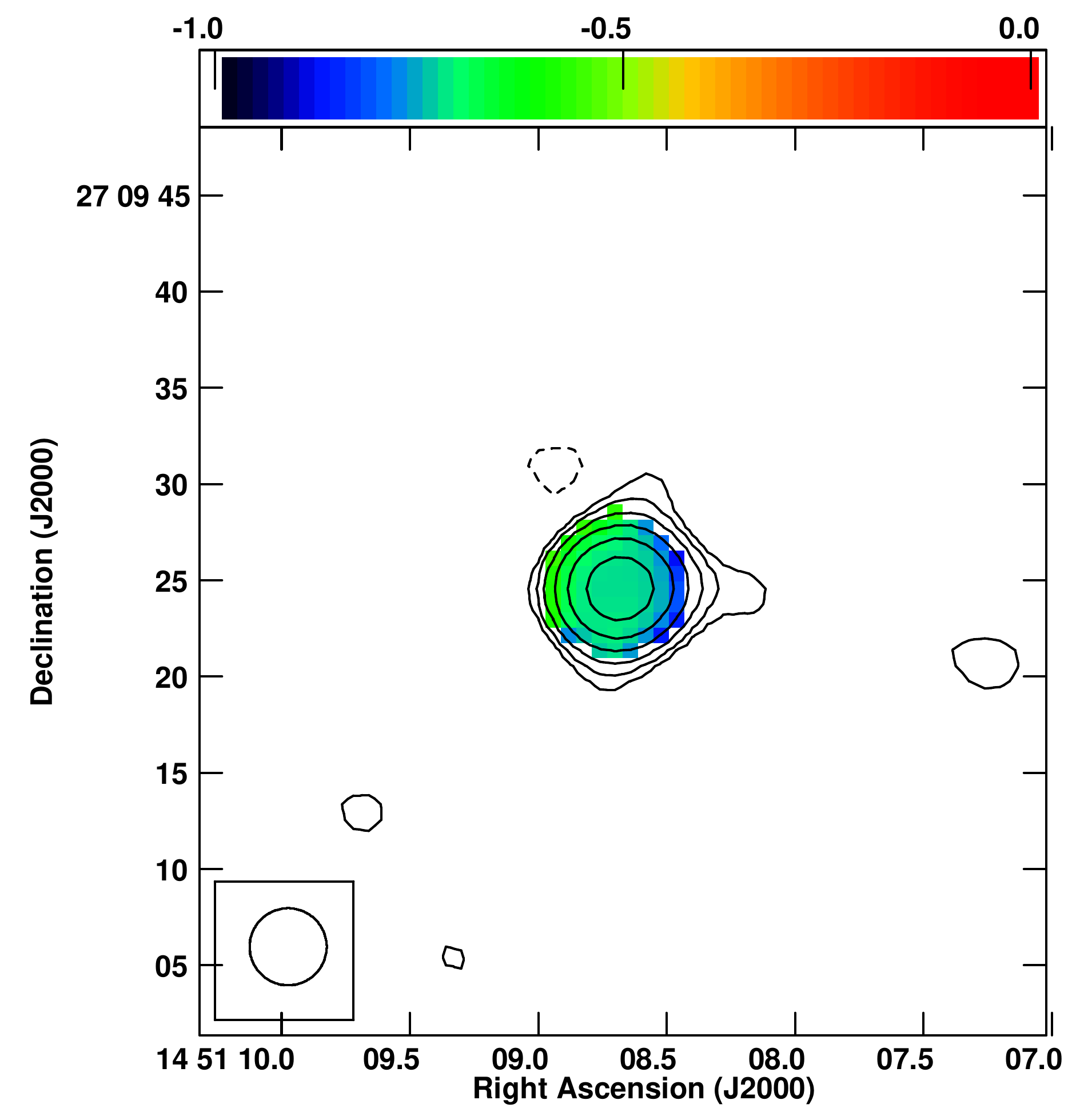}}
\end{figure*}

\begin{figure*}
\centerline{
\includegraphics[width=6cm]{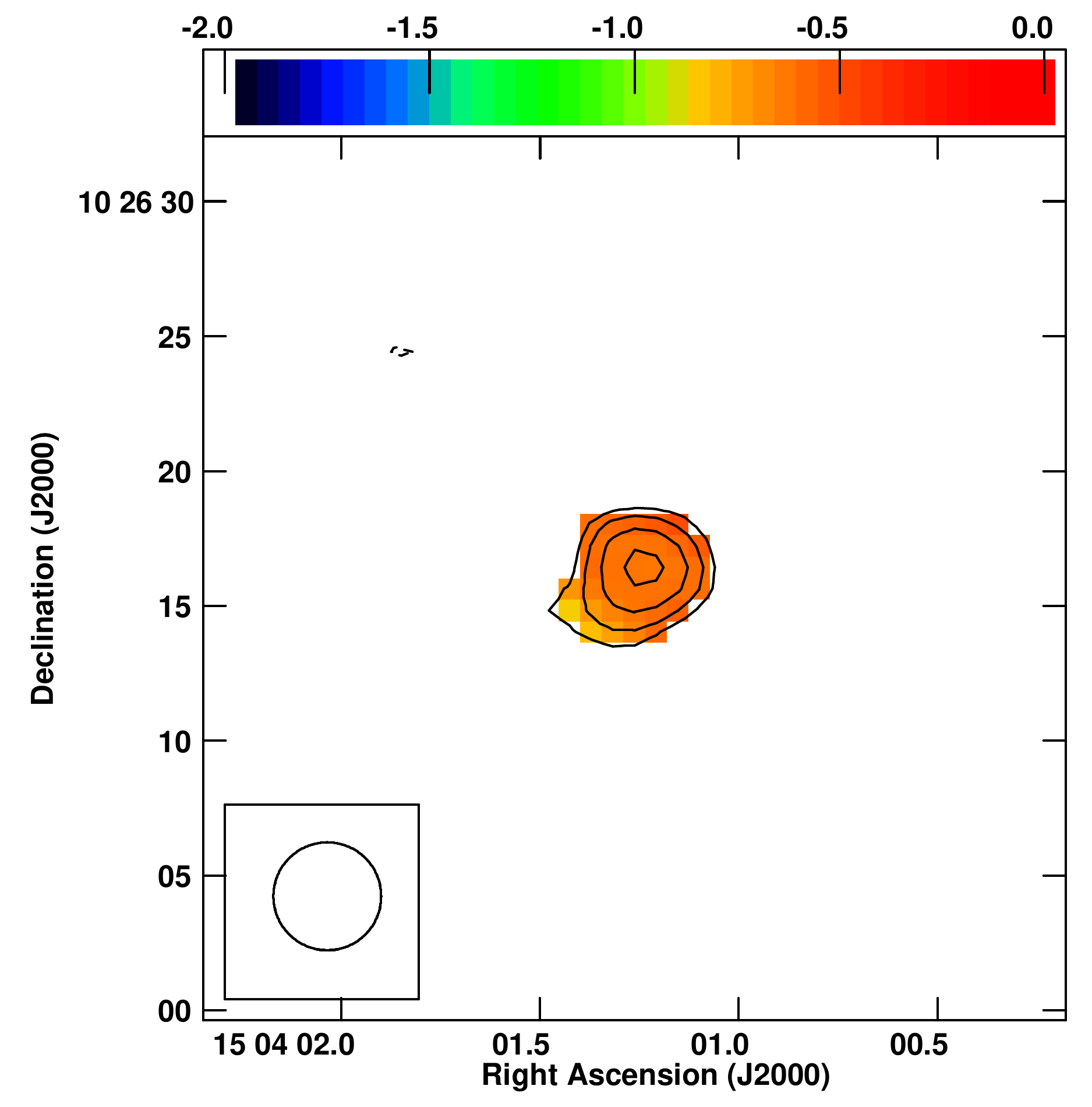}
\includegraphics[width=6cm]{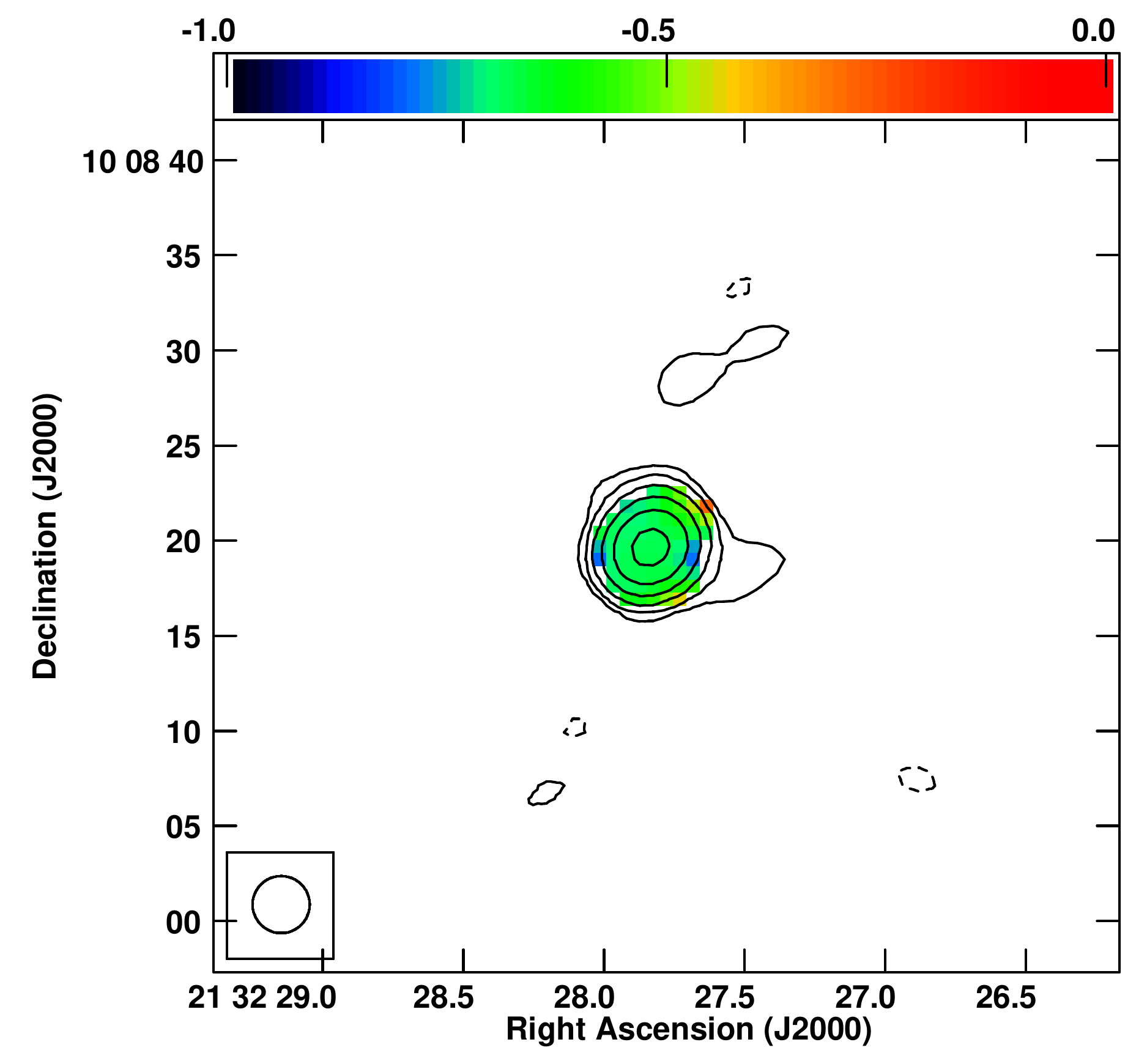}}
\caption{685 MHz total intensity contours in black superimposed on spectral index image in color for Top left: PG 1351+236; Top right: PG 1404+226; Middle left: PG 1426+015; Middle right: PG 1448+273; Bottom left: PG 1501+106; Bottom right: PG 2130+099. The peak contour flux is {\it x} mJy beam$^{-1}$ and the contour levels are {\it y} $\times$ (-1, 1, 2, 4, 8, 16, 32, 64, 128, 256, 512) mJy beam$^{-1}$, where {\it (x ; y)} for PG 1404+226, PG 1426+015, PG 1448+273 and PG 2130+099 are (132.2;0.10), (51.3;0.09), (17.8;0.10) and (56.9;0.17) respectively. The peak contour flux is {\it x*} mJy beam$^{-1}$ and the contour levels are {\it y*} $\times$ (-1, 1, 1.4, 2, 2.8, 4, 5.6, 8, 11.20, 16, 23, 32, 45, 64, 90, 128, 180, 256, 362, 512) mJy beam$^{-1}$, where {\it (x* ; y*)} for PG 1351+236 and PG 1501+106 are (36.8;0.18) and (1507.6;1.2) respectively. }
\end{figure*}

\begin{figure*}
\centerline{
\includegraphics[width=7cm]{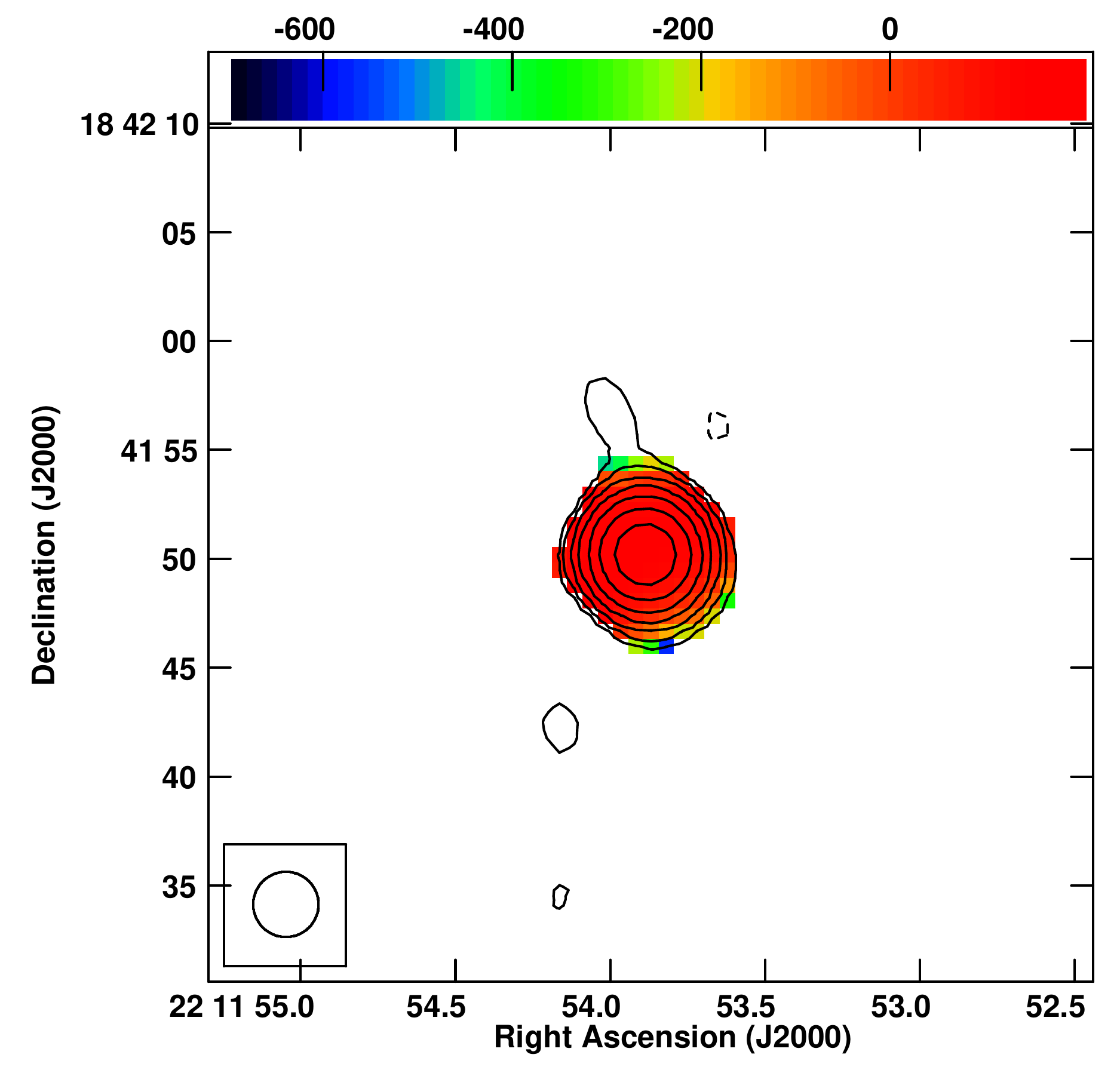}
\includegraphics[width=7cm]{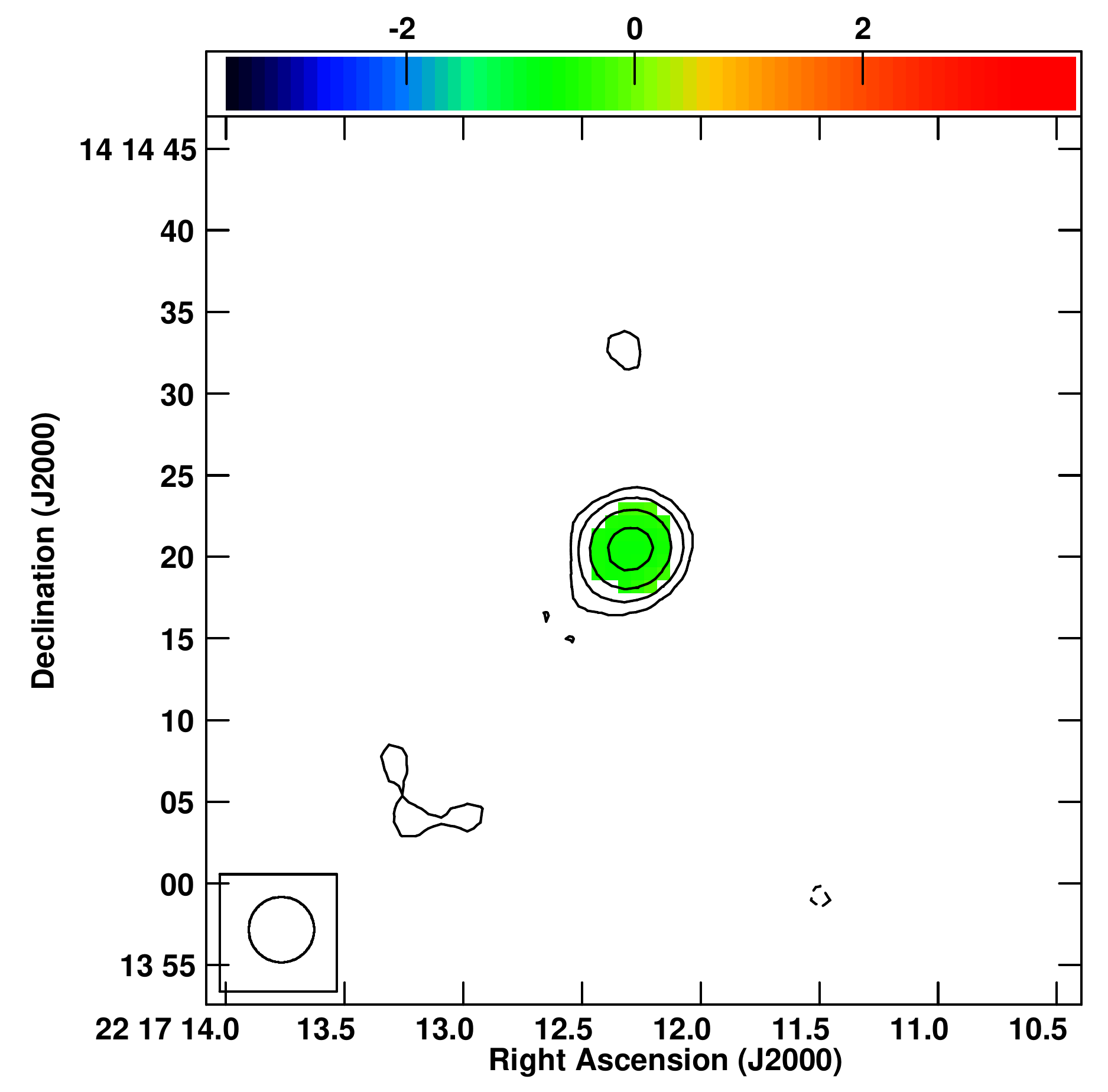}}
\end{figure*}

\begin{figure*}
\centerline{
\includegraphics[width=7cm]{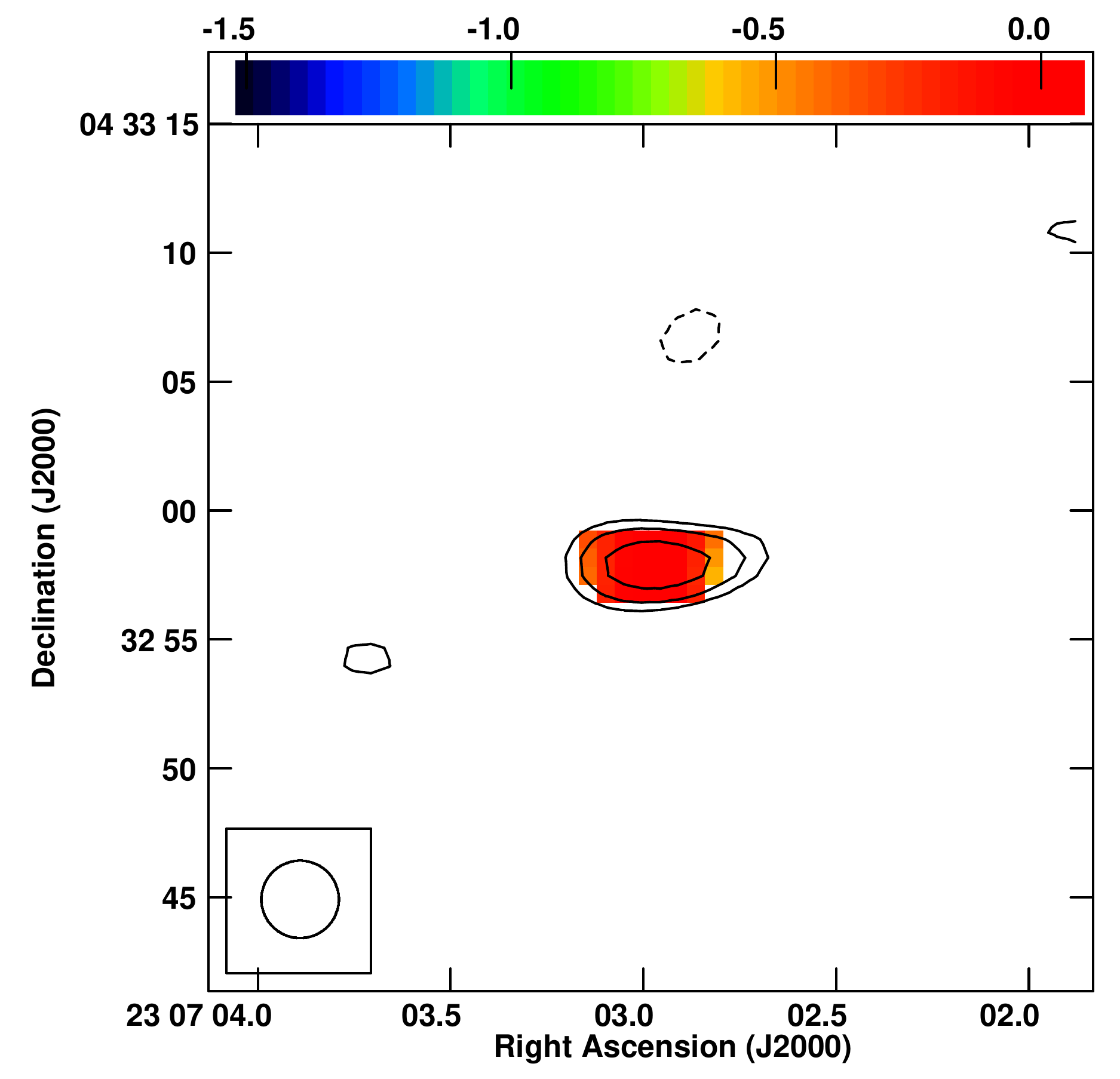}}
\caption{685 MHz total intensity contours in black superimposed on spectral index image in color for Top left: PG 2209+184; Top right: PG 2214+139; Bottom: PG 2304+042. The peak contour flux is {\it x} mJy beam$^{-1}$ and the contour levels are {\it y} $\times$ (-1, 1, 2, 4, 8, 16, 32, 64, 128, 256, 512) mJy beam$^{-1}$, where {\it (x ; y)} for PG 2209+184 and PG 2214+139 are (303;0.80) and (15.60;0.09) respectively. The peak contour flux for PG 2304+042 is 84.5 mJy beam$^{-1}$ and the contour levels are 0.35 $\times$ (-1, 1, 1.4, 2, 2.8, 4, 5.6, 8, 11.20, 16, 23, 32, 45, 64, 90, 128, 180, 256, 362, 512) mJy beam$^{-1}$.}
\end{figure*}

\bsp	
\label{lastpage}
\end{document}